\documentclass[epj]{svjour}
\usepackage{graphicx}
\usepackage{color}

\def\mb#1{\setbox0=\hbox{$#1$}\kern-.025em\copy0\kern-\wd0
\kern-0.05em\copy0\kern-\wd0\kern-.025em\raise.0233em\box0}

\begin{document}
   \title{Gravitational phase transitions with an exclusion constraint in position space}

 \author{Pierre-Henri Chavanis}

\institute{Laboratoire de Physique Th\'eorique (IRSAMC), CNRS and UPS, Universit\'e de Toulouse, F-31062 Toulouse, France\\
\email{chavanis@irsamc.ups-tlse.fr}
}

\titlerunning{Gravitational phase transitions with an exclusion constraint in position space}

   \date{To be included later }

   \abstract{We discuss the statistical mechanics of a system of
self-gravitating particles with an exclusion constraint in position space in a
space of dimension $d$. The exclusion constraint puts an upper bound on the
density of the system and can stabilize it against gravitational collapse. We
plot the caloric curves giving the temperature as a function of the energy and
investigate the nature of phase transitions as a function of the size of the
system and of the dimension of space in both microcanonical and canonical
ensembles. We consider stable and metastable states and emphasize the importance
of the latter for systems with long-range interactions. For $d\le 2$, there is
no phase transition. For $d>2$, phase transitions can take place between a
``gaseous'' phase  unaffected by the exclusion constraint and  a ``condensed''
phase dominated by this constraint. The condensed configurations
have a core-halo structure made of a ``rocky core'' surrounded by an
``atmosphere'', similar to a giant gaseous planet. For large systems there
exist microcanonical and canonical first order phase transitions. For
intermediate systems, only canonical first order phase transitions are present.
For small systems there is no phase transition at all. As a result, the phase
diagram exhibits two critical points, one in each ensemble. There also exist a
region of negative specific heats and a situation of ensemble inequivalence for
sufficiently large systems. We show that a statistical equilibrium state exists
for any values of energy and temperature in any dimension of space. This differs
from the case of the self-gravitating Fermi gas for which there is no
statistical equilibrium state at low energies and low temperatures when $d\ge 4$. By a
proper interpretation of the parameters, our results have application for the
chemotaxis of bacterial populations in biology described by a generalized
Keller-Segel model including an exclusion constraint in position space. They
also describe colloids at a fluid interface driven by attractive capillary
interactions when there is an excluded volume around the particles. Connexions
with two-dimensional turbulence are also mentioned.
 }

   \maketitle
%
%________________________________________________________________

\section{Introduction}
\label{sec_introduction}

The statistical mechanics of systems with long-range interactions is
currently a topic of active research \cite{houches,cdr}. Among long-range
interactions, the gravitational force plays a fundamental
role. As a result, the development of a statistical mechanics for
self-gravitating systems is of considerable interest. However, it is well-known that the gravitational interaction poses a lot of difficulties due to its long-range nature and its divergence at short distances. Therefore, the statistical mechanics of self-gravitating systems is very particular and must be addressed carefully \cite{paddy,ijmpb}.

Let us first consider a system of classical point masses in 
gravitational interaction in a space of dimension $d=3$. It is well-known that
no statistical equilibrium state exists in an unbounded domain. Indeed, the
density of states in the microcanonical ensemble or the partition function in
the canonical ensemble diverge because of the decreasing behavior of the
gravitational potential $u\propto 1/r$ at large distances $r\rightarrow +\infty$
and the fact that the accessible volume is infinite \cite{paddy}. Equivalently,
there is no maximum entropy state at fixed mass  and energy in the
microcanonical ensemble, and there is no minimum of  free energy at fixed mass
in the canonical ensemble.\footnote{We are referring here to the Boltzmann
entropy functional $S[f]$ and to the Boltzmann free energy functional
$F[f]=E[f]-TS[f]$ where $f({\bf r},{\bf v})$ is the distribution function (see
\cite{ijmpb} for details). We also recall that the microcanonical ensemble
describes an isolated Hamiltonian system (fixed energy) like a stellar system
\cite{paddy} while the canonical ensemble describes a dissipative system (fixed
temperature) like a self-gravitating Brownian gas \cite{sc}.} One can always
increase the entropy or decrease the free energy by spreading the system (see
Appendices A and B of \cite{sc}). There are not  even critical points of entropy
or free energy in an unbounded domain: the Boltzmann distribution coupled to the
Poisson equation has infinite mass \cite{paddy,ijmpb}. This absence of
equilibrium is related to the natural tendency of self-gravitating systems to
{\it evaporate} \cite{bt}. As a result, the statistical mechanics of
self-gravitating systems is essentially an out-of-equilibrium problem that must
be addressed with kinetic theories \cite{aakin}. However, evaporation is a very
slow process so that, on intermediate timescales, self-gravitating systems may
be considered as ``confined'' in a finite region of space. Furthermore, stellar
systems like globular clusters are never totally isolated from the surrounding.
In practice, they feel the tides of a nearby galaxy \cite{bt}. Globular clusters
are usually found in quasi stationary states described by the Michie-King model
which is a truncated isothermal distribution.\footnote{These  quasi stationary
states, reached by ``collisional'' stellar systems like globular clusters, are
physically distinct from the  quasi stationary states reached by
``collisionless'' stellar systems like elliptical galaxies described by the
Vlasov equation. In the first case, the distribution function (Michie-King
model) results from the competition between close encounters, trying to
establish an isothermal distribution, and the effect of evaporation
\cite{michie,king}. In the second case, the distribution is the result of an
incomplete violent relaxation \cite{lb}.} As a result, their density profile
vanishes at a finite radius $R$ (tidal radius) above which the stars are lost by
the system. These considerations are a motivation to consider the statistical
mechanics of self-gravitating systems enclosed within a ``box'', where the box
radius mimics  the tidal radius of more realistic systems.\footnote{The box is
also equivalent to a confining external potential that the system could
experience due to its interaction with other objects.} It is only with this
artifice that a rigorous statistical mechanics of self-gravitating systems can
be developed.

\begin{figure}
\begin{center}
\includegraphics[clip,scale=0.3]{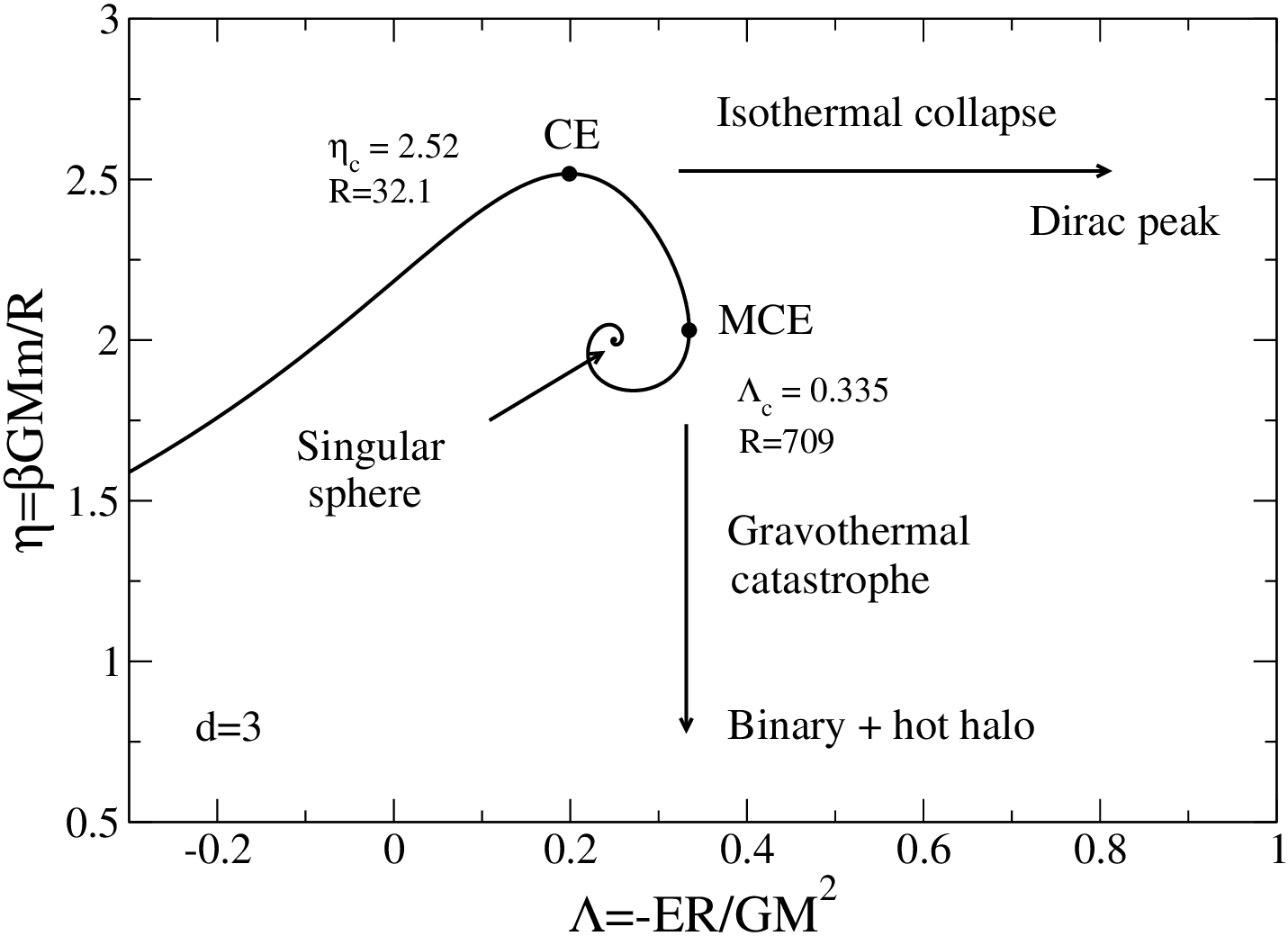}
\caption{Series of equilibria of isothermal spheres in $d=3$. It has a snail-like structure.}
\label{dim3}
\end{center}
\end{figure}

If we confine the system within a spherical box of radius $R$ in order to prevent evaporation, the following results are found \cite{paddy,ijmpb}. The density of states and the partition function diverge because of the singularity of the gravitational potential $u\propto 1/r$ at short distances $r\rightarrow 0$ and the fact that we can approach the particles at arbitrarily close distance \cite{paddy}. Equivalently, there is no global entropy maximum at fixed mass and energy  in the microcanonical ensemble, and there is no global minimum of  free energy  at fixed mass in the canonical ensemble. In the microcanonical ensemble, one can always increase the entropy by
forming a ``binary star surrounded by a hot halo'' and  by making the binary star tighter and tighter,
and the halo hotter and hotter in order to conserve the total energy (see
Appendix A of \cite{sc}). In the canonical ensemble, one can always decrease the
free energy by concentrating all the particles at the same point; the free
energy diverges when a ``Dirac peak'' containing all the particles is formed
(see Appendix B of \cite{sc}). This absence of equilibrium is related to the
natural tendency of self-gravitating systems to {\it collapse} \cite{bt}.
However, above  a critical energy $E_c=-0.335GM^2/R$ (Antonov energy)
\cite{antonov} in the microcanonical ensemble and above a critical temperature
$T_c=GMm/(2.52Rk_B)$ (Emden temperature) \cite{emden} in the canonical ensemble,
the system may be found in long-lived ``gaseous'' {\it metastable} states that
are local maxima of entropy at fixed mass and energy or local minima of free
energy at fixed mass. The series of equilibria of isothermal spheres has the
form of a spiral (see Fig. \ref{dim3}) but only the part of the branch up to CE
in the canonical ensemble and up to  MCE in the microcanonical ensemble is
stable (the region between CE and MCE where the specific heat is negative
corresponds to a region of ensemble inequivalence).  The lifetime of these
``gaseous'' metastable states is considerable since it scales as $e^N$ (except
close to the critical point) \cite{ijmpb,metastable} so these metastable states
are fully stable in practice.\footnote{This result has important applications in
astrophysics where the number of stars in a globular cluster is of the order of
$N\sim 10^6$ \cite{bt}. In that case, the lifetime of the ``gaseous'' metastable
states is of the order of $e^{10^6}$ dynamical times! This is so considerable
that these metastable states can be considered as fully stable states.  As a
result, self-gravitating systems at sufficiently high energies and high
temperatures are found in long-lived ``gaseous'' metastable states although
there exist structures with higher entropy or lower free energy. The probability
to form a ``binary $+$ a hot halo'' ($S\rightarrow +\infty$) in the
microcanonical ensemble or a ``Dirac peak'' ($F\rightarrow -\infty$) in the
canonical ensemble requires very particular correlations and 
takes too much time (see Sec. \ref{sec_lifetime}). Of course,
the lifetime of real globular clusters (that are not in boxes!) is ultimately
controlled by the evaporation time \cite{bt}.} Therefore, a self-gravitating
system in a box can reach a statistical equilibrium state described by the mean
field Maxwell-Boltzmann distribution at sufficiently high energies and at
sufficiently high temperatures, even if there is no statistical equilibrium
state in a strict sense \cite{paddy,ijmpb}.

However, below $E_c$ or $T_c$, there is no metastable state anymore,
and the system collapses. This is called {\it gravothermal catastrophe}
\cite{lbw} in the microcanonical ensemble and {\it isothermal collapse}
\cite{aa} in the canonical ensemble. In the microcanonical ensemble, the
gravothermal catastrophe leads to a ``binary star surrounded by a hot halo'' \cite{cohn}.
In the canonical ensemble, the  isothermal collapse leads to a ``Dirac peak'' containing all the particles \cite{post}. Therefore, the result of the
gravitational collapse is to form a singularity: a ``binary star $+$ hot halo''
in the microcanonical ensemble and a ``Dirac peak'' in the canonical ensemble
\cite{ijmpb}. However, in reality, some microscopic constraints will come into
play when the system becomes dense enough and they will prevent the formation of
these singularities. For example, if the particles are hard spheres (e.g., atoms
in a gas or dust particles
in the solar nebula), they cannot
be compressed indefinitely and a maximum density in physical space will be
reached when all the particles are packed together. Even if the particles are
point-like (e.g., electrons in white dwarf stars, neutrons in neutron stars,
massive neutrinos in dark matter halos...), quantum mechanics will put an upper
bound on the density in phase space on account of the Pauli exclusion principle
for fermions. In that case, the singularity (binary star or Dirac peak)
will be smoothed-out and replaced by
a compact object: a ``rocky core'' (proto-planet)
for hard spheres or a ``white dwarf'' for
fermions.\footnote{As first understood by
Fowler \cite{fowler} in his classical theory of white dwarf stars, the quantum pressure of the electrons is able to stabilize the star against gravitational
collapse.} At non-zero temperatures, this compact object will be surrounded by a
dilute atmosphere (vapor) so that the whole
configuration has a ``core-halo'' structure. For hard spheres,
this structure resembles a giant gaseous planet, and for fermions it resembles a
white dwarf star (with an envelope) or a dark matter halo. Therefore, when a
small-scale regularization is properly accounted for, the system
is stabilized and there exist an equilibrium state (global maximum of entropy
or global minimum of free energy) for each value of accessible energy and
temperature.   This can lead to phase transitions between a ``gaseous phase''
(unaffected by the small-scale regularization) at high energies and high
temperatures and a ``condensed phase'' (strongly dependent on the small-scale
regularization) at low energies and low temperatures.

These phase transitions have been analyzed in detail for a hard spheres gas \cite{aronson,stahl,paddy,pt,champion}, for a regularized potential of interaction \cite{follana,ispolatov,destri,nardini}, and for the self-gravitating Fermi gas \cite{htmath,ht,bilic,pt,chavcap,ispolatov,rieutord,fermionsd} (see a review in \cite{ijmpb}). The nature of the phase transitions depends on the size of the system $R$ relative to the size $R_*$ of the compact object. For large systems there exist microcanonical and canonical first order phase transitions. For intermediate systems, only canonical first order phase transitions are present. For small systems there is no phase transition at all. As a result, the phase diagram exhibits two critical points, one in each ensemble. There also exist a region of negative specific heats and a situation of ensemble inequivalence for sufficiently large systems. Although the details of these phase transitions depend on the specific form of the small-scale regularization (hard spheres, fermions,...), the phenomenology is expected to be relatively universal.

\begin{figure}
\begin{center}
\includegraphics[clip,scale=0.3]{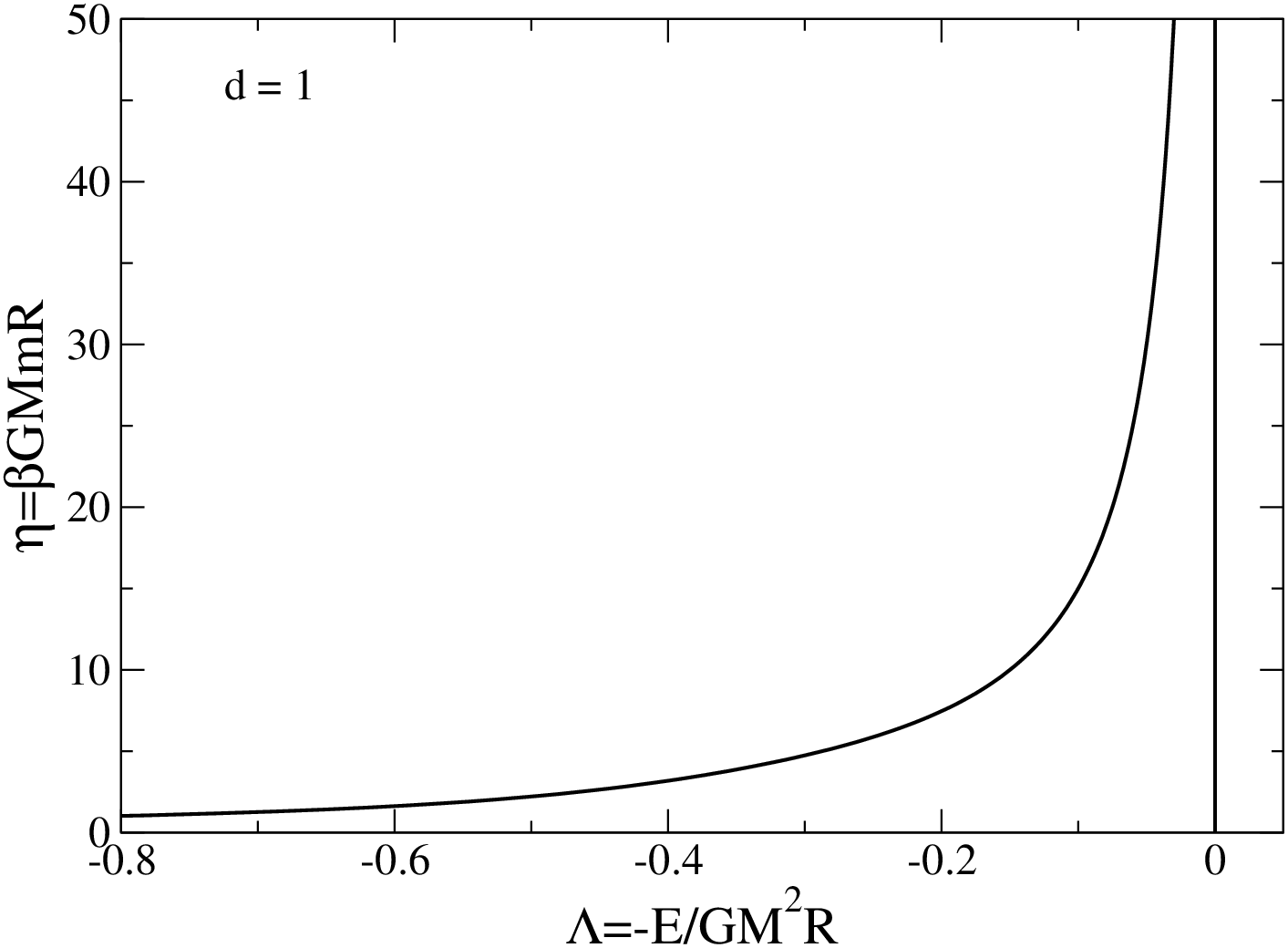}
\caption{Series of equilibria of isothermal spheres in $d=1$.}
\label{dim1}
\end{center}
\end{figure}

\begin{figure}
\begin{center}
\includegraphics[clip,scale=0.3]{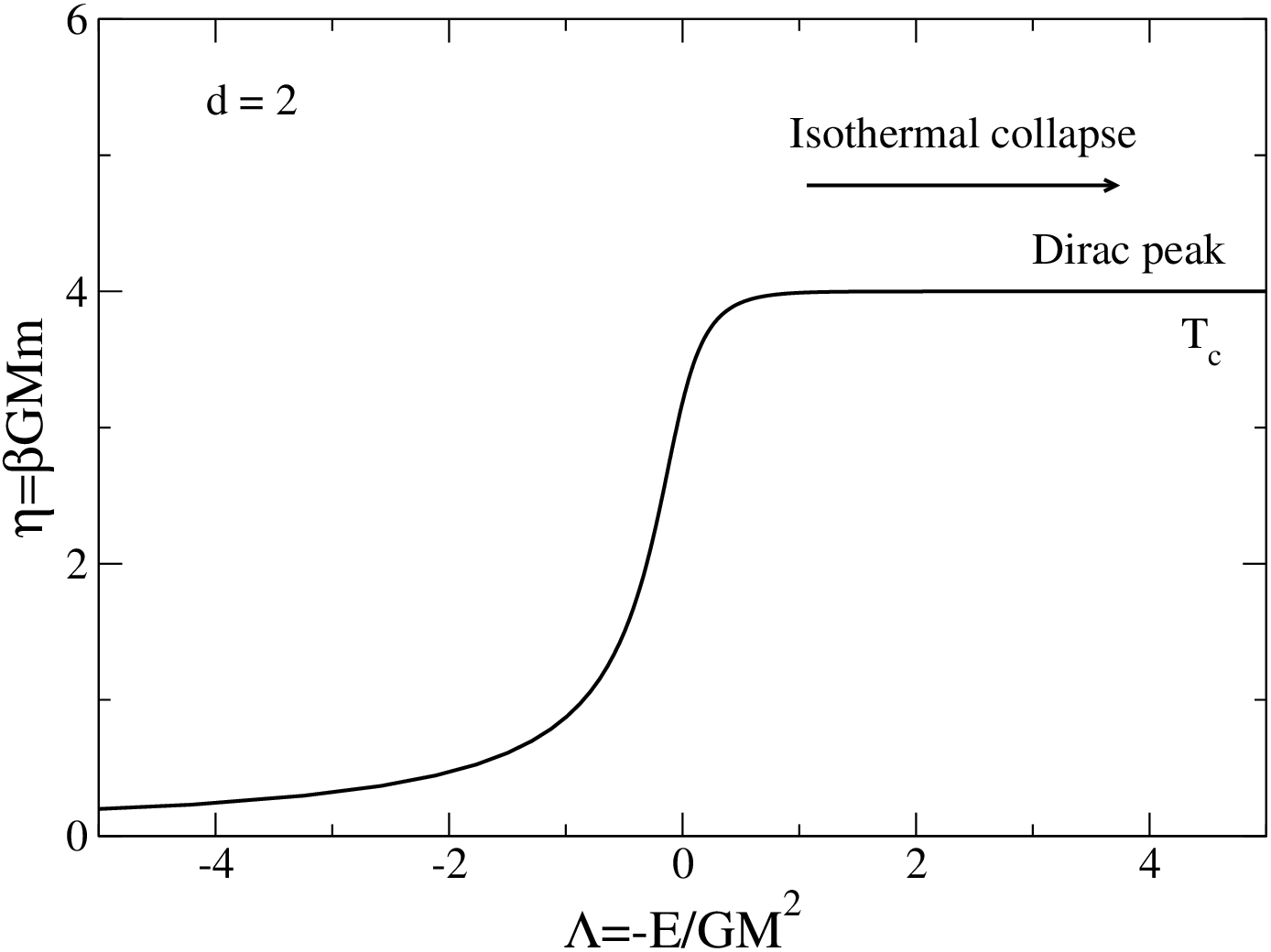}
\caption{Series of equilibria of isothermal spheres in $d=2$. It forms a plateau
at low energies.}
\label{dim2}
\end{center}
\end{figure}

\begin{figure}
\begin{center}
\includegraphics[clip,scale=0.3]{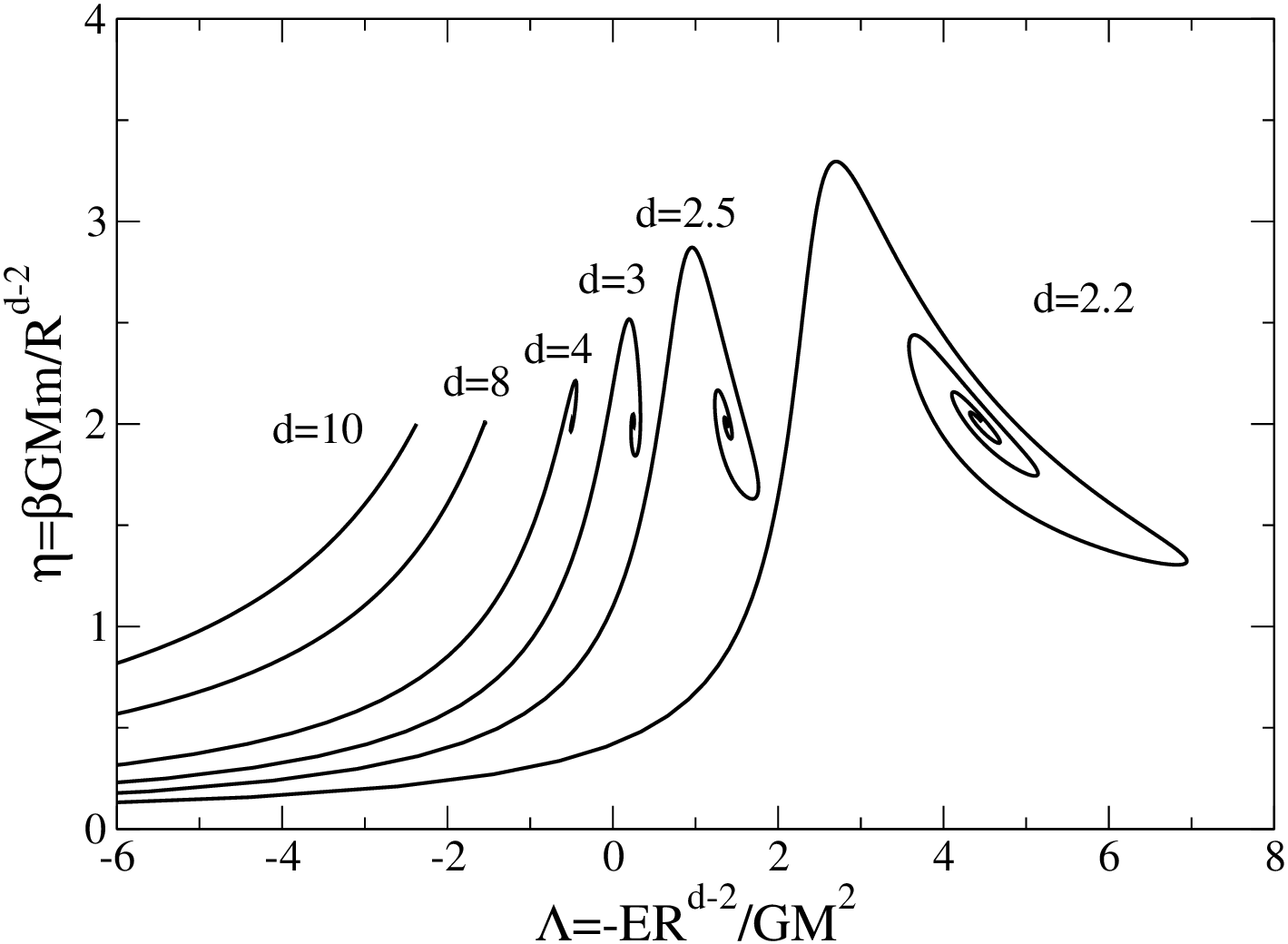}
\caption{Series of equilibria of isothermal spheres in different dimensions of
space. The spiral reduces to a point for $d\ge 10$.}
\label{calo2a10}
\end{center}
\end{figure}

It is interesting to study how these results depend on the dimension of space $d$ \cite{sc,wdd,fermionsd,cras}. We first consider classical point masses. In $d=1$, the caloric curve is monotonic and there exist an equilibrium state for all accessible values of energy ($E\ge 0$) and temperature (see Fig. \ref{dim1}). At $E=T=0$, the equilibrium state is a Dirac peak. The dimension $d=2$ is special. The caloric curve forms of a plateau (see Fig. \ref{dim2}). There exist an equilibrium state (global entropy maximum) for any value of energy in the microcanonical ensemble but an equilibrium state (global minimum of free energy) exists in the canonical ensemble only above a critical temperature $T_c=GMm/(4k_B)$ (see, e.g., \cite{sc,virialIJMPB}). Below  $T_c$, there is no critical point of free energy at fixed mass and the system undergoes an isothermal collapse leading to a Dirac peak containing all the particles. For $2<d<10$, the series of equilibria forms a spiral. The system undergoes a gravothermal catastrophe in the microcanonical ensemble and an isothermal collapse in the canonical ensemble. For $d\ge 10$ the spiral disappears \cite{sc} so the points of minimum energy $E_c$ and minimum temperature $T_c$ coincide (see Fig. \ref{calo2a10}). The effect of a small-scale regularization depends on the dimension of space. The case of self-gravitating fermions in different dimensions of space has been investigated specifically in \cite{wdd,fermionsd}. It is found that quantum mechanics stabilizes the system against gravitational collapse for $d<4$ but this is no more true for $d\ge 4$ at low energies and low temperatures. In other words, classical white dwarf stars (or dark matter halos made of massive neutrinos) would be unstable in a space of dimension $d\ge 4$!

The same kind of considerations apply to the chemotaxis of bacterial populations
in biology described by the Keller-Segel model \cite{ks}. In this
model, the bacteria experience a Brownian motion (diffusion) but they also
secrete a chemical substance (a pheromone) and are collectively attracted by it.
It turns out that this long-range interaction is similar to the gravitational
interaction in astrophysics. Indeed, the Keller-Segel model shows deep analogies
with the dynamics of self-gravitating Brownian particles described by the
Smoluchowski-Poisson system in the canonical ensemble \cite{tcrit}. In the
biological context, the consideration of a ``box'' in which the bacteria live is
completely justified (more than in astrophysics). For small values of the
control parameter (equivalent to an effective temperature)  in $d\ge 2$ the
Keller-Segel equations do not reach an equilibrium state and blow up. This leads
to Dirac peaks. Here again, we expect that these singularities will be
smoothed-out by small-scale constraints. Indeed, some generalized versions of
the Keller-Segel model have been proposed which include such regularization and
prevent the spatial density to diverge (see \cite{nfp} and references therein). One of the simplest
regularization is to assume that there is an exclusion constraint in position
space due to finite size effects. This is similar to the Pauli exclusion
principle for fermions except that
it occurs in position space instead of phase space. In other words, it puts a
bound on the spatial density ($\rho({\bf r})\le \sigma_0$) while the Pauli
exclusion principle puts a bound on the distribution function ($f({\bf r},{\bf
v})\le \eta_0$). A similar exclusion constraint occurs in the
 Miller-Robert-Sommeria (MRS) statistical theory of
2D turbulence \cite{miller,rs} whose aim is to describe large-scale
vortices such as Jupiter's great red spot. As a consequence of the 2D Euler
equation, the coarse-grained vorticity is necessarily smaller than its maximum
initial value leading to the constaint $\overline{\omega}({\bf r})\le
\sigma_0$. Finally, colloids
at a fluid interface driven by attractive capillary interactions present
analogies with self-gravitating Brownian particles \cite{dominguez}. Therefore, 
they can experience a form of gravitational collapse. However, if there
is an excluded volume around the particles the density must satisfy the bound
$\rho({\bf r})\le \sigma_0$. These analogies prompt us to consider the
case of self-gravitating particles with an exclusion constraint in position
space. This problem is interesting in astrophysics because it provides a simple
model of ``hard spheres'' with potential application to the formation of
planets (a more elaborated model of planet formation, using more
relevant equations of state, has been developed in \cite{stahl}).\footnote{In that
context, the radius of the ``box'' could represent the
Hill radius that accounts for the tidal effect of the Sun on the atmosphere of a
gaseous planet.} On the other
hand, with a proper reinterpretation of the parameters, this model also applies
to the chemotaxis of bacterial populations, to the formation of large-scale
vortices in 2D turbulence, and to colloids at a fluid interface. It is therefore
desirable to study this model in detail and describe the corresponding phase
transitions.

The paper is organized as follows. In Sec. \ref{sec_exclusion}, we
determine the statistical equilibrium state of a self-gravitating system
with an exclusion constraint in position space. The Fermi-Dirac entropy
in position space is introduced from a combinatorial analysis.
We derive the Fermi-Dirac distribution in position space and consider
asymptotic limits corresponding to the classical self-gravitating gas (isothermal sphere)
and to a completely degenerate structure (homogeneous sphere). We discuss the proper
thermodynamic limit of the self-gravitating gas with a small-scale
 regularization in position space. In Sec. \ref{sec_cc3}, we describe the phase transitions in
$d=3$. We emphasize the importance of metastable states in systems with long-range interactions.
In Sec. \ref{sec_ccother}, we consider other dimensions of space and show that no phase transition
 occurs in $d=1$ and $d=2$ dimensions. In Sec. \ref{sec_kinetic} we present dynamical models of self-gravitating systems including an  exclusion constraint in position space. These models can be used to describe hysteresis cycles and random transitions between gaseous and condensed states associated with first order phase transitions. These kinetic equations can also provide generalized Keller-Segel models of chemotaxis, and generalized models of colloidal suspensions with long-range interactions.

\section{Thermodynamics of self-gravitating systems with an exclusion constraint in position space}
\label{sec_exclusion}

\subsection{The Fermi-Dirac entropy}
\label{sec_fde}

We consider a system of $N$ particles interacting via Newtonian gravity
in a space of dimension $d$.  Let $\rho({\bf r},t)$ denote the smooth spatial density
of the system, i.e. $\rho({\bf r},t)d{\bf r}$
gives the mass of particles whose positions are in the
cell $({\bf r},{\bf r}+d{\bf r})$ at
time $t$. We assume that these particles are subjected to an
``exclusion constraint'' in position space (but not in velocity space). Specifically, we assume that
there cannot be more than one particle in a region of size $a$, where $a$ may be identified with the effective ``size'' of the particles. This may be viewed as a heuristic attempt to account for steric hindrance or finite size effects. This is also a very simple model of hard spheres.
To determine the statistical equilibrium state of the system (most probable macrostate) in the microcanonical and canonical ensembles, we proceed as explained in \cite{emergence}. We first have to determine the {\it a priori} probability of the macrostate $\rho({\bf r})$ and the corresponding entropy $S_0[\rho]$. To that purpose, we use  a combinatorial analysis which respects the exclusion contraint in position space.

We divide the domain into a very large number of microcells with size
$h$. We assume that the size $h$ is of the order of the size $a$ of a
particle so that a microcell is occupied either by $0$ or $1$
particle. This is how the exclusion constraint is introduced in our model.
This is similar to the Pauli exclusion principle for fermions except that it acts
in position space instead of phase space. We now group these microcells into macrocells each of
which contains many microcells but remains nevertheless small compared
to the spatial extension of the whole system. We call $\nu$ the number
of microcells in a macrocell. Consider the configuration $\lbrace n_i
\rbrace$ where there are $n_1$ particles in the $1^{\rm st}$
macrocell, $n_2$ in the $2^{\rm nd}$ macrocell etc..., each occupying
one of the $\nu$ microcells with no cohabitation. The number of ways
of assigning a microcell to the first element of a macrocell is $\nu$,
to the second $\nu -1$ etc. Assuming that the particles are
indistinguishable, the number of ways of assigning microcells to all
$n_i$ particles in a macrocell is thus
\begin{equation}
{1\over n_i!}{\times} {\nu!\over (\nu-n_i)!}. \label{phen1}
\end{equation}
To obtain the number of microstates corresponding to the macrostate
$\lbrace n_i \rbrace$ defined by the number of particles $n_i$ in each
macrocell (irrespective of their precise position in the cell), we
need to take the product of terms such as (\ref{phen1}) over all
macrocells. Thus, the number of microstates corresponding to the
macrostate $\lbrace n_i \rbrace$, which is proportional to the {\it a
priori} probability of the state $\lbrace n_i \rbrace$, is
\begin{equation}
W(\lbrace n_i \rbrace)=\prod_i {\nu!\over n_i!(\nu-n_i)!}.
\label{phen2}
\end{equation}
This is the Fermi-Dirac statistics in physical
space. As is customary, we define the entropy of the state $\lbrace
n_i \rbrace$ by
\begin{equation}
S_0(\lbrace n_i \rbrace)=k_B\ln W(\lbrace n_i \rbrace). \label{phen3}
\end{equation}
It is convenient here to return to a representation in terms of
the density in the
$i$-th macrocell
\begin{equation}
\rho_i=\rho({\bf r}_i)={n_i \ m\over \nu h^d}={n_i\sigma_0\over \nu}, \label{phen4}
\end{equation}
where we have defined $\sigma_0=m/h^d$, which represents the
maximum value of $\rho$ due to the exclusion constraint. Now,
using the Stirling formula, we have
\begin{eqnarray}
\ln W(\lbrace n_i \rbrace)\simeq \sum_i \nu
(\ln\nu-1)-\nu\biggl\lbrace {\rho_i\over \sigma_0}\biggl\lbrack
\ln\biggl ({\nu \rho_i\over \sigma_0}\biggr )-1\biggr\rbrack \nonumber\\
 +\biggl
(1-{\rho_i\over \sigma_0}\biggr )\biggl\lbrack\ln\biggl\lbrace
\nu\biggl (1-{\rho_i\over \sigma_0}\biggr
)\biggr\rbrace-1\biggr\rbrack\biggr\rbrace.\qquad  \label{phen5}
\end{eqnarray}
Passing to the continuum limit $\nu\rightarrow 0$, we obtain the
usual expression of the Fermi-Dirac entropy in position space
\begin{eqnarray}
S_0[\rho]=-k_B\frac{\sigma_0}{m}\int \biggl\lbrace \frac{\rho}{\sigma_0}\ln\left (\frac{\rho}{\sigma_0}\right )\nonumber\\
+\left (1-\frac{\rho}{\sigma_0}\right )\ln \left (1-\frac{\rho}{\sigma_0}\right )\biggr\rbrace\, d{\bf r}.\label{phen6}
\end{eqnarray}
In the dilute limit $\rho\ll \sigma_{0}$, it reduces to the
Boltzmann entropy
\begin{equation}
S_0=-k_B\int {\rho\over m}\biggl\lbrack \ln\biggl ({\rho\over \sigma_0}\biggr )-1\biggr \rbrack\,  d{\bf r}.
\label{phen7}
\end{equation}

The total mass and the total energy of the system in the mean field approximation are given by \cite{emergence}:
\begin{equation}
M=\int \rho \, d{\bf r},
\label{phen8}
\end{equation}
\begin{equation}
E=\frac{d}{2}Nk_B T+\frac{1}{2}\int\rho\Phi \, d{\bf r},
\label{phen9}
\end{equation}
where  $T$ is the temperature and $\Phi$ is the  gravitational potential. It is related to the density by the Poisson equation
\begin{equation}
\Delta\Phi=S_{d} G\rho,
\label{phen10}
\end{equation}
where $S_{d}=2\pi^{d/2}/\Gamma(d/2)$ is the surface of a unit
sphere in a space of dimension $d$ and $G$ is the constant of gravity
(which depends on the dimension of space). Finally, the entropy taking into account the exclusion constraint in position space is \cite{emergence}:
\begin{eqnarray}
\label{phen11}
S[\rho]=\frac{d}{2}Nk_B+\frac{d}{2}Nk_B\ln \left (\frac{2\pi k_B T}{m}\right )\nonumber\\
-k_B\frac{\sigma_0}{m}\int \biggl\lbrace \frac{\rho}{\sigma_0}\ln\left (\frac{\rho}{\sigma_0}\right )
+\left (1-\frac{\rho}{\sigma_0}\right )\ln \left (1-\frac{\rho}{\sigma_0}\right )\biggr\rbrace\, d{\bf r}.\nonumber\\
\end{eqnarray}
This Fermi-Dirac entropy can be interpreted as a generalized entropy in the framework of generalized thermodynamics \cite{nfp,emergence}. Generalized entropies arise when microscopic constraints act on the system and change the number of microstates associated with a given macrostate.

\subsection{The Fermi-Dirac distribution}
\label{sec_fdd}

Now that the entropy has been precisely justified from a combinatorial analysis, the statistical
equilibrium state (most probable macrostate) of the system in the microcanonical ensemble is obtained by maximizing the Fermi-Dirac entropy (\ref{phen11}) at fixed mass (\ref{phen8}) and energy (\ref{phen9}). Actually, the energy constraint (\ref{phen9}) can be used to express the temperature as a functional of the density. As a result, the entropy (\ref{phen11}) is a functional of the density $\rho$ and the equilibrium state is obtained by solving the maximization problem \cite{emergence}:
\begin{eqnarray}
S(E)=\max_{\rho}\lbrace S[\rho]\, |\, M[\rho]=M\rbrace.
\label{fdd1}
\end{eqnarray}
The critical points are determined by
\begin{equation}
\delta S/k_B-\alpha\delta M=0, \label{fdd2}
\end{equation}
where $\alpha$ is a Lagrange multiplier (chemical potential) associated with the conservation of mass (the conservation of energy has been taken into account in the expression of the entropy).  They lead to the Fermi-Dirac distribution in position space
\begin{eqnarray}
\label{fdd3}
\rho({\bf r})=\frac{\sigma_0}{1+\lambda e^{\beta m\Phi({\bf r})}},
\end{eqnarray}
where $\lambda=e^{m\alpha}$ is a strictly positive constant
(inverse fugacity) and $\beta=1/k_B T$ is the inverse
temperature. Clearly, the density satisfies $\rho({\bf r})\le \sigma_{0}$ which is a
consequence of the exclusion principle in position space. In the dilute limit
$\rho({\bf r})\ll\sigma_{0}$, the distribution function (\ref{fdd3}) reduces to the
Boltzmann formula
\begin{equation}
\rho({\bf r})={\sigma_{0}\over\lambda}e^{-\beta m\Phi({\bf r})}.
\label{fdd4}
\end{equation}

So far, we have assumed that the system is isolated so that the energy $E$
is conserved (microcanonical ensemble). If now the system is in contact with a thermal bath
fixing its temperature $T$, like for a Brownian gas, the statistical
equilibrium state (most probable macrostate) in the canonical ensemble is obtained by minimizing
the free energy
\begin{eqnarray}
\label{fdd5}
F[\rho]=E[\rho]-T S[\rho]
\end{eqnarray}
at fixed mass (\ref{phen8}). The equilibrium
state is obtained by solving the minimization problem \cite{emergence}:
\begin{eqnarray}
F(T)=\min_{\rho}\lbrace F[\rho]\, |\, M[\rho]=M\rbrace.
\label{fdd6}
\end{eqnarray}
The critical points are determined by
\begin{equation}
\delta F+k_B T \alpha\delta M=0, \label{fdd7}
\end{equation}
where $\alpha$ is a Lagrange multiplier (chemical potential) associated with the conservation of mass.  They lead to the Fermi-Dirac distribution in position space (\ref{fdd3}) as in the microcanonical ensemble. Therefore, the critical
points (first order variations) of the variational problems (\ref{fdd1}) and
(\ref{fdd6}) are the same. However, the stability of the system
(regarding the second order variations) can be different in the microcanonical
and canonical ensembles. When this happens, we speak of a situation of
{\it ensemble inequivalence} \cite{ellis,bb}. The set of solutions of (\ref{fdd1})
may not coincide with the set of solutions of (\ref{fdd6}). It can be shown that
a solution of a variational problem is always the solution of a more constrained
dual variational problem \cite{ellis}. Therefore, a solution of (\ref{fdd6})
with a given temperature $T$ is always a solution of (\ref{fdd1}) with the
corresponding energy $E$. Canonical stability implies microcanonical stability:
$(\ref{fdd6}) \Rightarrow (\ref{fdd1})$. However, the converse is wrong: a
solution of (\ref{fdd1}) is not necessarily a solution of (\ref{fdd6}).\footnote{For example,
the specific heat is necessarily positive in the canonical ensemble since it
measures the fluctuations of energy through the relation
$C=dE/dT=k_B\beta^2(\langle E^2\rangle-\langle E\rangle^2)\ge 0$, while it can
be positive or negative in the microcanonical ensemble. Therefore, a negative
specific heat is a sufficient condition of canonical instability. We note,
however, that it is not a necessary condition of canonical instability.} Ensemble inequivalence is generic for systems with long-range interactions but it is not compulsory. The stability
of the system can be determined by simply plotting the series of equilibria $\beta(E)$ and using the Poincar\'e theory on the linear series of equilibria (see \cite{poincare} and its application to self-gravitating systems \cite{lbw,katz,katz2,ijmpb}).

{\it Remark:} For isolated systems with long-range interactions that evolve at fixed energy, the proper statistical ensemble is the microcanonical ensemble. Since the energy is non-additive, the canonical ensemble is not physically justified to describe a subpart of the system (contrary to systems with short-range interactions) \cite{cdr}. However, since the canonical variational problem (\ref{fdd6}) provides a {\it sufficient} condition of microcanonical stability it can be useful in that respect. Indeed, if we can show that the system is canonically stable, then it is granted to be microcanonically stable. Therefore, we can start by studying the canonical stability problem, which is simpler, and consider the microcanonical stability problem only if the canonical ensemble does not cover the whole range of energies. On the other hand, the canonical ensemble is rigorously justified for dissipative systems with long-range interactions in contact with a thermal bath of another origin, like in the model of self-gravitating Brownian particles \cite{sc}.

\subsection{Equation of state}
\label{sec_eos}

The Fermi-Dirac distribution in position space (\ref{fdd3}) is equivalent to the condition of hydrostatic equilibrium
\begin{eqnarray}
\label{eos1}
\nabla p+\rho\nabla\Phi={\bf 0}
\end{eqnarray}
for a barotropic equation of state \cite{nfp,emergence}:
\begin{eqnarray}
\label{eos2}
p(\rho)=-\frac{k_B T}{m}\sigma_0\ln (1-\rho/\sigma_0).
\end{eqnarray}
Of course, the pressure becomes infinite when the density $\rho$ approaches the maximum value $\sigma_0$ (corresponding to a ``packing'' of the particles). In the dilute limit $\rho\ll\sigma_0$, Eq. (\ref{eos2}) reduces to the isothermal equation of state
\begin{equation}
p(\rho)=\rho\frac{k_B T}{m}
\label{eos3}
\end{equation}
which is associated with the Boltzmann distribution (\ref{fdd4}).

\subsection{Thermodynamical parameters}
\label{sec_para}

The density of particles is related to the gravitational potential by Eq. (\ref{fdd3}).
The gravitational potential is now obtained by substituting
Eq. (\ref{fdd3}) in the Poisson equation (\ref{phen10}). We assume spherical symmetry since maximum entropy states are necessarily spherically symmetric for non-rotating systems. We introduce the
rescaled distance $\xi=(S_d G\sigma_0\beta m)^{1/2}r$ and the variables $\psi=\beta m
(\Phi-\Phi_{0})$ and $k=\lambda e^{\beta m \Phi_{0}}$, where
$\Phi_{0}$ is the central potential. This leads to the equation
\begin{equation}
{1\over\xi^{d-1}}{d\over d\xi}\biggl (\xi^{d-1}{d\psi\over d\xi}\biggr )=\frac{1}{1+k e^{\psi}},
\label{p1}
\end{equation}
\begin{equation}
\psi(0)=\psi'(0)=0
\label{p2}
\end{equation}
which may be interpreted as the Emden equation \cite{chandra} with an exclusion constraint in position space. The density profile is given by
\begin{equation}
\rho(\xi)=\frac{\sigma_0}{1+k e^{\psi(\xi)}}.
\label{p3}
\end{equation}

As explained in the Introduction, we need to confine the system within a spherical box of radius $R$ in order to prevent evaporation and correctly define a statistical equilibrium state (the box typically represents the
size of the cluster under consideration).  In that case, the solution
of Eqs. (\ref{p1})-(\ref{p2}) is terminated by the box at the normalized radius
\begin{equation}
\alpha=(S_d G\sigma_0\beta m)^{1/2}R.
\label{p4}
\end{equation}
For a spherically symmetric configuration, the Gauss theorem can be
written as
\begin{equation}
{d\Phi\over dr}={GM(r)\over r^{d-1}},
\label{p5}
\end{equation}
where $M(r)=\int_{0}^{r}\rho S_{d}r^{d-1}dr$ is the mass within the
sphere of radius $r$. Applying this result at $r=R$ and using the
variables introduced previously we get
\begin{equation}
\eta\equiv {\beta GMm\over R^{d-2}}=\alpha\psi'(\alpha).
\label{p6}
\end{equation}
This equation relates the dimensionless box radius $\alpha$ and the variable $k$ to the dimensionless inverse temperature $\eta$. According to Eqs. (\ref{p4}) and (\ref{p6}), $\alpha$ and $k$ are related to each other by the relation $\alpha^{2}/\eta=d\mu$ or, explicitly,
\begin{equation}
\frac{\alpha}{\psi'(\alpha)}=d\mu,
\label{p7}
\end{equation}
where
\begin{equation}
\mu=\frac{S_d \sigma_0 R^d}{dM}
\label{p8}
\end{equation}
is the degeneracy parameter.  We shall give a physical interpretation of
this parameter in Sec. \ref{sec_mu}.

The energy is given by
\begin{equation}
E=\frac{d}{2}Nk_B T+\frac{1}{2}\int\rho\Phi\, d{\bf r}=K+W,
\label{p9}
\end{equation}
where the first term is the kinetic energy and the second term is the potential energy. Using the Poisson equation (\ref{phen10}) and integrating by parts, the potential energy may be written
as
\begin{equation}
W=\frac{1}{2 S_d G}\left\lbrack\oint \Phi \nabla\Phi\cdot d{\bf S}-\int (\nabla\Phi)^2\, d{\bf r}\right\rbrack.
\label{p10}
\end{equation}
Outside of the box, we have
\begin{equation}
{d\Phi\over dr}={GM\over r^{d-1}}
\label{p11}
\end{equation}
yielding
\begin{equation}
\Phi(r)=-\frac{1}{d-2}\frac{GM}{r^{d-2}},\qquad (d\neq 2),
\label{p12}
\end{equation}
\begin{equation}
\Phi(r)=GM\ln\left (\frac{r}{R}\right ),\qquad (d=2),
\label{p13}
\end{equation}
where we have used ordinary conventions to determine the constant of integration. These relations, applied at $r=R$, allow us to determine the surface term in Eq. (\ref{p10}). Introducing  the dimensionless variables defined previously, and noting that $r=\xi R/\alpha$, we obtain
\begin{equation}
\frac{W R^{d-2}}{GM^{2}}=-\frac{1}{2(d-2)}-\frac{1}{2\eta^2\alpha^{d-2}}\int_{0}^{\alpha}\left (\frac{d\psi}{d\xi}\right )^2\xi^{d-1}\, d\xi
\label{p14}
\end{equation}
in $d\neq 2$ and
\begin{equation}
\frac{W}{GM^{2}}=-\frac{1}{2\eta^2}\int_{0}^{\alpha}\left (\frac{d\psi}{d\xi}\right )^2\xi\, d\xi
\label{p15}
\end{equation}
in $d=2$. The kinetic energy may be written as
\begin{equation}
\frac{KR^{d-2}}{GM^{2}}=\frac{d}{2\eta}.
\label{p15b}
\end{equation}
Therefore, the  total energy is
\begin{eqnarray}
\Lambda\equiv -\frac{E R^{d-2}}{GM^{2}}=-\frac{d}{2\eta}+\frac{1}{2(d-2)}\nonumber\\
+\frac{1}{2\eta^2\alpha^{d-2}}\int_{0}^{\alpha}\left (\frac{d\psi}{d\xi}\right )^2\xi^{d-1}\, d\xi,\quad (d\neq 2),
\label{p16}
\end{eqnarray}
\begin{equation}
\Lambda\equiv -\frac{E R}{GM^{2}}=-\frac{1}{\eta}+\frac{1}{2\eta^2}\int_{0}^{\alpha}\left (\frac{d\psi}{d\xi}\right )^2\xi\, d\xi,\quad (d=2).
\label{p16b}
\end{equation}
For a given value of $\mu$ and $k$, we can
solve the ordinary differential equation (\ref{p1}) with the initial condition (\ref{p2})
until the value of
$\alpha$ at which the condition (\ref{p7}) is satisfied. Then,
Eqs. (\ref{p6}) and (\ref{p16})-(\ref{p16b}) determine the temperature and the
energy of the configuration. By varying the parameter $k$ (for a fixed
value of the degeneracy parameter $\mu$) between $0$ and $+\infty$, we can determine the full
series of equilibria $\beta(E)$.

The entropy of each
configuration is given by (see Appendix \ref{sec_entropy}):
\begin{eqnarray}
\frac{S}{Nk_{B}}=\ln(k)-\frac{d+2}{d}\eta\Lambda+\frac{1}{d-2}\eta+\psi(\alpha)\nonumber\\
+\mu\ln\left \lbrack 1+\frac{1}{k}e^{-\psi(\alpha)}\right \rbrack
-\frac{d}{2}\ln(\eta)-1\nonumber\\
+\frac{d}{2}\ln\left (\frac{2\pi GM}{R^{d-2}}\right ),\qquad (d\neq 2),
\label{p17}
\end{eqnarray}
\begin{eqnarray}
\frac{S}{Nk_{B}}=\ln(k)-2\eta\Lambda+\frac{\eta}{4}+\psi(\alpha)+\mu\ln\left \lbrack 1+\frac{1}{k}e^{-\psi(\alpha)}\right \rbrack\nonumber\\
-\ln(\eta)-1+\ln (2\pi GM),\qquad (d=2),\qquad
\label{p17b}
\end{eqnarray}
and the free energy by
\begin{equation}
F=E-TS.
\label{p18}
\end{equation}
Introducing the pressure at the box $P=p(R)$,
the global equation of state of the self-gravitating gas with an exclusion constraint in position space can be written as
\begin{equation}
{PV\over Nk_{B}T}=\mu\ln\left (1+\frac{1}{k}e^{-\psi(\alpha)}\right ).
\label{p19}
\end{equation}

Before considering the case of an arbitrary degree of degeneracy, it
may be useful to discuss first the non degenerate limit corresponding
to a classical isothermal gas without exclusion constraint, and the completely degenerate limit corresponding to a homogeneous self-gravitating sphere.

\subsection{The non degenerate limit: classical isothermal spheres}
\label{sec_class}

In the dilute limit $\rho\ll\sigma_{0}$, the distribution function (\ref{fdd3}) reduces to the
Boltzmann formula (\ref{fdd4}). Introducing the
rescaled distance $\xi=(S_{d} G\beta m\rho_{0})^{1/2}r$ where $\rho_{0}$ is the central density
and the variable $\psi=\beta m
(\Phi-\Phi_{0})$ where
$\Phi_{0}$ is the central potential, we obtain the Emden equation \cite{chandra}:
\begin{equation}
{1\over \xi^{d-1}}{d\over d\xi}\biggl (\xi^{d-1}{d\psi\over d\xi}\biggr )=e^{-\psi},
\label{class1}
\end{equation}
\begin{equation}
\psi(0)=\psi'(0)=0.
\label{class2}
\end{equation}
The density profile is given by
\begin{equation}
\rho(\xi)=\rho_{0}e^{-\psi(\xi)}. \label{class3}
\end{equation}
These equations can be obtained from Eqs. (\ref{p1})-(\ref{p3}) by taking the limit $k\rightarrow +\infty$ and redefining the variable $\xi$ using $\sigma_0=k\rho_0$. The dilute (non degenerate) limit
corresponds to high temperatures and high energies (for fixed $\mu$) or to high
values of $\mu$.

For $d\neq 2$, the thermodynamical
parameters are given by
\begin{equation}
\eta=\alpha\psi'(\alpha),
\label{class4}
\end{equation}
\begin{equation}
\Lambda={d(4-d)\over 2(d-2)}
{1\over\alpha\psi'(\alpha)}-{1\over
d-2}{e^{-\psi(\alpha)}\over\psi'(\alpha)^{2}},
\label{class5}
\end{equation}
\begin{eqnarray}
{S\over Nk_{B}}=-{d-2\over 2}\ln\eta-2\ln\alpha+\psi(\alpha)+{\eta\over d-2}-2\Lambda\eta\nonumber\\
+1-\frac{d}{2}+\ln(d\mu)+\frac{d}{2}\ln\left (\frac{2\pi GM}{R^{d-2}}\right ),\quad
\label{class6}
\end{eqnarray}
where $\alpha=(S_{d} G\beta m\rho_{0})^{1/2}R$ is the normalized box
radius. For $d=2$, the thermodynamical parameters can be calculated
analytically (see, e.g.,  \cite{sc,tcrit} and Appendix C of \cite{virialIJMPB}). Introducing the pressure at the box $P=p(R)$,
the global equation of state of the self-gravitating gas can be written as
\begin{equation}
{PV\over Nk_{B}T}={1\over d}{\alpha^{2}\over \eta}e^{-\psi(\alpha)}.
\label{class7}
\end{equation}
The structure and the stability of classical isothermal spheres in $d$
dimensions have been studied in \cite{sc}.

\subsection{The completely degenerate limit: homogeneous sphere}
\label{sec_deg}

In the completely degenerate limit $\beta\rightarrow +\infty$ (i.e. $T=0$), the density
reduces to a step function: $\rho=\sigma_{0}$ if
$r<R_*$ and $\rho=0$ if $r\ge R_*$ (this is the counterpart of the Fermi
distribution in phase space for fermions). This corresponds to a  homogeneous sphere in position space
with density $\sigma_0$ (the maximum value). Its radius is given by
\begin{equation}
R_*=\left (\frac{dM}{S_d\sigma_0}\right )^{1/d}.
\label{deg1}
\end{equation}
This is the counterpart of the mass-radius relation of white dwarfs. The pressure (\ref{eos2}) is infinite. A self-gravitating homogeneous sphere is equivalent to a
polytrope of index $n=0$. It is stable in any dimension of space. By contrast,
white dwarfs are equivalent to polytropes of index $n=d/2$ and they are stable 
only for $d<4$ \cite{wdd,fermionsd}. The completely degenerate
limit corresponds to $k=0$ and the Emden equation with an exclusion constraint
in position space becomes
\begin{equation}
{1\over\xi^{d-1}}{d\over d\xi}\biggl (\xi^{d-1}{d\psi\over d\xi}\biggr )=1.
\label{deg1b}
\end{equation}
It has the analytical solution $\psi=\xi^2/(2d)$.

For $d\neq 2$, the gravitational potential and the  energy of a self-gravitating homogeneous sphere are
\begin{equation}
\Phi(r)=\left\lbrack\left (\frac{r}{R_*}\right )^2-\frac{d}{d-2}\right\rbrack \frac{GM}{2R_*^{d-2}},
\label{deg2}
\end{equation}
\begin{equation}
E=-\frac{d}{d^2-4}\frac{GM^{2}}{R_{*}^{d-2}}.
\label{deg3}
\end{equation}
From Eqs. (\ref{deg1}) and (\ref{deg3}), we obtain
\begin{equation}
E=-\frac{d}{d^2-4}\left (\frac{S_d\sigma_0}{d}\right )^{\frac{d-2}{d}}GM^{\frac{d+2}{d}}.
\label{deg4}
\end{equation}

For $d=2$, the gravitational potential and the  energy of a self-gravitating homogeneous sphere are
\begin{equation}
\Phi(r)=\left\lbrack\left (\frac{r}{R_*}\right )^2-1\right\rbrack \frac{GM}{2}+GM\ln\left (\frac{R_*}{R}\right ),
\label{deg5}
\end{equation}
\begin{equation}
E=-\frac{GM^{2}}{8}+\frac{GM^2}{2}\ln\left (\frac{R_*}{R}\right ).
\label{deg6}
\end{equation}

In the completely degenerate limit, a self-gravitating gas with an exclusion constraint in position space is equivalent to a cold homogeneous sphere of density $\sigma_0$. The inverse temperature is infinite ($\eta=+\infty$) and the entropy is zero ($S=0$). The ground state energy is $\Lambda_{max}=[{d}/(d^2-4)]\mu^{(d-2)/d}$ for $d\neq 2$ and $\Lambda_{max}=1/8+(1/4)\ln\mu$ for $d=2$.

\subsection{The degeneracy parameter}
\label{sec_mu}

When self-gravitating particles experience an exclusion constraint in position space, there is an additional parameter in the problem, namely the maximum value $\sigma_0$ of the spatial density. This leads to the dimensionless parameter $\mu$ defined by Eq. (\ref{p8}) called the degeneracy parameter.  We can give different physical
interpretations to this parameter.

The degeneracy parameter can be written as
\begin{eqnarray}
\mu=\frac{\sigma_{0}}{\langle \rho\rangle}, \qquad  \langle \rho\rangle=\frac{d M}{S_d R^d}.\label{f3a}
\end{eqnarray}
Therefore, it can be viewed as the ratio between
the maximum value of the density $\sigma_{0}$ and the typical ``average''  density $\langle
\rho\rangle$ of a system of mass $M$ enclosed in a sphere of radius $R$. The classical limit is recovered when $\langle \rho\rangle\ll \sigma_0$, i.e. for dilute systems in position space.

Using the results of Sec. \ref{sec_deg}, we can write the degeneracy
parameter in the form
\begin{eqnarray}
\mu=\left (\frac{R}{R_{*}}\right )^{d}, \qquad R_*=\left (\frac{dM}{S_d\sigma_0}\right )^{1/d}.\label{f3b}
\end{eqnarray}
Therefore, it can be viewed as the ratio, up to
the power $d$, of the system's size $R$ to the size $R_{*}$ of a
completely degenerate structure (a cold homogeneous sphere at $T=0$) with mass $M$. Therefore, the case of large $\mu$ corresponds to
large systems and the case of small $\mu$ corresponds to small
systems. In particular, the classical limit corresponds to large systems as compared to the size of the compact object: $R\gg R_*$.

It is also relevant to express $\mu$ as a function of an excluded volume $(1/d)S_d a^d$ where $a$ may be regarded as the effective ``size'' of the particles. The maximum density can then be written as
\begin{equation}
\sigma_0=\frac{dm}{S_d a^d}.
\label{cc11}
\end{equation}
In terms of the ``size'' $a$, the degeneracy parameter may be written as
\begin{equation}
\mu=\frac{1}{N}\left (\frac{R}{a}\right )^d.
\label{cc11b}
\end{equation}
We note that $1/\mu$ is usually called the filling factor.

The classical gas without small-scale constraint is recovered in the limit $R\gg R_*$, or $\langle \rho\rangle\ll\sigma_0$, or $a\ll R$. In terms of the (dimensionless) degeneracy parameter, the classical limit corresponds to $\mu\rightarrow \infty$. Degeneracy effects (close packing) will come into play for small values of $\mu$. We note that $\mu\ge 1$ is required in order to have $R_*\le R$.

\subsection{The thermodynamic limit}
\label{sec_thermo}

For self-gravitating systems, the mean field approximation is exact (except close to a critical point) in the
thermodynamic limit $N\rightarrow +\infty$ with \cite{ijmpb}:
\begin{equation}
\Lambda=-\frac{ER^{d-2}}{GM^2}, \quad \eta=\frac{\beta GMm}{R^{d-2}},\quad \mu=\frac{1}{N}\left (\frac{R}{a}\right )^d
\label{thermo1}
\end{equation}
of order $O(1)$. In this limit, we always have $S\sim E/T\sim  \Lambda\eta Nk_B\sim N$.

Physically, it is logical to take $m\sim G\sim a\sim \sigma_0\sim 1$ since these quantities do not depend on $N$ in principle. In that case, we find the scalings
\begin{equation}
R\sim N^{1/d}, \quad E\sim N^{\frac{d+2}{d}},\quad T\sim N^{2/d}.
\label{thermo2}
\end{equation}
We note that the scaling $N/R^d\sim 1$ corresponds to the mass-radius relation of a homogeneous sphere. On the other hand, the scaling $E\sim N^{(d+2)/d}$ corresponds to the energy (\ref{deg4}) of a self-gravitating homogeneous sphere. Defining a velocity scale by $v\sim (k_B T/m)^{1/2}$ and a dynamical time by $t_D\sim R/v\sim 1/\sqrt{G\rho}$ we find that $v\sim N^{1/d}$ and $t_D\sim 1$.

On the other hand, it is convenient to use a system of units such that $m\sim R \sim v\sim t_D\sim 1$. In that case, we find the scalings
\begin{equation}
G\sim \frac{1}{N}, \quad E\sim N,\quad T\sim 1,\quad a\sim \frac{1}{N^{1/d}},\quad \sigma_0\sim N.
\label{thermo3}
\end{equation}
This is the Kac scaling. With this scaling, the energy
is extensive but it remains fundamentally non-additive. The temperature is
intensive. This scaling is very convenient since the length, velocity and time
scales are of order unity. Furthermore, since the coupling constant $G$ scales
as $1/N$, we immediately see that the limit $N\rightarrow +\infty$ corresponds
to a regime of weak coupling.

Another possible scaling is $R \sim v\sim t_D\sim G\sim 1$. In that case, we find
\begin{equation}
m\sim \frac{1}{N}, \quad E\sim 1,\quad T\sim \frac{1}{N},\quad a\sim \frac{1}{N^{1/d}},\quad \sigma_0\sim 1.
\label{thermo4}
\end{equation}

If we impose $E\sim N$, $T\sim 1$, $G\sim 1$ and $m\sim 1$, we obtain
\begin{equation}
R\sim N^{\frac{1}{d-2}}, \quad a\sim N^{\frac{2}{d(d-2)}},\quad t_D\sim N^{\frac{1}{d-2}},\quad \sigma_0\sim \frac{1}{N^{\frac{2}{d-2}}}.
\label{thermo5}
\end{equation}
However, in that case, the dynamical time diverges with the number of particles for $d>2$ and tends to zero for $d<2$. In addition, this scaling is not defined in $d=2$.

We can also impose $E\sim N$, $T\sim 1$, $G\sim 1$, and $t_D\sim 1$. In that case, we obtain
\begin{equation}
R\sim N^{\frac{1}{d+2}}, \quad a\sim \frac{1}{N^{\frac{2}{d(d+2)}}},\quad m\sim \frac{1}{N^{\frac{2}{d+2}}},\quad \sigma_0\sim 1.
\label{thermo6}
\end{equation}

Of course, many other scalings can be obtained that are more or less convenient (see the discussion in \cite{dvega,rieutord,ijmpb,aakin}). The important thing is that they respect the fundamental scaling given by Eq. (\ref{thermo1}).  It is also interesting to compare the above scalings to those obtained  for self-gravitating fermions \cite{htmath,pt,rieutord,fermionsd,ijmpb}.

\section{Caloric curves in $d=3$}
\label{sec_cc3}

\subsection{The series of equilibria of Fermi-Dirac spheres in position space}
\label{sec_sef}

The critical points of the
Fermi-Dirac entropy $S[\rho]$ at fixed energy and mass (i.e., the
distributions $\rho({\bf r})$ which cancel the first
order variations of $S$ at fixed $E$ and $M$) form a series of equilibria
parameterized by the variable $k$. At each point in the
series of equilibria corresponds a temperature $T$ and an energy
$E$ determined by Eqs. (\ref{p6}) and (\ref{p16})-(\ref{p16b}). We can thus plot $T(E)$ along the
series of equilibria by varying $k$ between $k=0$ (completely degenerate limit) and $k\rightarrow +\infty$ (classical limit). There can be several values of temperature
$T$ for the same energy $E$ because the variational problem
(\ref{fdd1}) can have several solutions: a local entropy maximum
(metastable state), a global entropy maximum (fully stable state), and one or several
saddle points (unstable states). We shall represent all these solutions on the series of equilibria
because local entropy maxima (metastable states) are in general
more physical than global entropy maxima for the timescales achieved
in astrophysics (see \cite{metastable,ijmpb} and Sec. \ref{sec_lifetime}). The same discussion applies in the canonical ensemble. We recall that the series of equilibria is the same in microcanonical and canonical ensembles since the variational problems (\ref{fdd1}) and (\ref{fdd6}) have the same critical points (first order variations).

\begin{figure}
\begin{center}
\includegraphics[clip,scale=0.3]{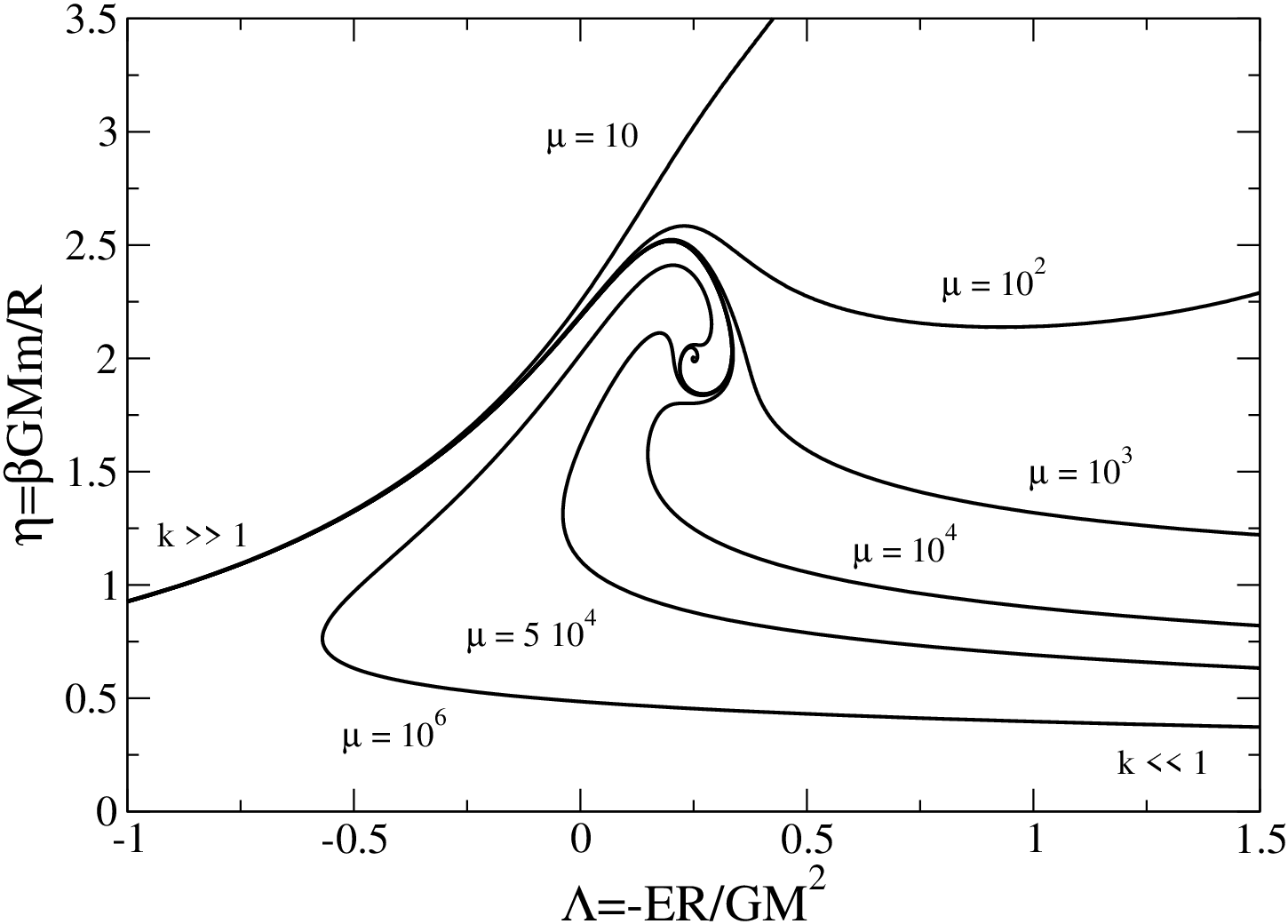}
\caption{Series of equilibria of self-gravitating systems with an exclusion constraint in position space for different values of the degeneracy parameter $\mu$ (note that for
large values of $\mu$, the minimum energy $E_{min}(\mu)$ corresponding to
$T=0$ is outside the frame of the figure). For $\mu\gg 1$,
the spiral makes several rotations before unwinding. }
\label{calomulti}
\end{center}
\end{figure}

When the exclusion constraint in position space is taken into account, the structure of the
series of equilibria $\beta(E)$ depends on the value of the degeneracy
parameter $\mu$ as shown in Fig. \ref{calomulti}. For $\mu\rightarrow
+\infty$, we recover the classical spiral of
Fig. \ref{dim3}. However, for smaller values of $\mu$, we see
that the effect of the exclusion constraint  is to unwind the spiral.  Depending on
the value of the degeneracy parameter, the series of equilibria can
have different shapes. In the following, we shall consider two typical
series of equilibria corresponding to a relatively large value $\mu=5\, 10^4$
(Sec. \ref{sec_large}) and a relatively small
value $\mu=10^3$ (Secs. \ref{sec_small} and \ref{sec_small2}) of the
degeneracy parameter.

\subsection{The case of large systems (or weak small-scale cut-offs) in the microcanonical ensemble: $\mu=5\, 10^4$} \label{sec_large}

\begin{figure}
\begin{center}
\includegraphics[clip,scale=0.3]{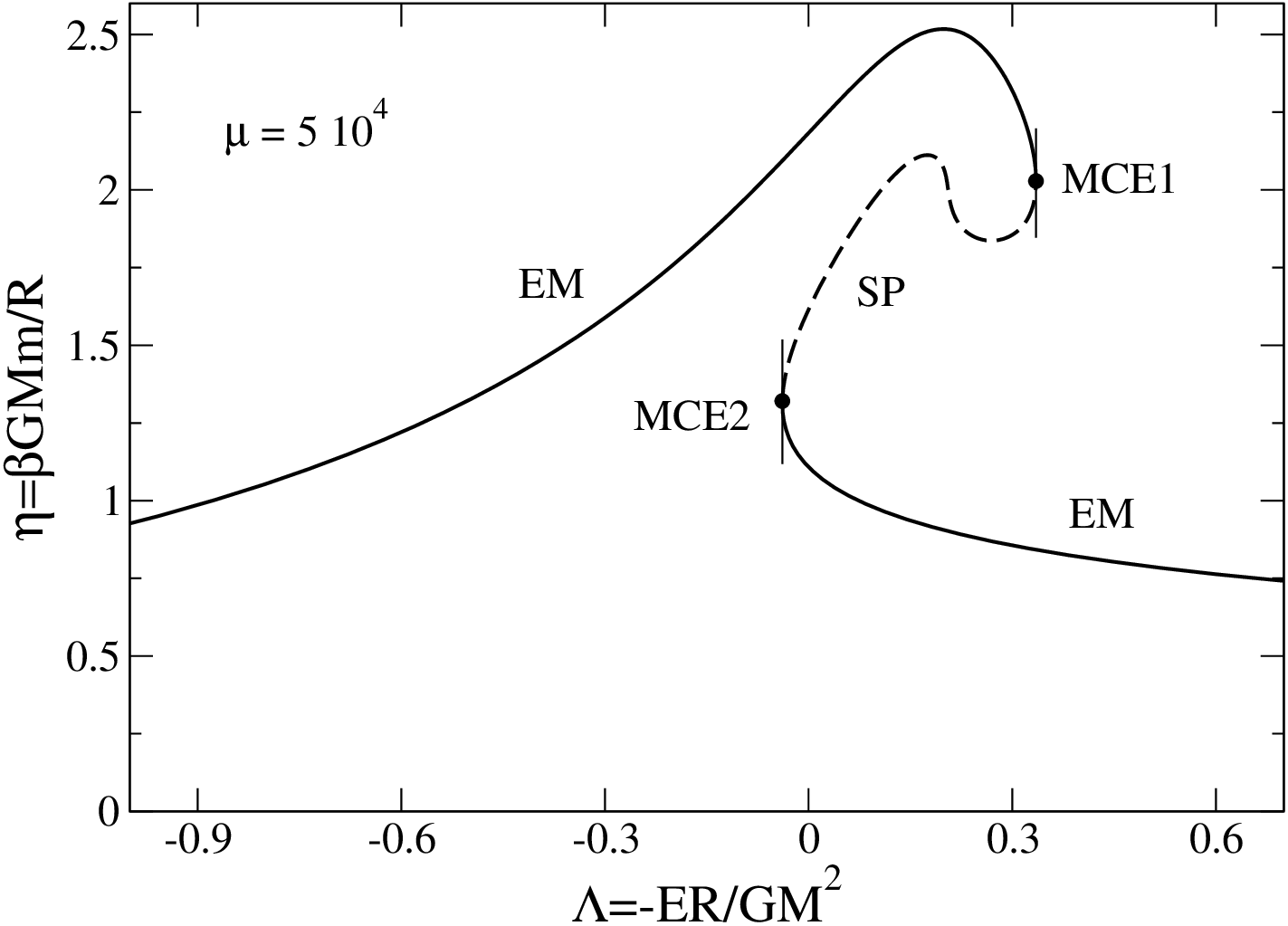}
\caption{Series of equilibria of self-gravitating systems
with an exclusion constraint in position space for $\mu=5\, 10^4$. It has a
$Z$-shape structure (dinosaur's neck). In the microcanonical ensemble,
the stable states (entropy maxima EM) are located on the full
curve. The upper branch corresponds to the ``gaseous phase'' (unaffected by the exclusion
constraint) and the lower branch to the ``condensed phase'' (stabilized by the exclusion constraint). The mode of stability lost at MCE1 is regained at MCE2. The dashed curve corresponds to unstable
saddle points of entropy. They are similar to the gaseous states
(upper branch) but they possess a small nucleus (``germ'').}
\label{ETmicro5instable}
\end{center}
\end{figure}

For $\mu=5\, 10^4$, the series of equilibria is represented in Fig.
\ref{ETmicro5instable}. It has a $Z$-shape structure resembling a {\it
dinosaur's neck} \cite{ijmpb}. This typical value of
the degeneracy parameter corresponds to a relatively large system size $R$, or a relatively small excluded volume $a^3$.  Indeed, we clearly
see the trace of the classical spiral ($a\rightarrow 0$) which is not completely
unwound.

\begin{figure}
\begin{center}
\includegraphics[clip,scale=0.3]{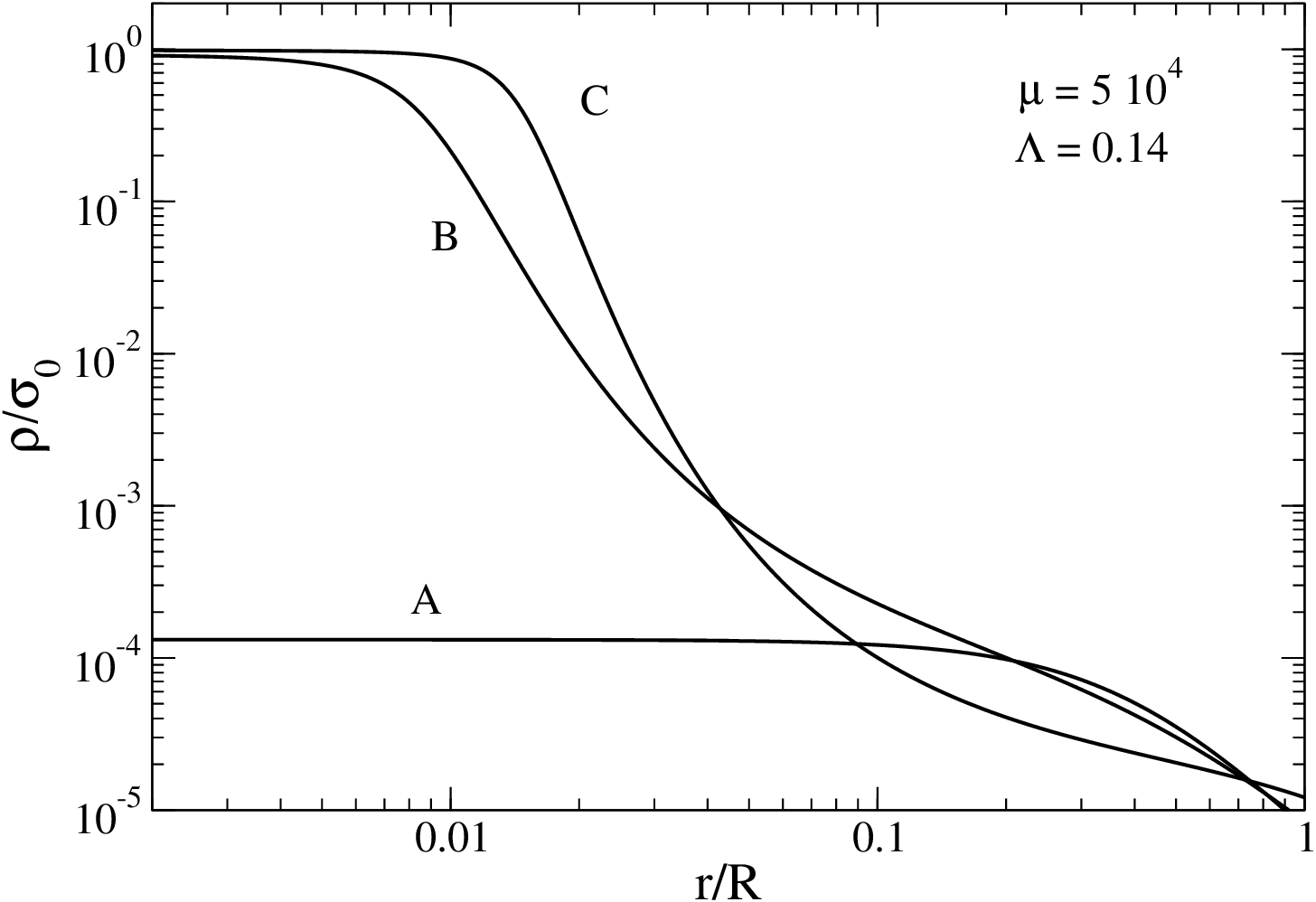}
\caption{Typical equilibrium density profiles for a
degeneracy parameter $\mu=5\, 10^{4}$ and an energy $\Lambda=0.14$. The profile A corresponds to the gaseous phase (EM), the profile C to the condensed phase (EM) and the profile B to the unstable phase (SP). In the gaseous phase, the density decreases as $r^{-2}$ at large distances. In the condensed phase, there is a core of density $\sigma_0$ (with large potential energy) surrounded by an almost homogeneous (hot) halo. The unstable density profile is similar to the gaseous density profile except that it contains an embryonic nucleus (with weak potential energy).}
\label{profiles5}
\end{center}
\end{figure}

In this section, we consider an isolated system.
In that case, the control parameter is
the energy $E$ and the relevant statistical ensemble is the
microcanonical ensemble.  In the
microcanonical ensemble, we must determine maxima of entropy at
fixed mass and energy. The stability of the solutions can be settled by using the theory of Poincar\'e on the linear series of equilibria \cite{poincare,lbw,katz,katz2,ijmpb}. The conserved quantity is the energy $E$ and the parameter conjugate to the energy with respect to the entropy $S$ (thermodynamical potential) is the inverse temperature $T^{-1}=\partial S/\partial E$. From the Poincar\'e theorem, a change of stability can occur only at a turning point of energy. A mode of stability is lost if the curve $\beta(-E)$ rotates clockwise and gained if it turns anti-clockwise. At high energies self-gravity is negligible. The system is equivalent to a non-interacting gas in a box and we know from ordinary thermodynamics that it is stable (entropy maximum). From the turning
point criterion, we conclude that all the states on the upper branch
of the series of equilibria are stable (entropy maxima EM) until the
first turning point of energy MCE1. For sufficiently large values of
$\mu$, this corresponds to the Antonov energy $E_{c}$. At that
point, the curve turns clockwise so that a mode of stability is
lost. This mode of stability is regained at the second turning point
of energy MCE2 at which the curve turns anti-clockwise. The
corresponding energy $E_{*}(\mu)$ depends on the value of the
degeneracy parameter  and tends to $E_{*}(\mu)\rightarrow
+\infty$ for $\mu\rightarrow +\infty$ (some analytical estimates
of this energy
are given in \cite{pt}). The solutions on the  branch
between MCE1 and MCE2 are unstable saddle points (SP) of entropy
while the solutions on the lower branch after MCE2 are entropy
maxima (EM). The solutions on the upper branch are non degenerate
and have a smooth density profile; they form the ``gaseous phase''.
The solutions on the lower branch have a core-halo structure; they
form the ``condensed phase''. They consist of a degenerate nucleus
surrounded by a ``vapor''. The nucleus (condensate) is equivalent to a self-gravitating
homogeneous sphere  at $T=0$ with the maximum density $\sigma_0$. At nonzero temperatures,
this compact object is surrounded by a dilute atmosphere. This
structure resembles a giant gaseous planet, with a rocky core and an
atmosphere. Typical
density profiles are shown in Fig. \ref{profiles5}. The degenerate
nucleus (with a constant density $\sigma_0$) of the condensed configuration  is clearly visible.

\begin{figure}
\begin{center}
\includegraphics[clip,scale=0.3]{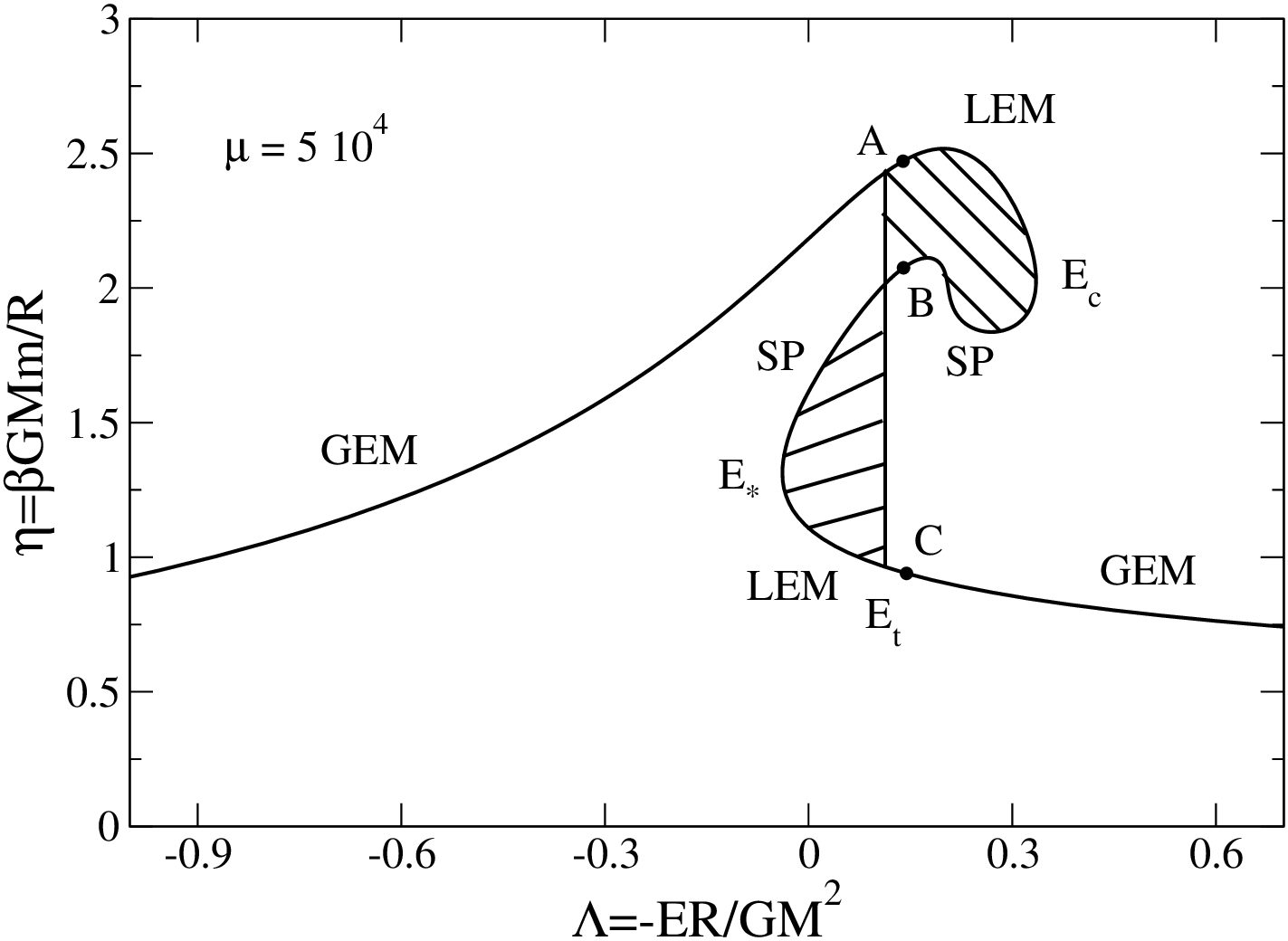}
\caption{Maxwell construction in the microcanonical ensemble determining the transition energy $E_{t}(\mu)$ at which the gaseous states pass from global entropy maxima (GEM) to local entropy maxima (LEM) while the condensed states pass from local entropy maxima to global entropy maxima.}
\label{ETmicro5}
\end{center}
\end{figure}

\begin{figure}
\begin{center}
\includegraphics[clip,scale=0.3]{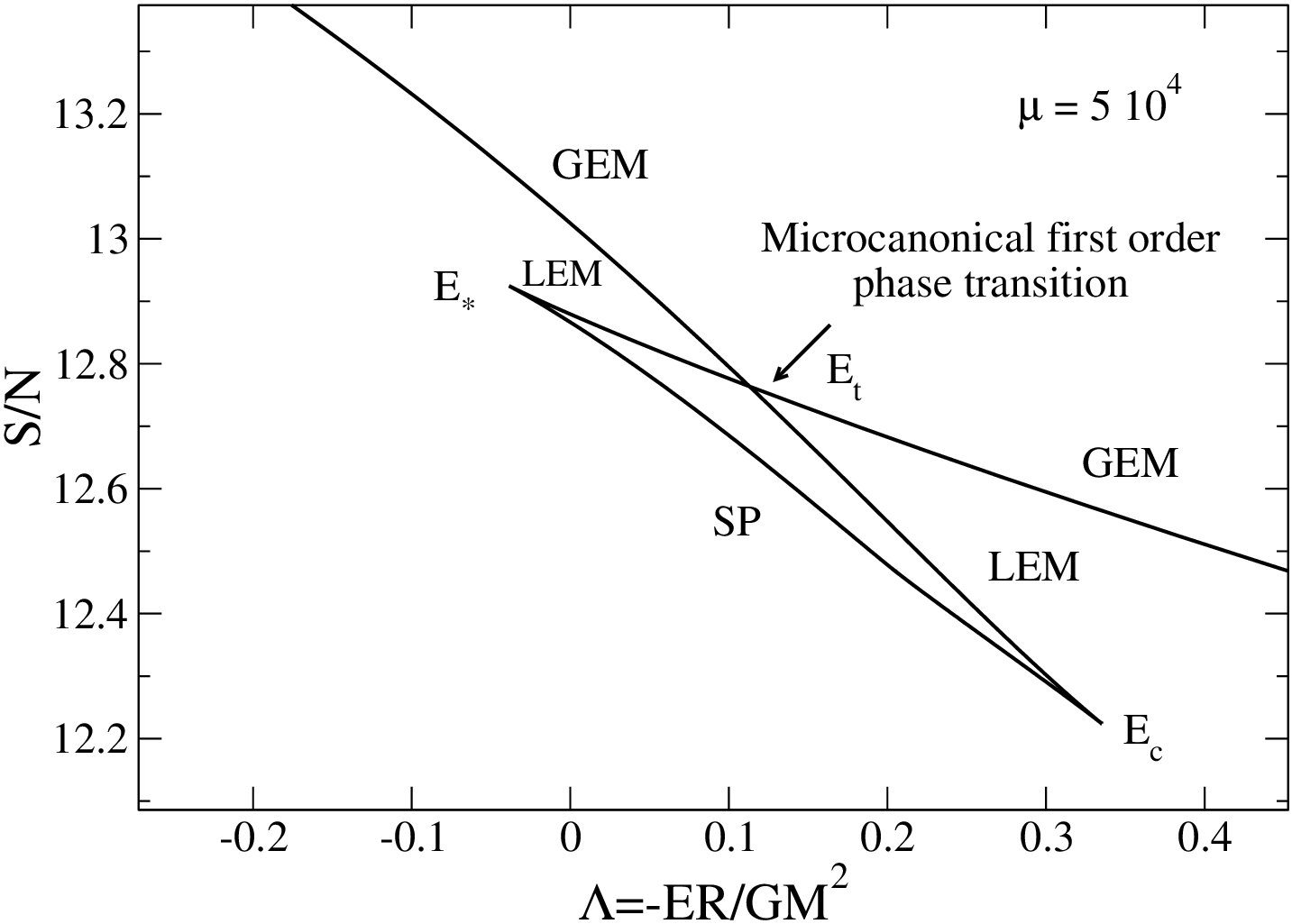}
\caption{Entropy of each phase versus
energy for $\mu=5\, 10^{4}$. A
microcanonical first order phase transition is expected at
$E_{t}(\mu)$ at which the two stable branches (solutions A and C)
intersect.  The ``kink'' in the curve $S(E)$ at the transition point
where the two branches intersect corresponds to a discontinuity (jump) of
temperature in the caloric curve $\beta(E)$. However, the entropic
barrier played by the solution $B$ (saddle point) prevents this phase transition to
occur in practice \cite{metastable,ijmpb}.}
\label{Smicro5}
\end{center}
\end{figure}

\begin{figure}
\begin{center}
\includegraphics[clip,scale=0.3]{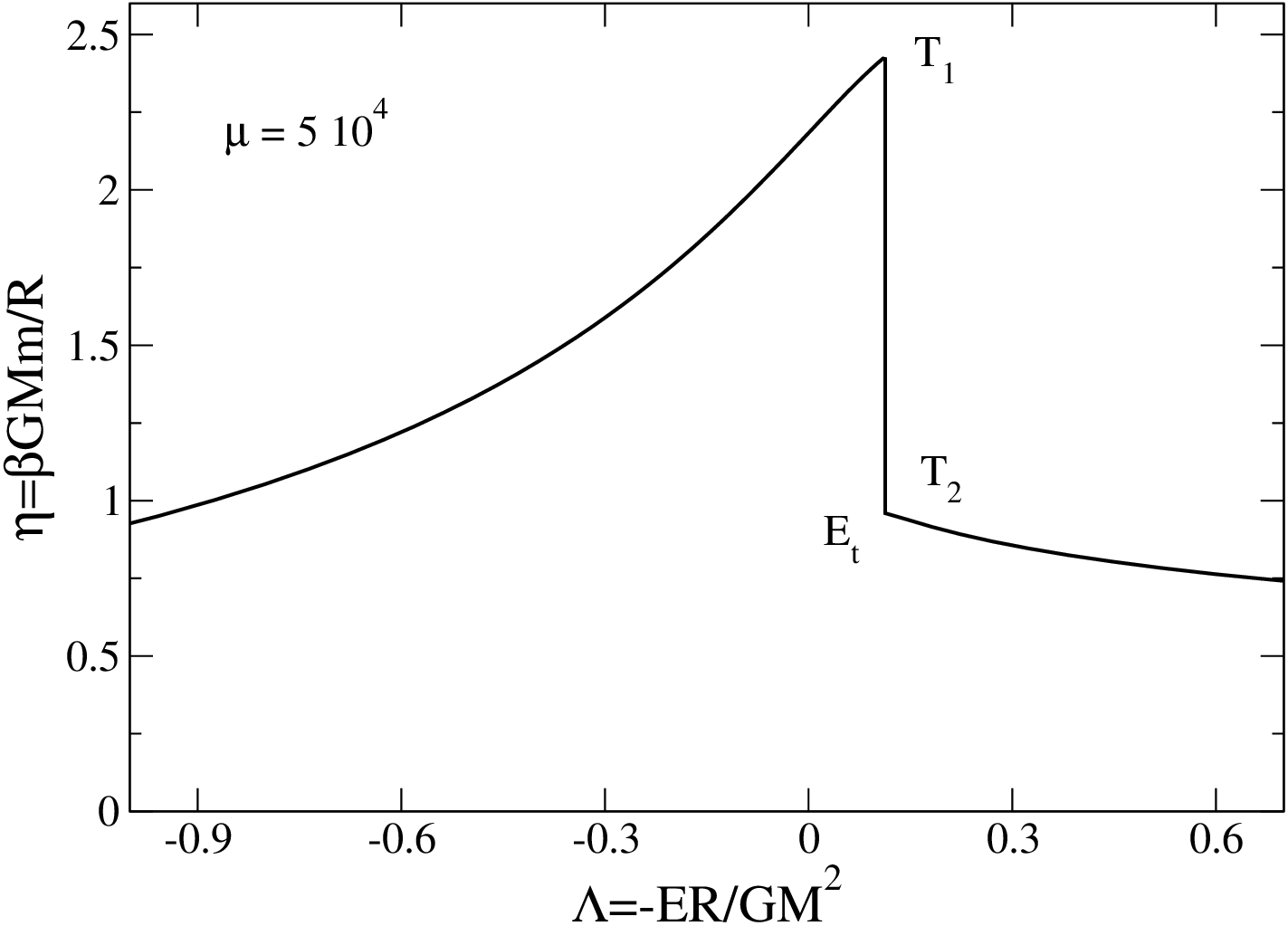}
\caption{Strict microcanonical caloric curve for $\mu=5\, 10^{4}$. This figure is obtained from Fig. \ref{ETmicro5} by keeping only global entropy maxima (GEM). It corresponds therefore to the exact microcanonical caloric curve $\beta_{micro}(E)=dS/dE(E)$ which  is univalued \cite{ijmpb}. For $N\rightarrow +\infty$, there is a discontinuity of temperature at the transition energy $E_{t}(\mu)$. For finite $N$ systems, this discontinuity is smoothed-out \cite{ispolatov}. Although this caloric curve is correct in a strict sense, it is {\it not} physical because it ignores metastable states that have an infinite lifetime in the thermodynamic limit $N\rightarrow +\infty$. The physical caloric curve (see Fig. \ref{ETmicro5meta}) is obtained from Fig. \ref{ETmicro5instable} by discarding the unstable saddle points of entropy that form the intermediate branch.}
\label{ETmicro5strict}
\end{center}
\end{figure}

We must now determine which states are local entropy maxima and which
states are global entropy maxima. This can be done by performing a
vertical Maxwell construction \cite{ijmpb} or by
plotting the entropy of the two phases and determining at which
energy they become equal. The Maxwell construction is shown in
Fig. \ref{ETmicro5}. The entropy versus energy plot is represented
in Fig.
\ref{Smicro5}. There is a crossover at the transition energy
$E_{t}$. For $E>E_{t}(\mu)$ the gaseous states are global entropy
maxima (GEM) and for $E_{c}<E<E_{t}$ the gaseous states are local
entropy maxima (LEM), i.e. metastable states. Inversely, for
$E_{t}<E<E_{*}$ the condensed states are local entropy maxima
(metastable states) and for $E<E_{t}$ the condensed states are global
entropy maxima. In a strict sense, the statistical equilibrium states
correspond to the global entropy maxima. Therefore, the strict caloric
curve in the microcanonical ensemble for $\mu=5\, 10^4$ is the one
represented in Fig. \ref{ETmicro5strict}. It contains only global maxima of entropy at fixed mass and
energy. From this curve, we expect that a {\it microcanonical first order}
phase transition occurs at $E=E_t$, connecting the
gaseous phase to the condensed phase. It is accompanied by a
discontinuity of temperature $\Delta T=T_1-T_2$ (this is the counterpart
of the latent heat in the canonical ensemble) and specific heat. If $E_{t}>E_{gas}$
(where $E_{gas}$ is the energy corresponding to the first turning
point of temperature) the specific heat passes from a positive to
a negative value. If $E_{t}<E_{gas}$, the specific heat is always
negative at the transition (the crossover occurs for $\mu\simeq
19045$; see the intersection between $E_t$ and $E_{gas}$ in Fig.
\ref{phasemicro}). The discontinuity of temperature is due to the
fact that our mean field treatment is valid in the $N\rightarrow +\infty$ limit. For finite values of $N$ the discontinuity of
temperature is smoothed out because close to $E_{t}(\mu)$ the
metastable states contribute significantly to the density of states
\cite{ispolatov,metastable}.

\begin{figure}
\begin{center}
\includegraphics[clip,scale=0.3]{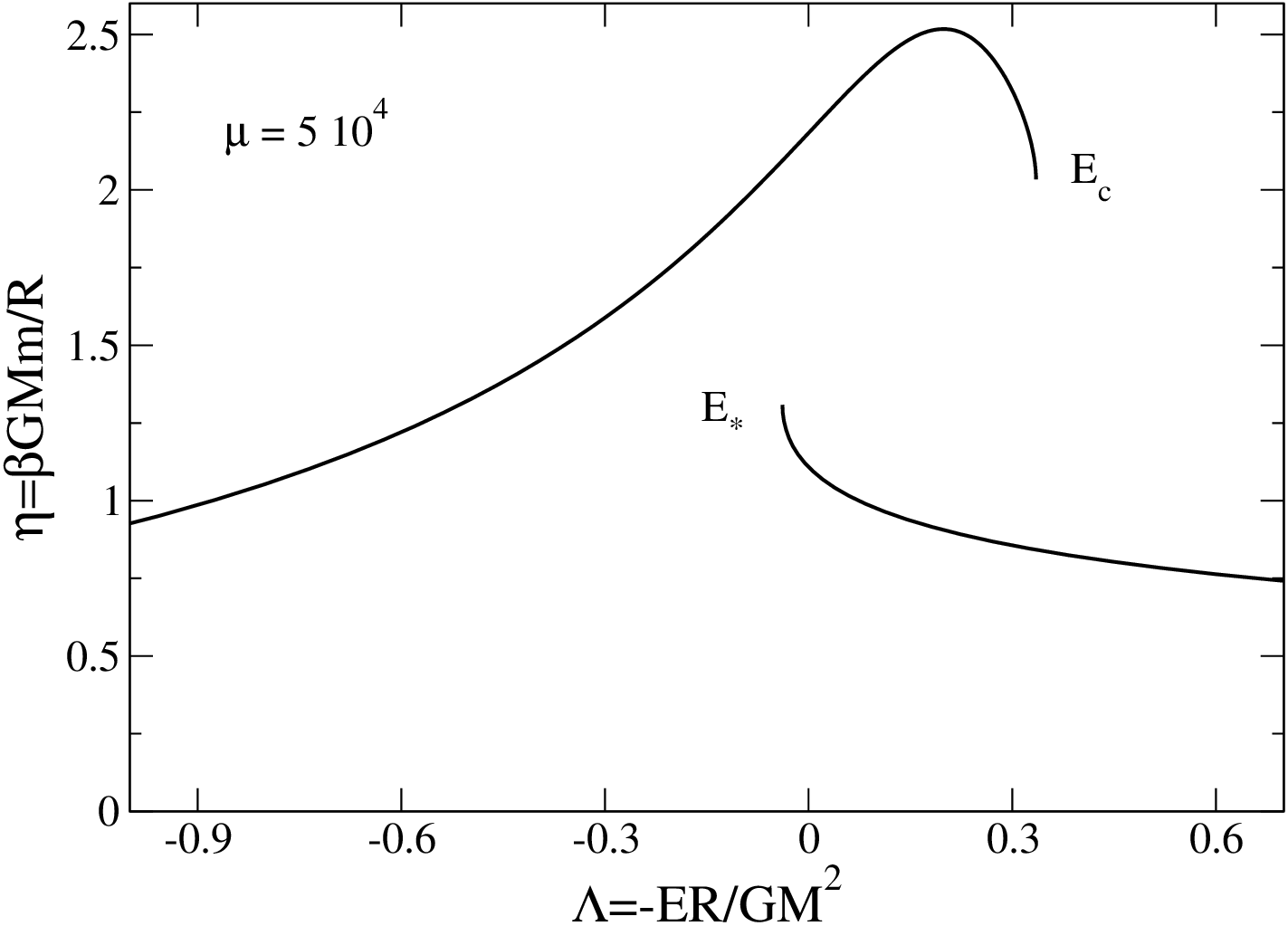}
\caption{Physical microcanonical caloric curve for $\mu=5\, 10^{4}$. This figure displays stable and metastable equilibrium states. The metastable states have infinite lifetime in the thermodynamic limit so they are physically relevant. The microcanonical first order  phase transition of Fig. \ref{ETmicro5strict} does not take place in practice. Note that the structure of the physical canonical caloric curve remains the same for larger values of $\mu$. Therefore, although the series of equilibria becomes more and more complex because of the appearance of the spiral, this does not affect the physical microcanonical caloric curve (the spiral corresponds to unstable states).}
\label{ETmicro5meta}
\end{center}
\end{figure}

The preceding discussion suggests the occurrence of a microcanonical first order
 phase transition at $E_{t}$.  However, for
self-gravitating systems, the metastable states are long-lived  because the probability of a fluctuation able to trigger the phase transition is extremely weak.  Indeed, the
system has to cross the entropic barrier played by the solution on
the intermediate branch (point B). This state is similar to
point A except that it contains a small embryonic nucleus which plays the role of a ``germ'' in the
language of phase transitions (see Fig. \ref{profiles5}). The
important point is that the entropic barrier scales like $N$ for systems with long-range interactions and consequently the probability of transition scales like $e^{-N}$ (see
Sec. \ref{sec_lifetime}).  Therefore, the metastable states are
extremely robust. In practice, the microcanonical first order phase
transition at $E_t$ does {\it not} take place because, for
sufficiently large values of $N$, the system remains frozen in
the metastable phase. Therefore, the strict caloric curve of Fig.
\ref{ETmicro5strict} is not physical. The physical microcanonical caloric
curve is the one shown in Fig. \ref{ETmicro5meta} which takes the metastable states
into account . It is obtained from the series of
equilibria of Fig. \ref{ETmicro5instable} by discarding only the
unstable saddle points.

\begin{figure}
\begin{center}
\includegraphics[clip,scale=0.3]{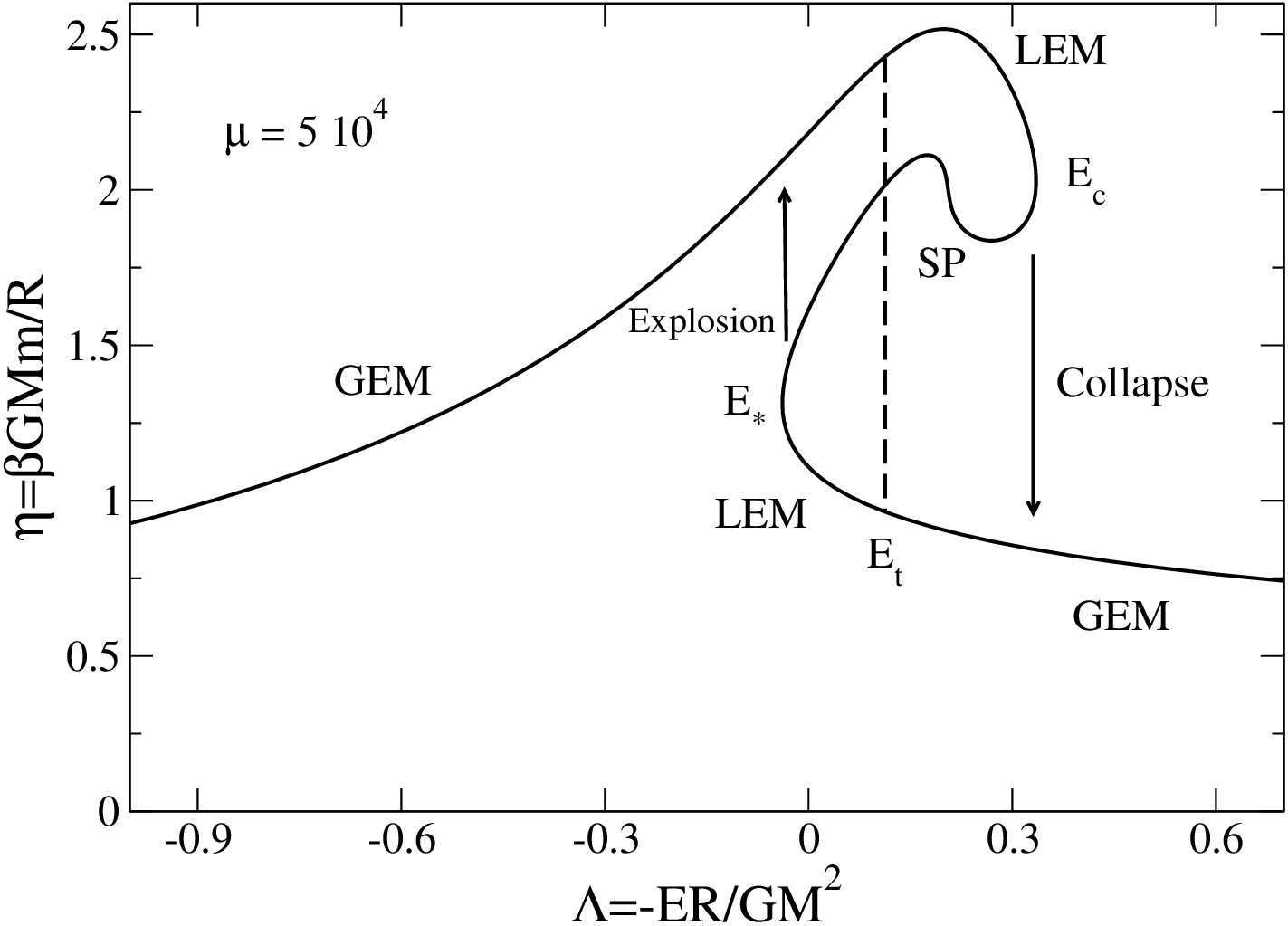}
\caption{Summary of the phase transitions for the self-gravitating
 gas with a degeneracy parameter $\mu=5\, 10^{4}$ in the
microcanonical ensemble. The points on the upper branch  form the ``gaseous'' phase. They
are global entropy maxima (GEM) for $E>E_{t}(\mu)$ and local entropy
maxima (LEM), i.e.  metastable states, for $E<E_{t}(\mu)$. The points on the lower branch 
form the ``condensed'' phase. They are LEM for $E>E_{t}(\mu)$ and
GEM for $E<E_{t}(\mu)$. The points on the intermediate branch are unstable saddle points (SP).
Due to the existence of metastable states, the system displays a
microcanonical hysteretic cycle marked by a ``collapse'' and an
``explosion'' at the spinodal points where the branch of metastable
states disappears: for $E\le E_c$, a gaseous configuration undergoes
a gravothermal catastrophe (collapse) and for $E\ge E_*(\mu)$, a
condensed configuration undergoes an explosion. These microcanonical
phase transitions exist only above a critical point $\mu_{MCP}=1750$
(see Sec. \ref{sec_maxwell}). This corresponds to a system size
$R>12.05R_{*}$.}
\label{ETmicro5summary}
\end{center}
\end{figure}

The true phase transition occurs at the critical energy $E_{c}$ at
which the metastable branch disappears (see Fig. \ref{ETmicro5summary}). This
point MCE1 is similar to a spinodal point in the language of phase
transitions. Below $E_{c}$, the system undergoes a gravothermal
catastrophe (collapse). However, in the presence of an exclusion constraint, the core ceases
to shrink when it becomes degenerate. Since this collapse is
accompanied by a discontinuous jump of entropy (see Fig. \ref{Smicro5}),
this is sometimes called a {\it microcanonical zeroth order} phase
transition. The resulting equilibrium state possesses a
small degenerate nucleus which contains a moderate fraction of the
total mass but a large potential energy.
For $\mu=5\, 10^{4}$, the degenerate core following the collapse
at $E=E_c^{-}$ contains about
$2.10\%$ of the total mass (its radius defined as the distance at which the density is equal to $0.95\sigma_0$ is about $7.98\ 10^{-3}R$) and this
fraction of mass decreases for larger values of $\mu$.
\footnote{This is expected because in the classical limit
$\mu\rightarrow +\infty$ the degenerate nucleus should be replaced a
``binary star'' with a negligible mass but a huge binding energy surrounded by a hot halo
(see the Introduction).}  The rest of the mass is diluted in a hot
envelope with an almost uniform density held by the box (the condensed states are hotter than the corresponding gaseous states because the gravitational energy created by the collapse of the core is released in the halo in the  form of kinetic energy). In an open system, the halo
would be dispersed at infinity so that only the degenerate core (equivalent to a self-gravitating homogeneous sphere) would
remain.

For $E<E_{t}$, the condensed states are global
entropy
maxima. If now energy is increased, the system remains in the
condensed phase past the transition energy $E_t$. Indeed, for the same
reason as before the condensed states with $E_{t}<E<E_{*}$ are
long-lived metastable states so that the first order phase transition
from the condensed states to the gaseous states does not take
place. However, above $E_{*}$ the condensed metastable branch
disappears and the system undergoes a discontinuous transition
reversed to the collapse at $E_c$ (see Fig. \ref{ETmicro5summary}). This
transition is sometimes called an ``explosion'' since it transforms
the dense core into a relatively uniform mass distribution. 
Since the collapse and the explosion occur at
different values of the energy, due to the presence of metastable states,
we can generate an {\it hysteretic cycle} by
varying the energy between $E_{*}$ and $E_{c}$.  This hysteretic cycle
has been followed numerically by
Ispolatov \& Karttunen \cite{ik1,ik2} by using Molecular Dynamical
Methods. In their study, the small-scale regularization is played by a
soften potential.

\subsection{The case of small systems (or large small-scale cut-offs) in the microcanonical ensemble: $\mu=10^3$} \label{sec_small}

\begin{figure}
\begin{center}
\includegraphics[clip,scale=0.3]{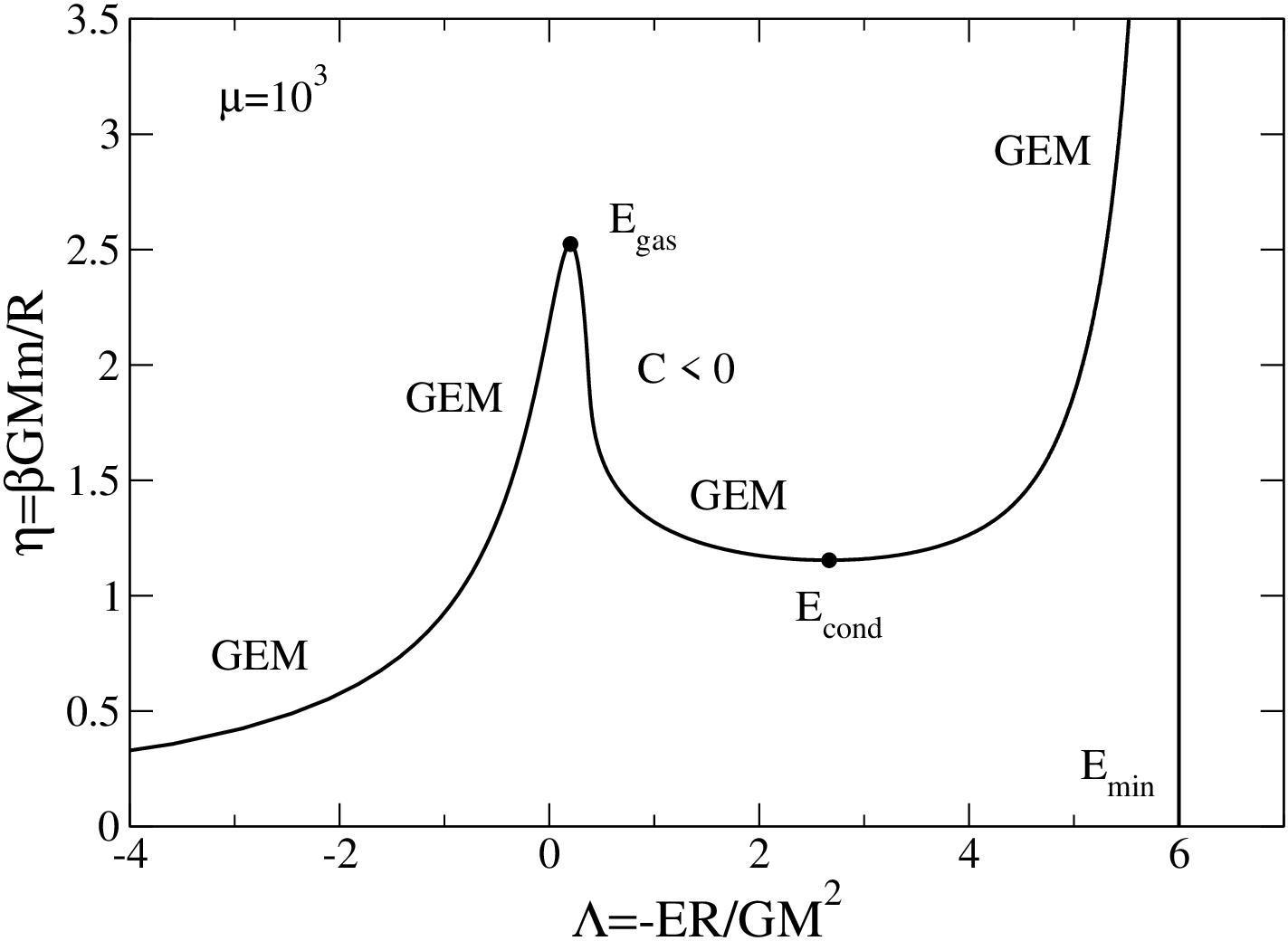}
\caption{Series of equilibria of self-gravitating systems with an exclusion constraint
in position space  with
$\mu=10^{3}$. It has a $N$-shape structure. All the solutions are
global entropy maxima (GEM) so that this figure also represents the
exact caloric curve in the microcanonical ensemble. The part of the
curve between the extrema of temperature (at $E_{gas}$ and $E_{cond}$)
has negative specific heat $C<0$.}
\label{ETmicro3}
\end{center}
\end{figure}

For $\mu=10^3$, the series of equilibria is represented in Fig.
\ref{ETmicro3}. It has a $N$-shape structure. This typical value of
the degeneracy parameter corresponds to a relatively small system size $R$ or
a relatively large excluded volume $a^3$. Indeed, the trace of the
classical spiral has disappeared and the $T(E)$ curve is now
univalued.

\begin{figure}
\begin{center}
\includegraphics[clip,scale=0.3]{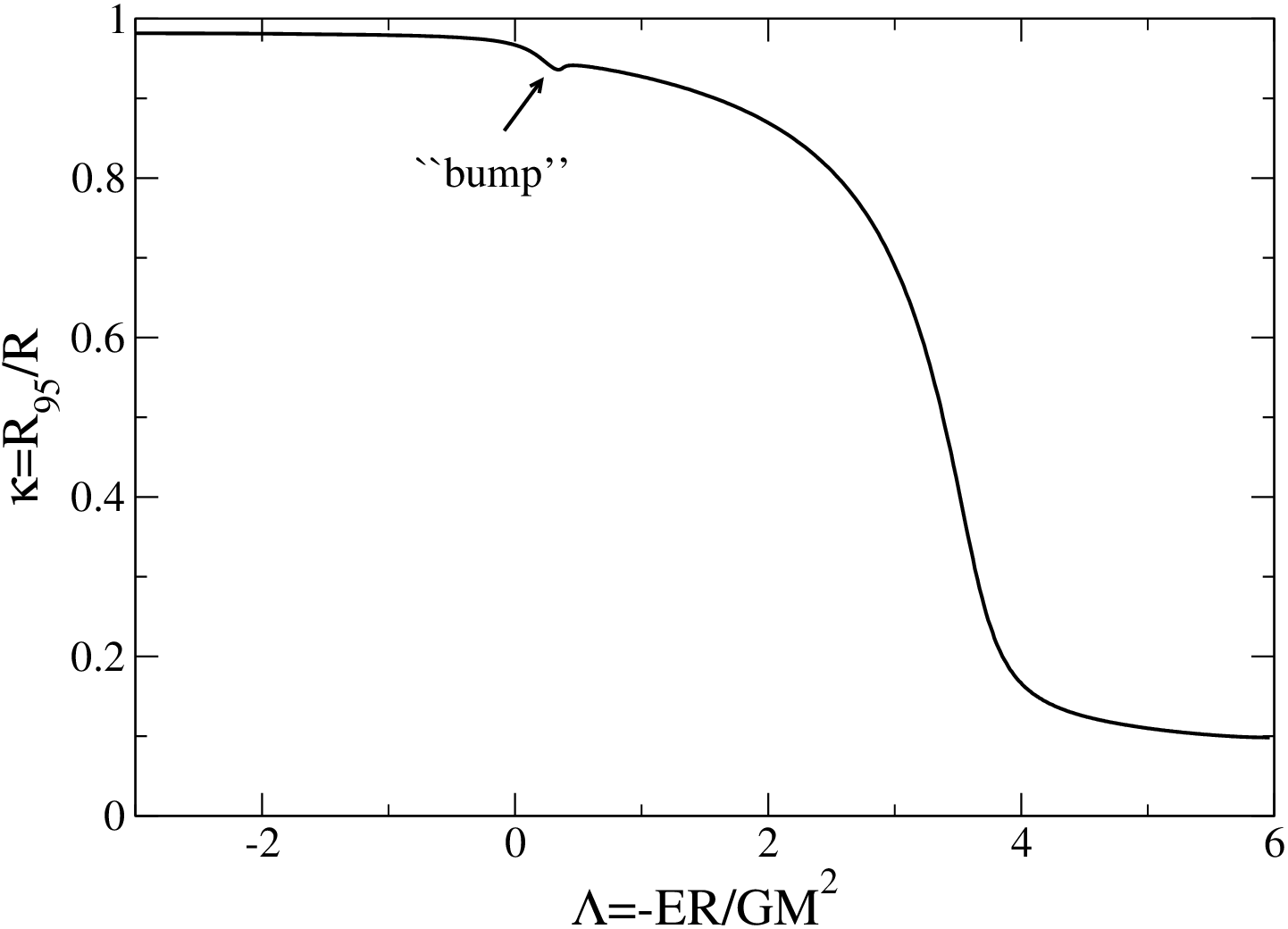}
\caption{Evolution of the order parameter $\kappa=R_{95}/R$ with
the energy ($R_{95}$ is the radius containing $95\%$ of the mass
and $R$ is the radius of the whole configuration). The figure
shows the smooth passage from gaseous states to condensed states
as the energy is progressively decreased. We note the presence of a small
``bump'', where $\kappa$ increases before decreasing again. The bump occurs near the Antonov
energy $\Lambda_c=0.335$.}
\label{rayon95}
\end{center}
\end{figure}

In this section, we consider an isolated system. The control parameter is the energy and
the relevant statistical ensemble is the microcanonical ensemble. We
note that the series of equilibria does not present turning points
of energy so that, according to the Poincar\'e theorem, all the series of equilibria is stable and
corresponds to global entropy maxima (GEM) at fixed mass and energy.
Therefore, for sufficiently small values of the degeneracy
parameter, there is no phase transition in the
microcanonical ensemble. The microcanonical first order phase
transition at $E_t$, as well as the gravothermal catastrophe at
$E_c$, are suppressed. However, there is a sort of condensation
(clustering) as the energy is progressively decreased. At large
energies, the equilibrium states are almost homogeneous or slightly
inhomogeneous. They coincide with classical isothermal
self-gravitating systems. At
smaller energies, in the region of negative specific heats, they
have a ``core-halo'' structure with a partially degenerate nucleus
and a non-degenerate envelope. This is like a solid condensate
embedded in a vapor.
As energy is further decreased, the nucleus becomes more and more
degenerate and contains more and more mass. At the minimum energy
$E_{min}$, corresponding to $T=0$, all the mass is in the completely
degenerate nucleus.  In that case, the atmosphere has been
swallowed and the system reduces to a homogeneous self-gravitating sphere of
density $\sigma_0$ (the maximum value) corresponding to the close packing of all
the particles. Therefore,
depending on the degree of degeneracy and on the value of the energy, a
wide variety of nuclear concentrations can be achieved. In Fig.
\ref{rayon95}, we have plotted the radius containing $95\%$ of the
mass as a function of the energy. The quantity $\kappa=R_{95}/R$ which
measures the degree of concentration of the system can serve as an
{\it order parameter}. At high energies, the particles are uniformly distributed
in the sphere of radius $R$ so that $\kappa=(0.95)^{1/3}\simeq 0.983\sim 1$. At
low energies, the particles are concentrated in a homogeneous core of density
$\sigma_0$ so that $\kappa=(0.95/\mu)^{1/3}\ll 1$.

\begin{figure}
\begin{center}
\includegraphics[clip,scale=0.3]{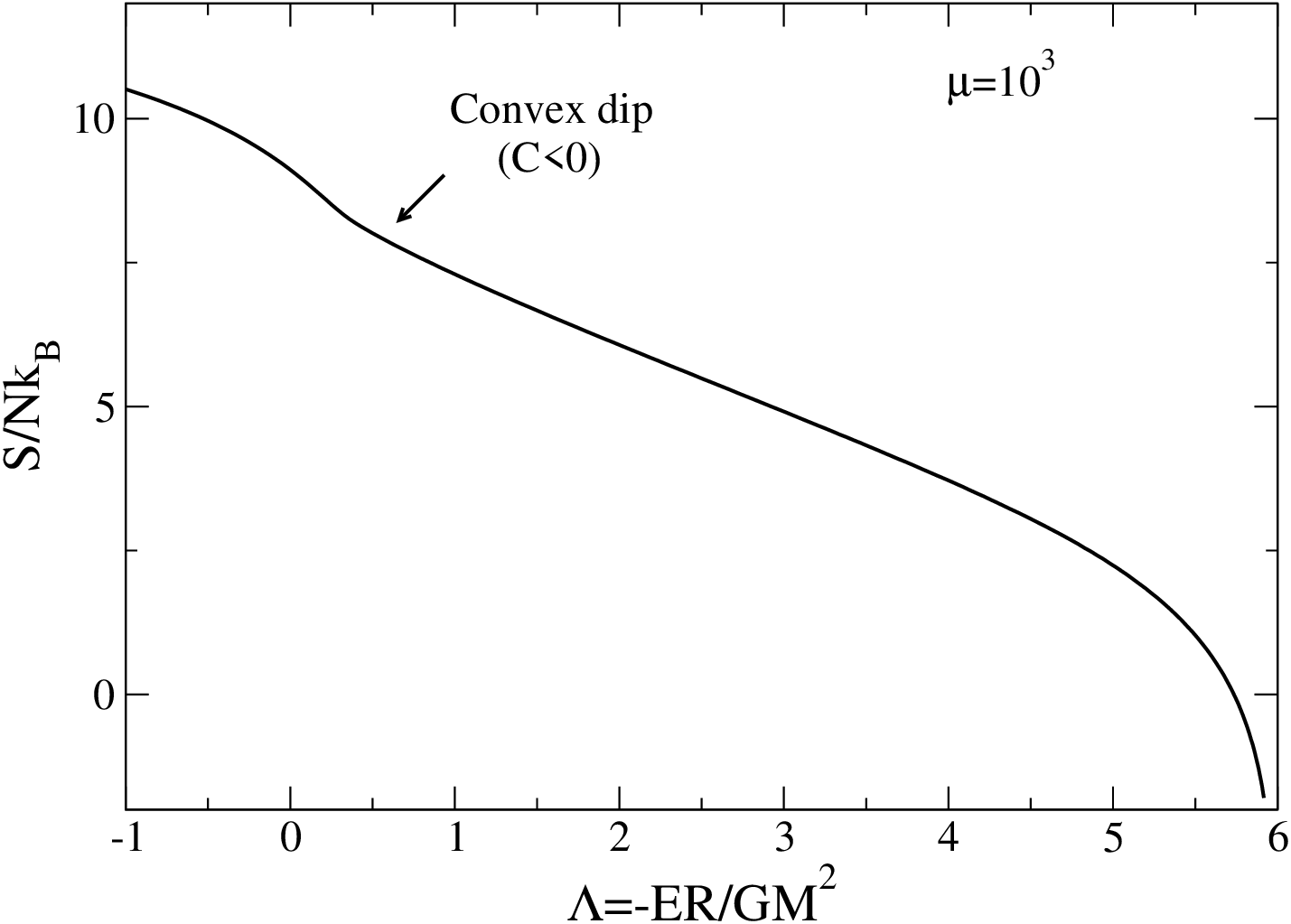}
\caption{Plot of entropy $S$ vs energy $E$ for $\mu=10^3$. Since
$S''(E)=-1/(CT^{2})$, the entropy presents a convex intruder
($S''>0$) in the region of negative specific heat ($C<0$) between
$E_{gas}$ and $E_{cond}$.}
\label{Smicro3}
\end{center}
\end{figure}

We have chosen to plot the series of equilibria $\beta(E)$ in Fig. \ref{ETmicro3}
because this is the proper representation to apply the Poincar\'e theorem.
We note in particular that, for $\mu=10^3$, the region of negative
specific heats $C<0$ is stable in the microcanonical ensemble
(it corresponds to global entropy maxima at fixed mass and energy). Another
interesting curve is the entropy $S(E)$
represented in Fig. \ref{Smicro3}. It displays  a convex
intruder in the region of negative specific heats. For extensive
systems, this convex intruder does not exist because the system can
gain entropy by splitting in two co-existing phases. However, this
phase separation is forbidden for systems with long-range interactions
because the energy is non-additive.  Therefore, the region of negative
specific heats in the caloric curve (see Fig. \ref{ETmicro3}) and the convex intruder
in the entropy vs energy curve (see Fig. \ref{Smicro3})
are allowed in the microcanonical ensemble for such systems \cite{gross}.

\subsection{The case of small systems (or large small-scale cut-offs) in the canonical ensemble:
$\mu=10^3$} \label{sec_small2}

\begin{figure}
\begin{center}
\includegraphics[clip,scale=0.3]{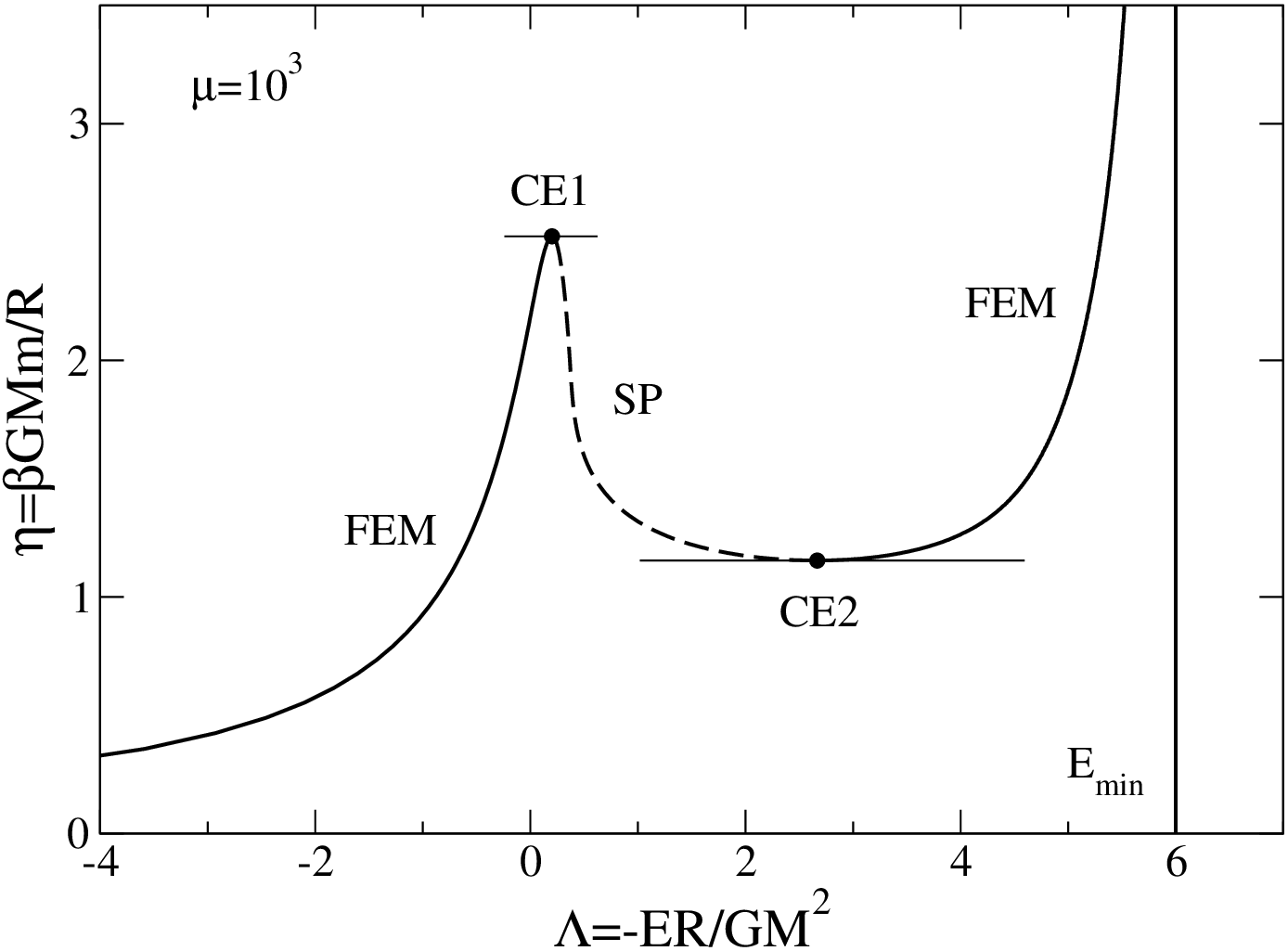}
\caption{Series of equilibria of self-gravitating system with an
exclusion constraint in physical space for $\mu=10^3$. In the
canonical ensemble, the stable states (minima of free energy FEM) are
located on the full curve. The left branch corresponds to the
``gaseous phase'' and the right branch to the ``condensed phase''. The
mode of stability lost at CE1 is regained at CE2. The dashed curve
with negative specific heats corresponds to unstable saddle points of
free energy.}
\label{ETcano3instable}
\end{center}
\end{figure}

We now consider the canonical situation in which the system is in
contact with a heat bath imposing its temperature $T$. In that case,
the control parameter is the temperature $T$ and the relevant statistical
ensemble is the canonical ensemble. This
situation can describe a gas of self-gravitating Brownian particles
\cite{ijmpb,sc}. In the canonical ensemble we must determine minima of free energy at fixed mass.
Considering again the case $\mu=10^{3}$, we note that the series of
equilibria $E(T)$ represented in Fig. \ref{ETcano3instable} is multi-valued.\footnote{In the canonical
description, where the control parameter is the temperature, it would be
more convenient to rotate the figures by $90^{o}$ to represent $E$ as
a function of $T$. However, in this paper, we shall always show the
curves in the $T(E)$ representation.} This gives rise to
canonical phase transitions.   The stability of the solutions can be settled
by using the Poincar\'e theorem taking now the
temperature $T$ as the control parameter and the free energy $F$ as the
thermodynamical potential (the roles of $E$ and $T$ are reversed with respect to the microcanonical situation) \cite{ijmpb}. At high temperatures, self-gravity is negligible
with respect to thermal motion and the system is equivalent to a
non-interacting gas in a box. Therefore, according to ordinary
thermodynamics, we know that it is stable (minimum of free energy at
fixed mass). From the turning point criterion, we conclude that all
the solutions on the left branch of Fig. \ref{ETcano3instable} are stable
(free energy minima FEM) until the first turning point of
temperature CE1. For sufficiently large values of $\mu$, this
corresponds to the Emden temperature $T_{c}$. At that point, the
curve turns clockwise so that a mode of stability is lost. This mode
of stability is regained at the second turning point of temperature
CE2 at which the curve turns anti-clockwise. The corresponding
temperature $T_{*}(\mu)$ depends on the value of the degeneracy
parameter and tends to $T_{*}(\mu)\rightarrow +\infty$ for
$\mu\rightarrow +\infty$ (some analytical estimates of this
temperature are given in \cite{pt}). Therefore, the solutions on the  branch
between CE1 and CE2 are unstable saddle points (SP) of free energy.
They lie in the region of negative specific heats which is forbidden
(unstable) in the canonical ensemble. Finally, the solutions on the
right branch of Fig. \ref{ETcano3instable} after CE2 are maxima of free
energy (FEM).  The solutions on the left branch are non degenerate
and have a smooth density profile; they form the ``gaseous phase''.
The solutions on the right branch have a ``core-halo'' structure
with a massive degenerate nucleus and a dilute atmosphere; they form
the ``condensed phase''. These condensed configurations with a
``rocky core'' and an ``atmosphere'' resemble giant gaseous planets.

\begin{figure}
\begin{center}
\includegraphics[clip,scale=0.3]{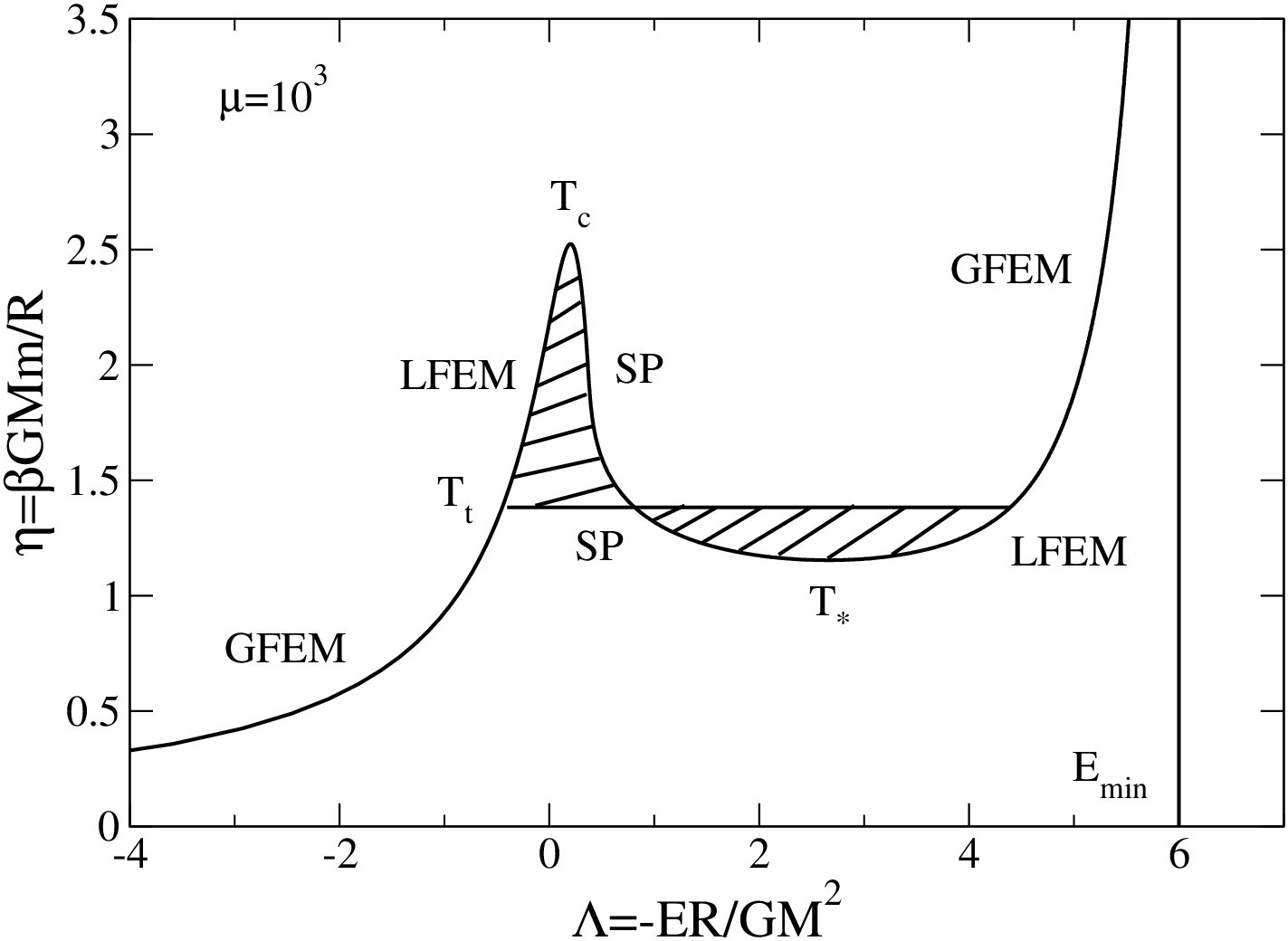}
\caption{Maxwell construction in the canonical ensemble determining the
transition temperature $T_{t}(\mu)$ at which the gaseous states pass from global
minima of free energy (GFEM) to local minima of free energy (LFEM) while the
condensed states pass from local minima
of free energy  to global minima of free energy. }
\label{ETcano3}
\end{center}
\end{figure}

\begin{figure}
\begin{center}
\includegraphics[clip,scale=0.3]{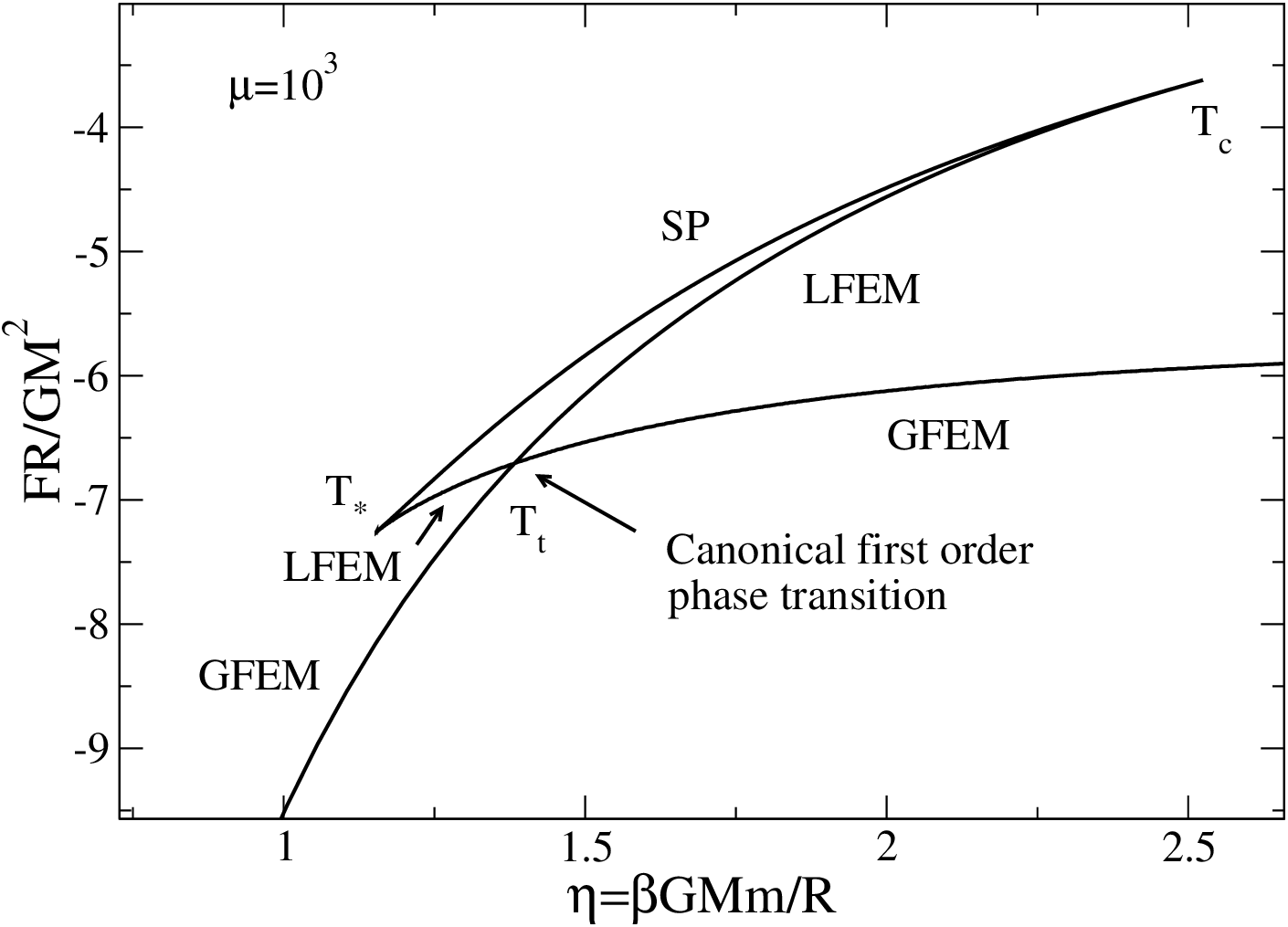}
\caption{Free energy of the gaseous and condensed phases versus
temperature for $\mu=10^3$. In the canonical ensemble, a first order
phase transition is expected at $T_t$ at which the two branches
intersect. The kink in the curve $F(\beta)$ corresponds to a jump in
energy in the caloric curve $E(\beta)$. However, the barrier of free energy played by the saddle point (SP) prevents this phase transition to occur in practice \cite{ijmpb,metastable}.}
\label{Fcano3}
\end{center}
\end{figure}

\begin{figure}
\begin{center}
\includegraphics[clip,scale=0.3]{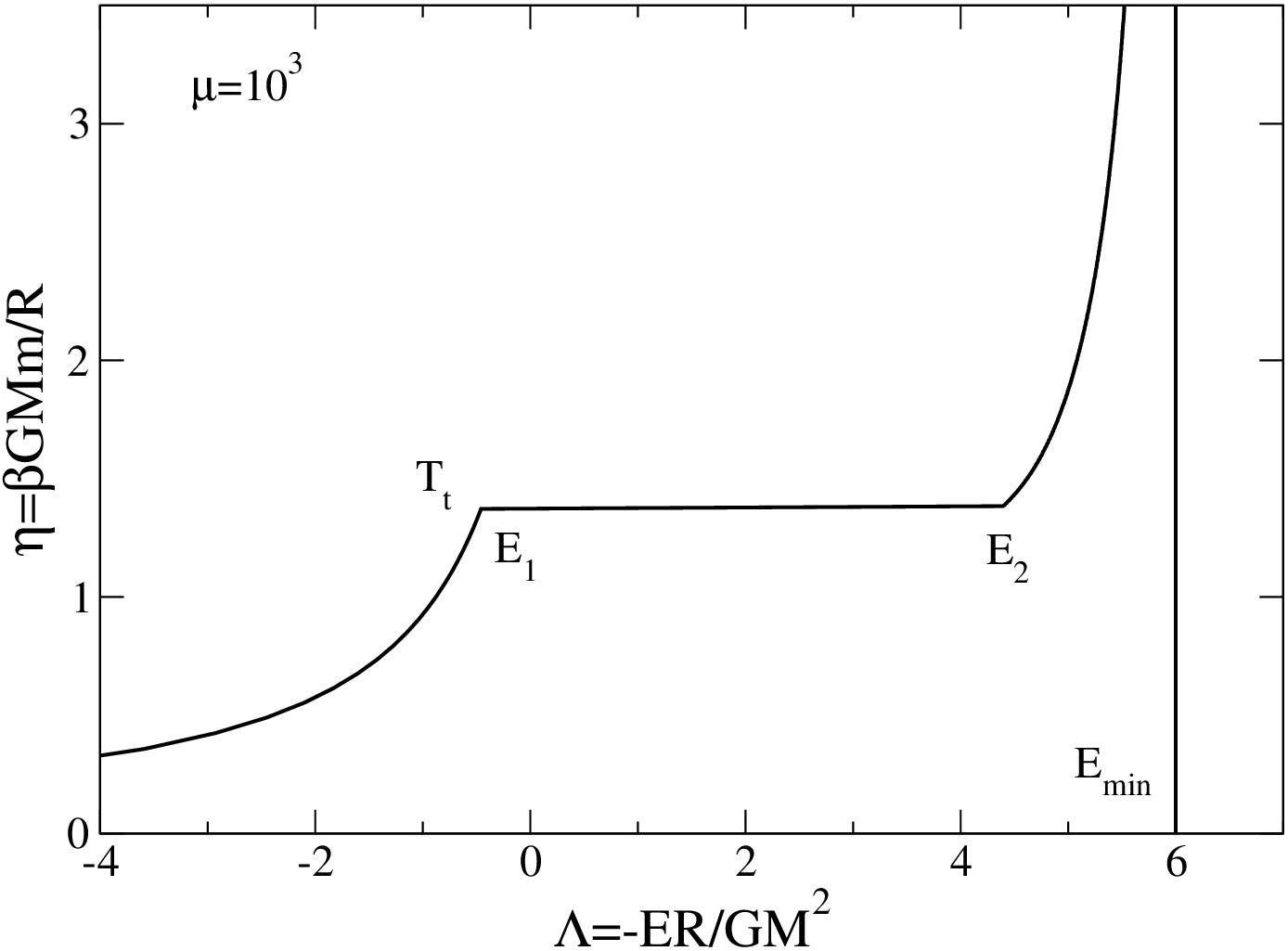}
\caption{Strict canonical caloric curve for $\mu=10^{3}$.
This figure is obtained from Fig. \ref{ETcano3} by keeping only global minima of
free energy (GFEM). It corresponds therefore to the exact canonical caloric
curve $\langle E\rangle_{cano}(T)=-\partial\ln Z/\partial\beta$ which  is
univalued \cite{ijmpb}. It does not display negative specific heats contrary to
the
microcanonical caloric curve with the same value of the degeneracy parameter
(see Fig. \ref{ETmicro3}). For $N\rightarrow +\infty$, there is a discontinuity
of energy (latent heat) at the transition temperature $T_{t}(\mu)$. For finite
$N$ systems, this discontinuity is smoothed-out \cite{ispolatov}. Although this
canonical caloric
curve is correct in a strict sense, it is {\it not} physical because it ignores
metastable states that have an infinite lifetime in the thermodynamic limit
$N\rightarrow +\infty$. The physical canonical caloric curve (see Fig.
\ref{ETcano3meta}) is obtained from Fig. \ref{ETcano3instable} by discarding the
unstable saddle points of free energy that form the intermediate branch with
negative specific heats.}
\label{ETcano3strict}
\end{center}
\end{figure}

We must now determine which states are local minima of free energy
and which states are global minima of free energy. This can be done
by performing a horizontal Maxwell construction \cite{ijmpb}
or by plotting the
free energy of the two phases
and determining at which temperature they become equal. The Maxwell
construction is shown in Fig. \ref{ETcano3}. The free energy
versus temperature plot is represented in Fig. \ref{Fcano3}. There
is a crossover at the transition temperature $T_{t}(\mu)$. For
$T>T_{t}(\mu)$ the gaseous states are global minima of free energy
(GFEM) and for $T_{c}<T<T_{t}(\mu)$ the gaseous states are local
minima of free energy (LFEM), i.e. metastable states. Inversely, for
$T_{t}<T<T_{*}$ the condensed states are local minima of free energy
(metastable states) and for $T<T_{t}$ they are global minima of free
energy. In a strict sense, the statistical equilibrium states
correspond to the global minima of free energy.  Therefore, the
strict caloric curve in the canonical ensemble for $\mu=10^{3}$ is
the one represented in Fig. \ref{ETcano3strict}. It is obtained by
keeping only the global maxima of free energy at fixed mass. From
this curve, we expect the occurrence of a {\it canonical
first order} phase transition at $T=T_t$ connecting the gaseous phase
to the condensed phase.  It is accompanied by a discontinuity
of energy and specific heat, i.e. a huge release of latent heat $\Delta E=E_1-E_2$. For
finite values of $N$, the discontinuity is smoothed-out
because close to $T_{t}(\mu)$ the metastable states contribute
significantly to the partition function \cite{ispolatov,metastable}.
It is instructive to compare the strict canonical caloric curve of
Fig. \ref{ETcano3strict} with the microcanonical caloric curve of Fig.
\ref{ETmicro3} for the same value of the degeneracy parameter
$\mu=10^3$. We see that the region of negative specific heats in the
microcanonical ensemble is replaced by an isothermal phase transition (plateau) in the
canonical ensemble that connects the gaseous phase to the condensed phase. This corresponds to a situation of ensemble inequivalence: the energies between $E_1$ and $E_2$ are accessible in the microcanonical ensemble but not in the canonical ensemble.
We also note that these two curves are strikingly
analogous to those obtained with the simple model of Padmanabhan
\cite{paddy} with $N=2$ particles (see also Fig. 1 in \cite{ijmpb}), although in that case the plateau is smoothed out due to finite $N$ effects.

\begin{figure}
\begin{center}
\includegraphics[clip,scale=0.3]{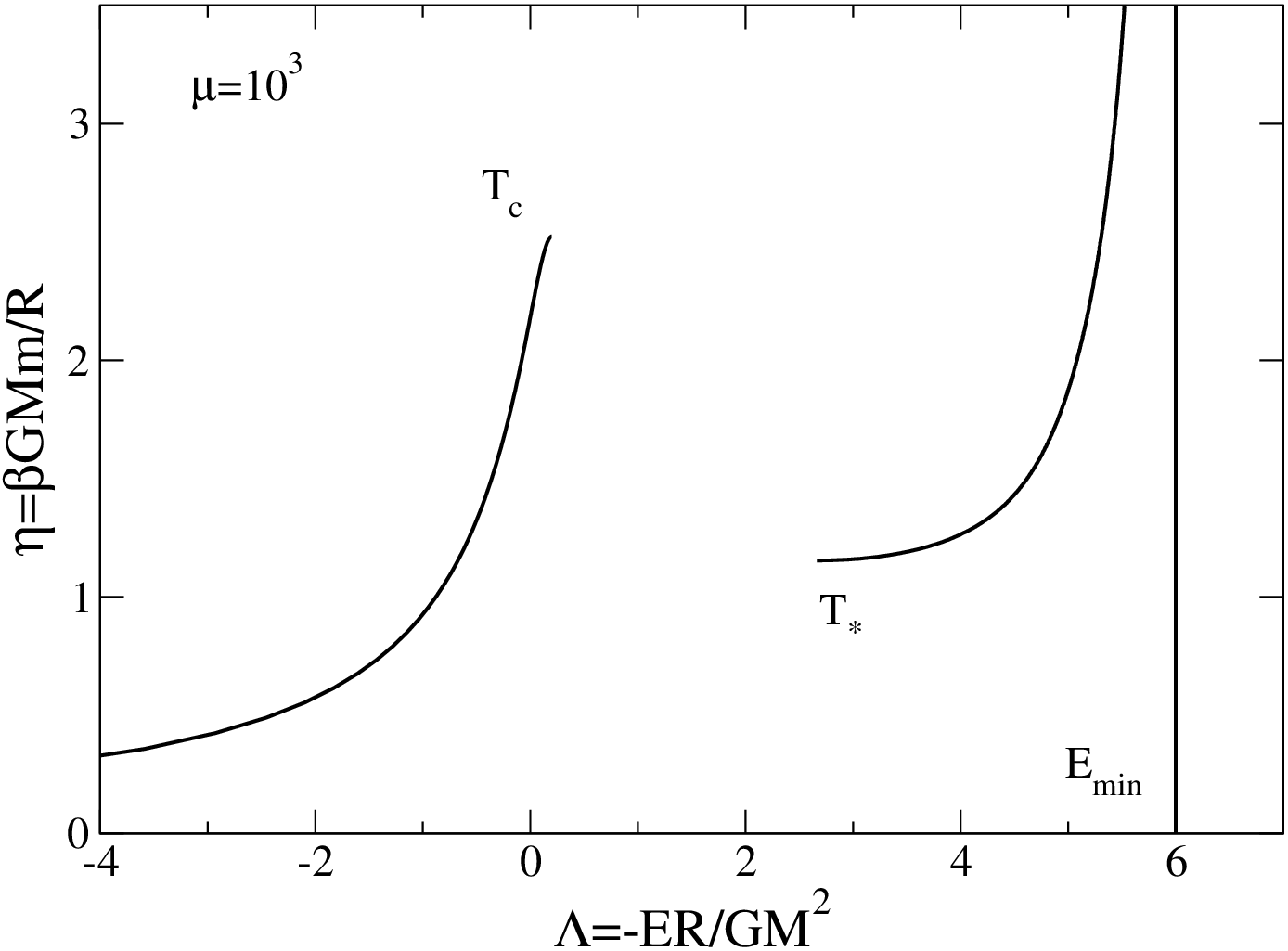}
\caption{Physical canonical caloric curve for $\mu=10^{3}$. This figure displays stable and metastable equilibrium states (since the control parameter in the canonical ensemble is the temperature, it is more logical to rotate the figure by $90^{o}$ to look at it). The metastable states have infinite lifetime in the thermodynamic limit so they are physically relevant. The canonical first order  phase transition of Fig. \ref{ETcano3strict} does not take place in practice. Note that the structure of the physical canonical caloric curve remains the same for larger values of $\mu$. Therefore, although the series of equilibria becomes more and more complex because of the appearance of the spiral, this does not affect the physical canonical caloric curve (the spiral corresponds to unstable states). }
\label{ETcano3meta}
\end{center}
\end{figure}

\begin{figure}
\begin{center}
\includegraphics[clip,scale=0.3]{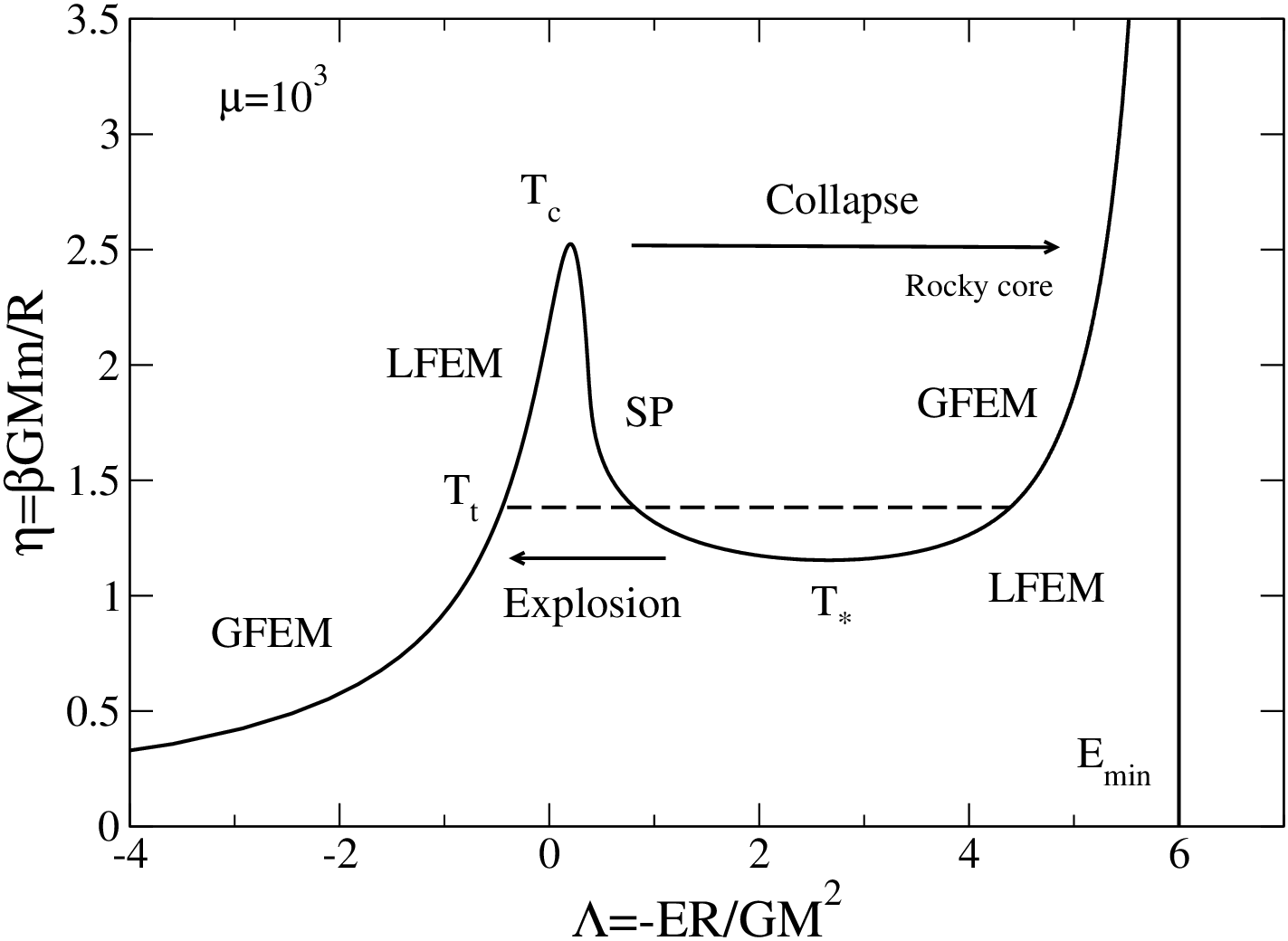}
\caption{Summary of the phase transitions for the self-gravitating
gas with a degeneracy parameter $\mu=10^{3}$ in the canonical
ensemble. The points on the left branch form the ``gaseous'' phase. They are global
minima of free energy (GFEM) for $T>T_{t}(\mu)$ and local minima of
free energy (LFEM), i.e.  metastable states, for $T<T_{t}(\mu)$. The
points on the right branch form the
``condensed'' phase. They are LFEM for $T>T_{t}(\mu)$ and GFEM for
$T<T_{t}(\mu)$. The points on the intermediate branch are
unstable saddle points (SP). Due to the existence of metastable
states, the system displays a canonical hysteretic cycle marked by a
``collapse'' and an ``explosion'' at the points where the branch of
metastable states disappears: for $T\le T_c$, a gaseous
configuration undergoes an isothermal collapse and for $T\ge
T_*(\mu)$, a condensed configuration undergoes an explosion. These canonical phase
transitions exist only above a critical point $\mu_{CCP}=32.4$ (see
Sec. \ref{sec_maxwell}). This corresponds to a system size $R>3.19R_{*}$. }
\label{ETcano3summary}
\end{center}
\end{figure}

The preceding discussion suggests the occurrence of a canonical first order
 phase transition at $T_{t}$. However, for self-gravitating systems, the
metastable states are long-lived 
because the probability of a fluctuation able to trigger the phase
transition is extremely weak. Indeed, the system has to cross the
free energy barrier played by the solutions on the intermediate
branch. For long-range systems, this barrier of potential increases
linearly with the number $N$ of particles so that the probability of
transition scales as $e^{-N}$ (see Sec. \ref{sec_lifetime}).  Therefore,
the metastable states are extremely robust and the canonical first
order phase transition at $T_t$ does not take place in practice. For
$N\gg 1$, the system remains on the metastable branch past the
transition temperature $T_t$. Therefore the strict canonical caloric
curve of Fig. \ref{ETcano3strict} is not physical. The physical
canonical caloric curve is the one shown in Fig. \ref{ETcano3meta} which
takes  metastable states into account. It is obtained from the
series of equilibria of Fig. \ref{ETcano3instable} by discarding only the
unstable saddle points of free energy at fixed mass. As a result, the physical region of ensemble inequivalence corresponds to the energies between $E_{gas}$ and $E_{cond}$. They are accessible in the microcanonical ensemble but not in the canonical ensemble.

The true phase transition occurs at the critical temperature
$T_{c}$ (Emden temperature) at which the metastable branch disappears (see
Fig. \ref{ETcano3summary}).
This point CE1 corresponds to a canonical
spinodal point. Below $T_{c}$ the system undergoes an isothermal
collapse.  However, for self-gravitating particles with an exclusion constraint in position space, the core ceases to
shrink when it becomes degenerate.  Since this collapse is accompanied
by a discontinuous jump of free energy (see Fig. \ref{Fcano3}), this
is sometimes called a {\it canonical zeroth order} phase transition. The
resulting equilibrium state possesses a small degenerate nucleus which
contains a large fraction of the total mass. For $\mu=10^3$ the core following
the collapse at $T=T_c^-$ contains about $63.5\%$ of the total mass (its radius defined as the distance
at which the density is equal to $0.95\sigma_0$ is about $8.59\
10^{-2}R$) and this fraction of mass increases for larger values of $\mu$.
\footnote{This is expected because in the classical limit
$\mu\rightarrow +\infty$ the degenerate nucleus should be replaced by
a ``Dirac peak'' containing the whole mass (see the Introduction).} The
rest of the mass is diluted in a halo. In an
open system, the halo would be dispersed at infinity so that only the
degenerate core (equivalent to a self-gravitating homogeneous sphere) would remain.

For $T<T_{t}(\mu)$, the condensed states are global
minima of free energy. If now temperature is increased, the system remains in the
condensed phase past the transition temperature $T_t$. Indeed, for the
same reason as before, the condensed states with $T_{t}<T<T_{*}$ are
long-lived metastable states so that the first order phase transition
from the condensed phase to the gaseous phase does not take
place. However, above $T_{*}$ the condensed metastable branch
disappears and the system undergoes a discontinuous transition
reversed to the collapse at $T_c$ (see Fig. \ref{ETcano3summary}). This
transition is sometimes called an ``explosion'' since it transforms
the dense core into a relatively uniform mass distribution. Since the
collapse and the explosion occur at different values of temperature,
due to the presence of metastable states, we can generate an {\it
hysteretic cycle} in the canonical ensemble
by varying the temperature between $T_{*}$ and $T_{c}$.  This
hysteretic cycle has been followed
numerically by Chavanis {\it et al.} \cite{crrs} considering a model of
self-gravitating Brownian fermions and solving the
Smoluchowski-Poisson system with a fermionic equation of state. Similar results
would be obtained for a self-gravitating Brownian gas with an exclusion constraint
in position space by solving the Smoluchowski-Poisson system with
the equation of state (\ref{eos2}) (see Sec. \ref{sec_kinetic}).

\subsection{The classical limit $\mu\rightarrow +\infty$}
\label{sec_muinf}

\begin{figure}
\begin{center}
\includegraphics[clip,scale=0.3]{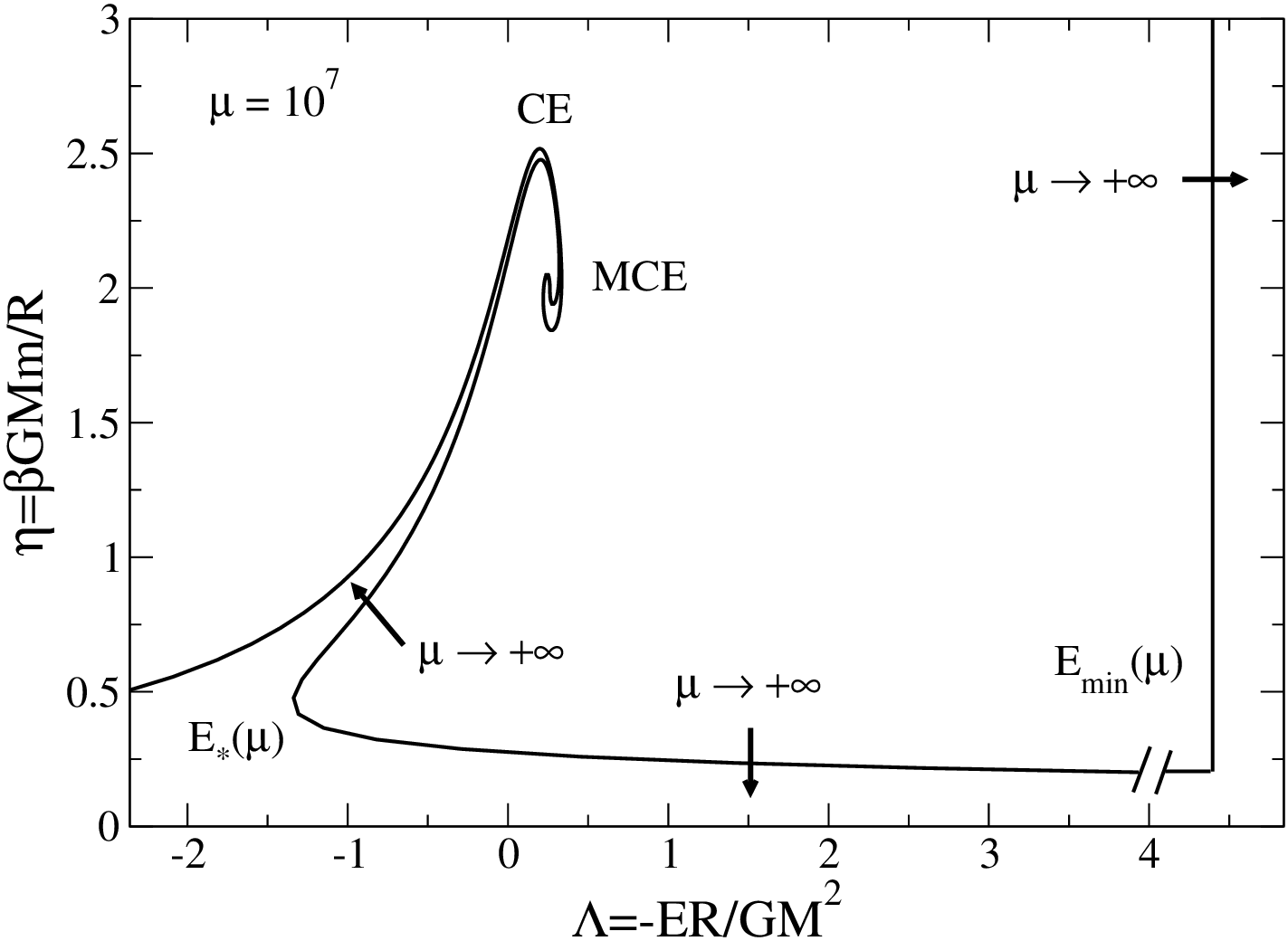}
\caption{Caloric curve of the self-gravitating gas with an exclusion constraint in position space in the classical limit $\mu\rightarrow +\infty$ (here $\mu=10^{7}$). The characteristic
energies and the characteristic temperatures behave like $\Lambda_{max}(\mu)=(3/5)\ \mu^{1/3}$,
$\Lambda_{*}(\mu)\sim \Lambda_{t}(\mu)\sim -\mu^{1/3}/(\ln\mu)^{5/3}$, and $\eta_{*}(\mu)\sim \eta_{t}(\mu)\sim \ln\mu/\mu^{1/3}$ \cite{pt}.}
\label{infinity}
\end{center}
\end{figure}

\begin{figure}
\begin{center}
\includegraphics[clip,scale=0.3]{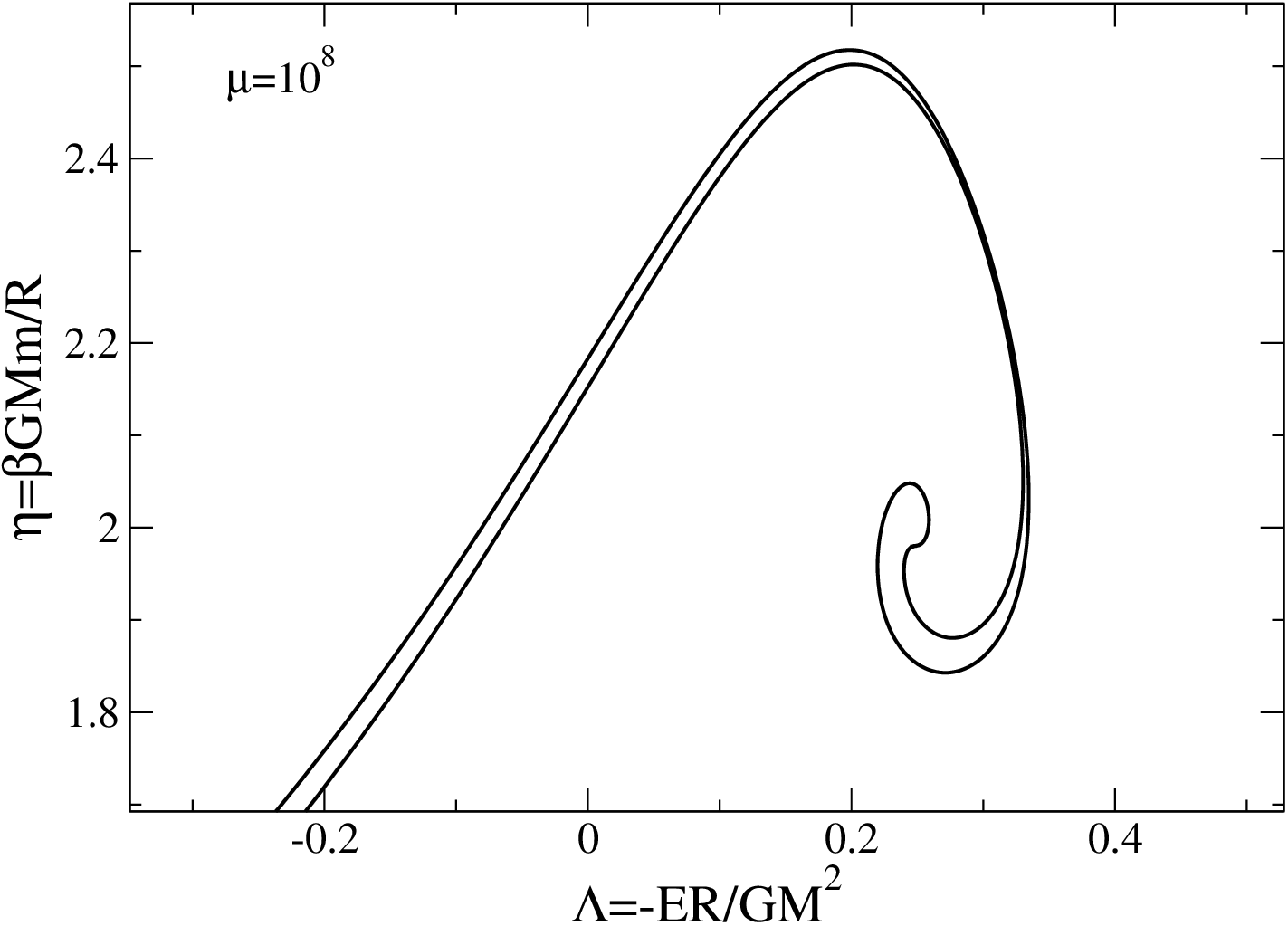}
\caption{Zoom on the spiral for a large value of $\mu$ (specifically $\mu=10^8$). The spiral rotates several times before unwinding. However, this is essentially a mathematical curiosity since the states on the spiral are unphysical (unstable) as discussed in the text.}
\label{doublespirale}
\end{center}
\end{figure}

It is of interest to discuss the classical limit $\mu\rightarrow
+\infty$ specifically so as to make the connection with the
results recalled in the Introduction. For large but finite values of $\mu$,
the series of equilibria winds up and makes several turns before
finally unwinding (see Figs. \ref{infinity} and \ref{doublespirale}).
A mode of stability is lost each time the curve winds up (rotates clockwise)
and a mode of stability is regained each time the curve unwinds (rotates
anti-clockwise). Therefore, only the part of the series of equilibria before
the first turning point and after the last turning point is stable. The states on the spiral are unstable, hence unphysical. Let us discuss the limit $\mu\rightarrow +\infty$ in the microcanonical and canonical ensembles more precisely.

For $\mu\rightarrow +\infty$, in the microcanonical ensemble, the transition energy $E_{t}(\mu)$ is rejected to $+\infty$ so that the gaseous states up to MCE are metastable (local entropy maxima) while the condensed states are fully stable (global entropy maxima). The minimum energy (ground state) is rejected to $-\infty$.
Therefore, the branch of condensed states coincides with the $x$-axis at
$\beta=0$. They are made of a ``binary star'' (i.e. a core with a small mass but a huge potential energy) surrounded by a hot halo with $T\rightarrow +\infty$. Therefore, the $\mu\rightarrow +\infty$ limit of the caloric curve (Fig. \ref{infinity}) is formed by
the metastable gaseous branch of Fig. \ref{dim3} up to MCE  plus a singular stable
condensed branch at $\beta=0$ coinciding with the $x$-axis (binary star $+$ hot halo). On the other hand, the saddle points are superposed to the spiral and to the branch of gaseous states although they have a very different structure (germ).

For $\mu\rightarrow +\infty$, in the canonical ensemble, the transition temperature $T_{t}(\mu)$ is rejected to $+\infty$ so that the gaseous states up to CE are metastable (local minima of free energy) while the condensed states are fully stable (global minima of free energy). The branch of condensed states (vertical line) is rejected to $E\rightarrow -\infty$. They are made of ``Dirac peaks''. Therefore, the $\mu\rightarrow +\infty$ limit of the caloric curve (Fig. \ref{infinity}) is formed by
the metastable gaseous branch of Fig. \ref{dim3} up to CE  plus a singular stable
condensed branch at $E=-\infty$ (Dirac peak). On the other hand, the saddle points are superposed to the spiral and to the branch of gaseous states although they have a very different structure (germ).

\subsection{Microcanonical and canonical critical points}
\label{sec_maxwell}

\begin{figure}
\begin{center}
\includegraphics[clip,scale=0.3]{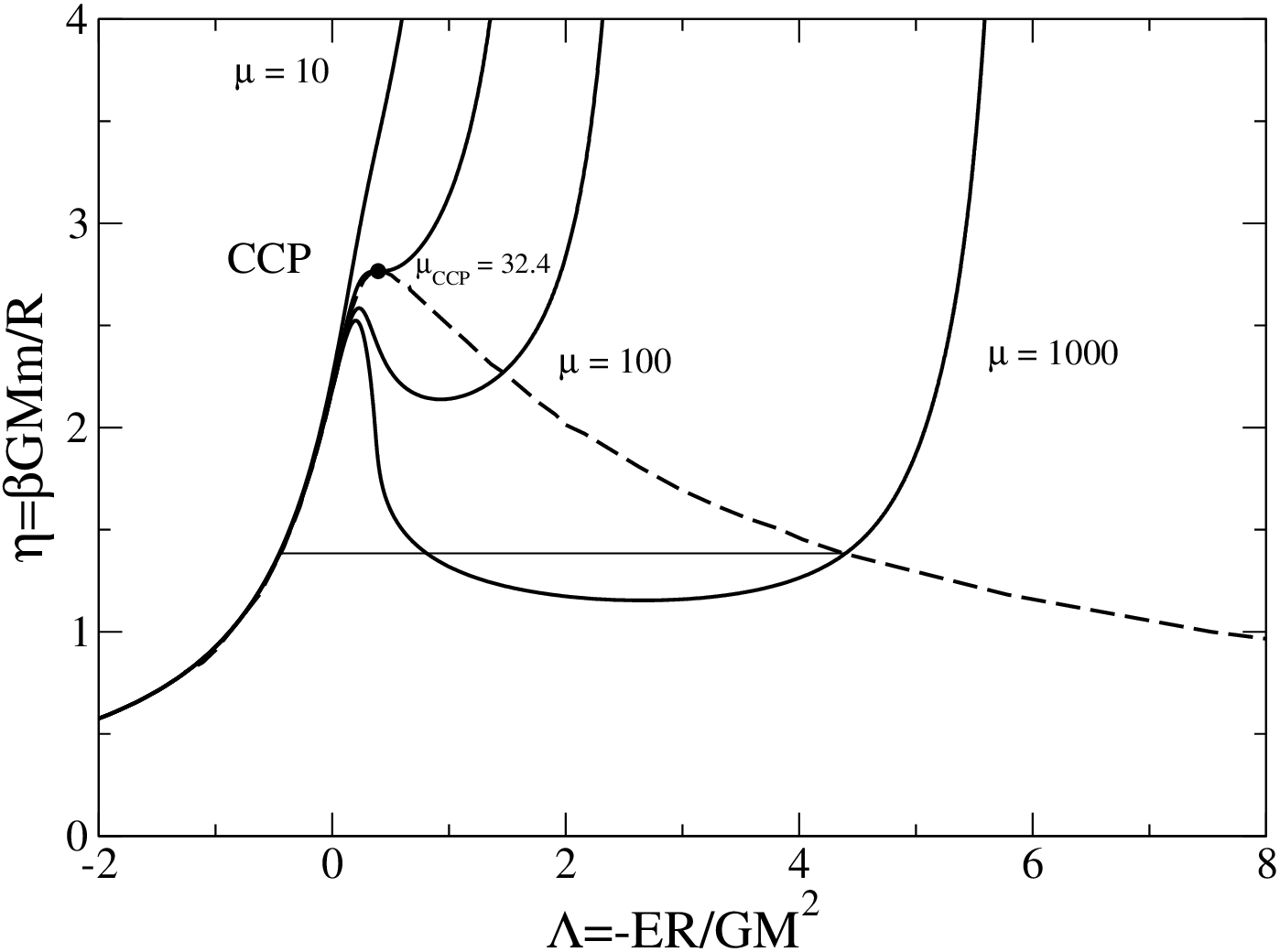}
\caption{Enlargement of the caloric curve near the critical
point in the canonical ensemble ($\mu_{CCP}=32.4$, 
$\Lambda_{CCP}\simeq 0.393$, $\eta_{CCP}\simeq 2.7657$).}
\label{ccp}
\end{center}
\end{figure}

The deformation of the series of equilibria as a function of the
degeneracy parameter $\mu$ is represented in
Fig. \ref{calomulti}. There exist two critical points in the problem,
one in each ensemble. For $\mu<\mu_{CCP}\simeq 32.4$, the curve
$\beta(E)$ is monotonic, so there is no phase transition. For
$\mu>\mu_{CCP}\simeq 32.4$, the curve $E(\beta)$ is multi-valued so that
a {\it canonical first order} phase transition is expected. The
temperature of transition $T_t(\mu)$ in the canonical ensemble can be obtained by
a horizontal Maxwell construction \cite{ijmpb}.
If we keep only global minima of free energy, the $N$-curve has to be
replaced by a horizontal plateau. We see that the extent of the
plateau decreases as $\mu$ decreases (see the dashed line in Fig. \ref{ccp}).  At the canonical
critical point $\mu_{CCP}$, the plateau disappears and the caloric  curve $E(T)$
presents an inflexion point. At that point the specific heat is infinite.

\begin{figure}
\begin{center}
\includegraphics[clip,scale=0.3]{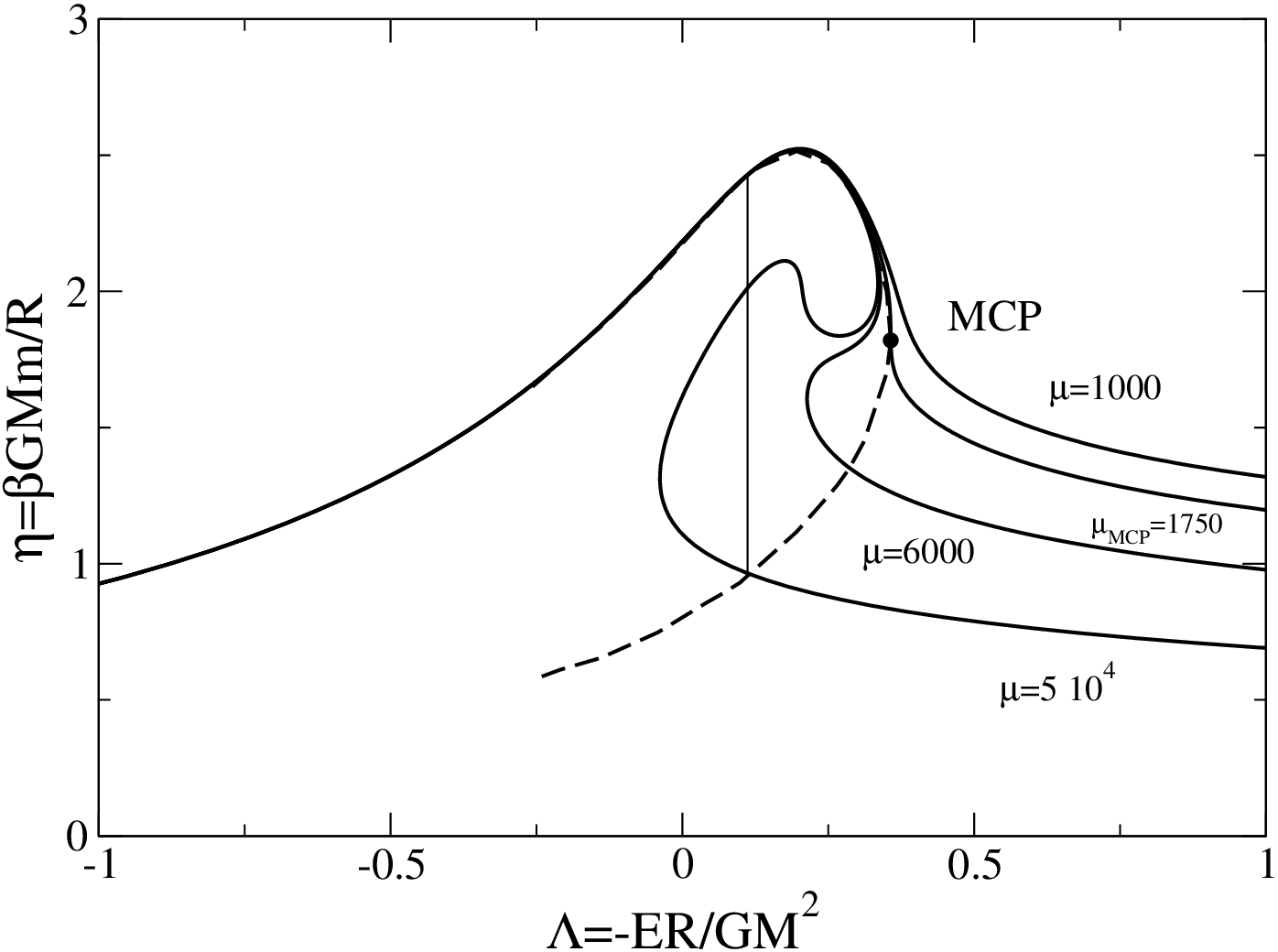}
\caption{Enlargement of the caloric curve near the critical
point in the microcanonical ensemble ($\mu_{MCP}=1750$,
$\Lambda_{MCP}\simeq 0.357$, $\eta_{MCP}\simeq 1.82$).}
\label{mcp}
\end{center}
\end{figure}

For $\mu>\mu_{MCP}\simeq 1750$, the curve $\beta(E)$ is multivalued so
that a {\it microcanonical first order} phase transition is expected
(in addition to the canonical first order phase transition that exists
for any $\mu>\mu_{CCP}$). The energy of transition $E_t(\mu)$ can be obtained by
a vertical Maxwell construction \cite{ijmpb}.
If we keep only global maxima of entropy, the $Z$-curve has to be
replaced by a vertical plateau. We see that the extent of the plateau
decreases as $\mu$ decreases (see the dashed line in Fig. \ref{mcp}). At the
microcanonical critical point $\mu=\mu_{MCP}$, the plateau disappears
and the caloric curve $T(E)$ presents an inflexion point. At that point 
the specific heat vanishes.

Therefore, for $\mu>\mu_{MCP}$, the system exhibits a
microcanonical and a canonical phase transition, for
$\mu_{CCP}<\mu<\mu_{MCP}$ the system exhibits  only a canonical
phase transition, and for $\mu<\mu_{CCP}$ the system does not exhibit
any phase transition. We recall, however, that due to the presence of long-lived
metastable states, the first order phase transitions and the
``plateaux'' are not physically relevant. Only the zeroth order phase
transitions that occur at $E_{c}$ in MCE and at $T_{c}$ in CE
(spinodal points) are relevant.

\subsection{Phase diagrams and ensemble inequivalence}
\label{sec_pd}

\begin{figure}
\begin{center}
\includegraphics[clip,scale=0.3]{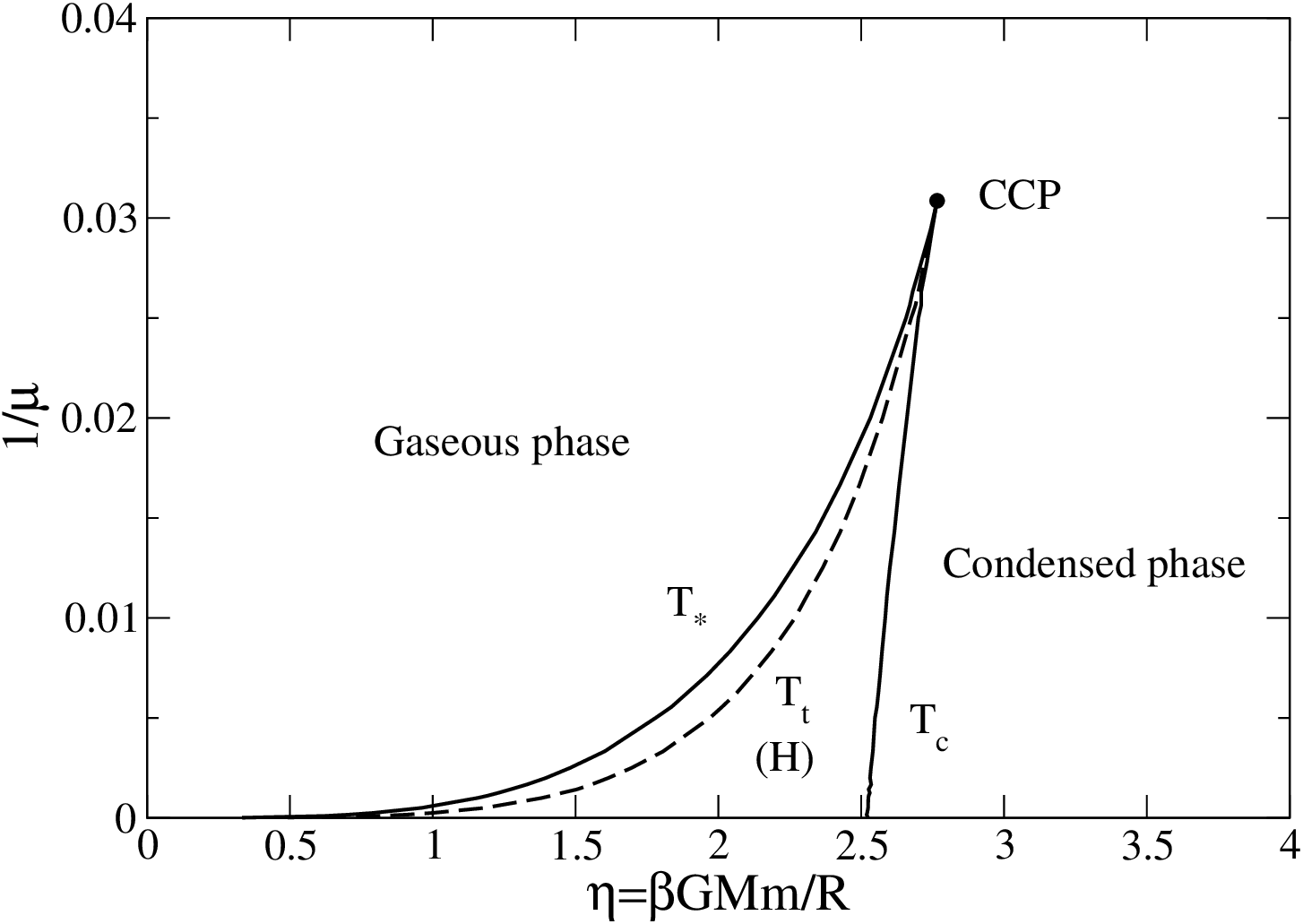}
\caption{Canonical phase diagram in the $(T,\mu)$ plane. The $H$-zone
between $T_{c}$ and $T_{*}$ corresponds to an hysteretic zone where the actual phase depends on
the history of the system. If the system is initially prepared in a gaseous
state, it will remain gaseous until the minimum temperature $T_{c}$ at which it
will collapse and become condensed. Inversely, if the system is initially
prepared in a condensed state, it will remain condensed until the maximum
temperature $T_{*}$ at which it will explode and become gaseous.}
\label{phasecano}
\end{center}
\end{figure}

Typical caloric curves illustrating microcanonical first order  and canonical
first order phase transitions are shown in Figs. \ref{ETmicro5}
and \ref{ETcano3} respectively. The equilibrium phase diagram of self-gravitating
particles with an exclusion constraint in position space
can be directly deduced from these curves by identifying
characteristic energies and characteristic temperatures.  In the canonical ensemble,
we note $T_{t}$ the temperature of transition (determined by the
equality of the free energies of the two phases), $T_{c}$ (Emden
temperature) the end point of the metastable gaseous phase (first
turning point of temperature) and $T_{*}$ the end point of the
metastable condensed phase (last turning point of temperature). The
canonical phase diagram is represented in Fig. \ref{phasecano}. It shows in
particular the canonical critical point $\mu_{CCP}=32.4$ at which the
canonical first order phase transition disappears.

\begin{figure}
\begin{center}
\includegraphics[clip,scale=0.3]{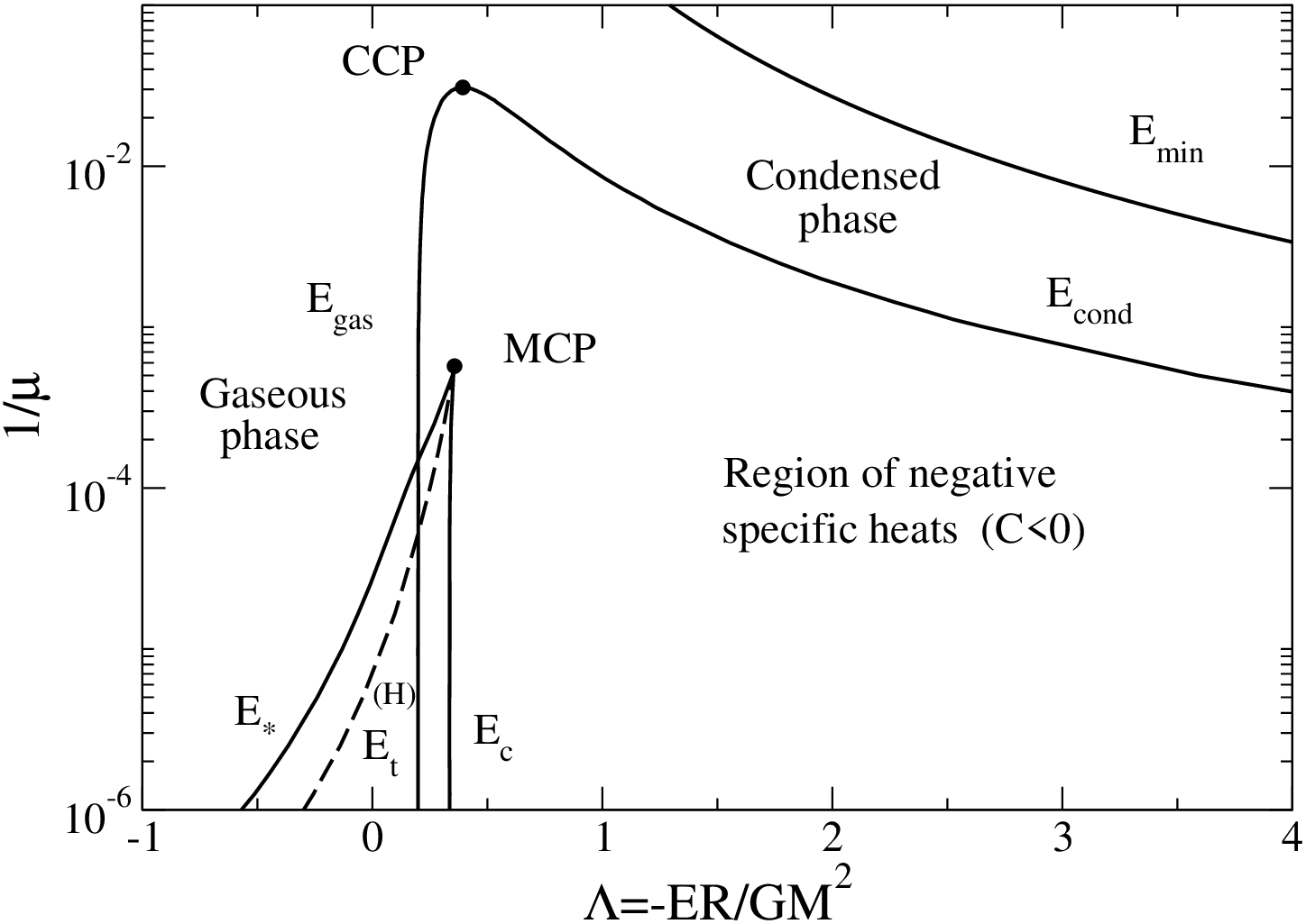}
\caption{Microcanonical phase diagram in the $(E,\mu)$ plane.  The
$H$-zone corresponds to an hysteretic zone where the actual phase
depends on the history of the system. The phase diagram in MCE is more
complex than in CE due to the existence of the negative specific heat
region that is forbidden in CE. This corresponds to a region of ensemble
inequivalence.}
\label{phasemicro}
\end{center}
\end{figure}

In the microcanonical ensemble, we note $E_{t}$ the energy of
transition (determined by the equality of the entropy of the two
phases), $E_{c}$ (Antonov energy) the end point of the metastable
gaseous phase (first turning point of energy) and $E_{*}$ the end
point of the metastable condensed phase (last turning point of
energy). We also denote by $E_{gas}$ the energy at which we enter in
the zone with negative specific heat (first turning point of
temperature) and $E_{cond}$ the energy at which we leave the zone of negative specific heat
(last turning point of
temperature). We also introduce the minimum energy $E_{min}$ (ground
state). The microcanonical phase diagram is represented in Fig.
\ref{phasemicro}. It shows in particular the microcanonical critical
point $\mu_{MCP}=1750$ at which the microcanonical first order phase
transition disappears. The structure of the equilibrium phase diagrams
can be easily understood in the light of the preceding discussion.

We note that the microcanonical phase diagram is more complex than the
canonical one due to the existence of a negative specific heat region.
We recall that canonical stability (minimum of $F$ at fixed $M$) is a
sufficient, but not necessary, condition of microcanonical stability
(maximum of $S$ at fixed $M$ and $E$) \cite{ellis}.
Hence, canonical stability implies microcanonical
stability but not the opposite. Since canonical equilibria are always
realized as microcanonical equilibria, they constitute a sub-domain of
the microcanonical phase diagram. As a result, the microcanonical
ensemble is wider than the canonical one. For example, it can contain configurations
with negative specific heats that are forbidden in the canonical ensemble.
More precisely, the region of ensemble inequivalence is as follows. For $\mu<\mu_{CCP}$ the series of equilibria is monotonic so the ensembles are equivalent. For $\mu>\mu_{CCP}$, the states with energy between $E_{gas}$ and $E_{cond}$ are accessible in the microcanonical ensemble but not in the canonical ensemble (compare Figs. \ref{ETmicro3} and \ref{ETcano3meta}). This corresponds to the physical region of ensemble inequivalence taking metastable states into account. If we consider only fully stable states, the strict region of ensemble inequivalence corresponds to the states with energy
between  $E_1$ and $E_2$ (compare Figs. \ref{ETmicro3} and \ref{ETcano3strict}). However, we have explained that the metastable states must be considered as stable states.

\subsection{The lifetime of metastable states}
\label{sec_lifetime}

\begin{figure}
\begin{center}
\includegraphics[clip,scale=0.3]{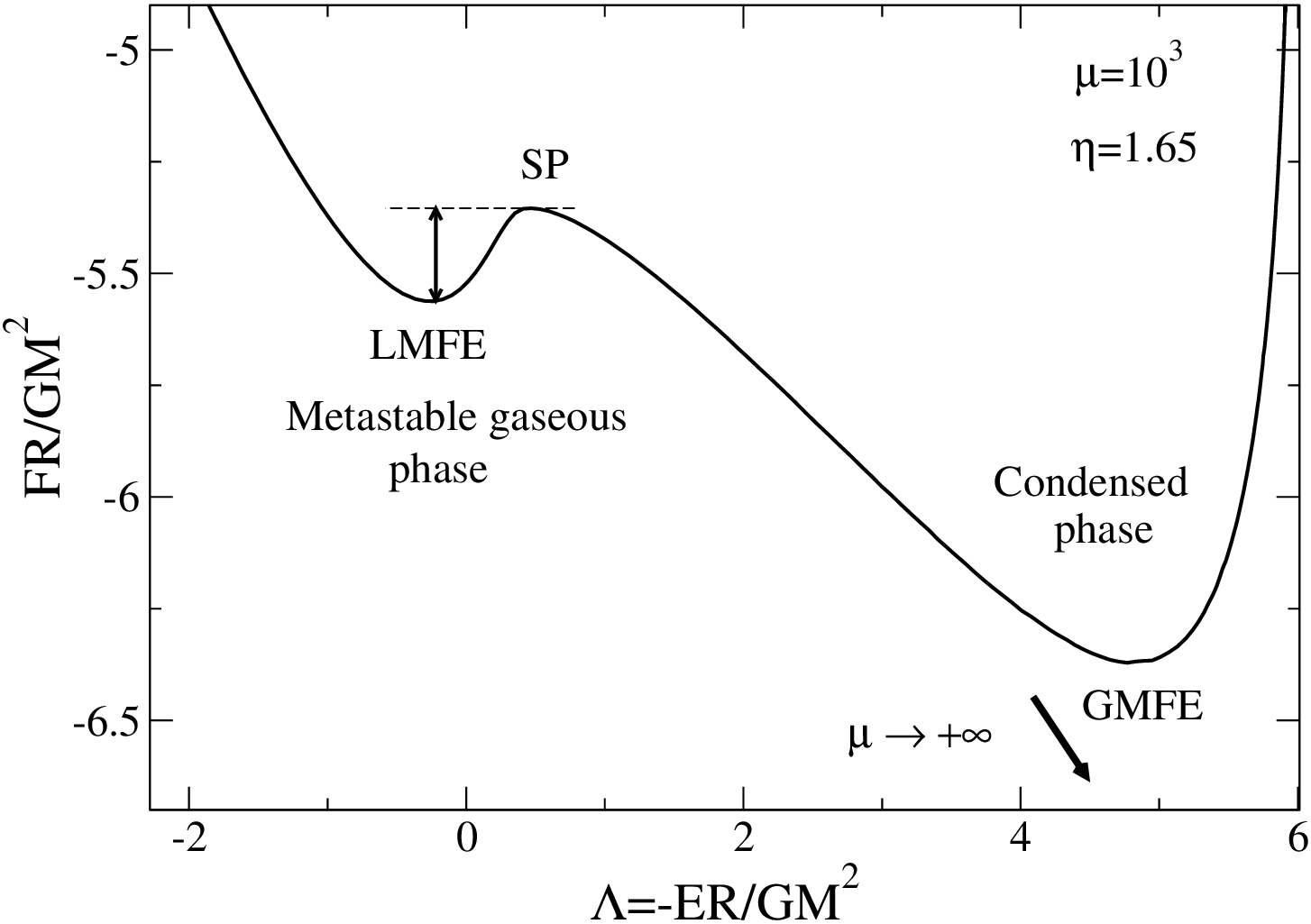}
\caption{Free energy
as a function of the energy for a self-gravitating Brownian gas at temperature
$T$ (the figure corresponds to $\eta=1.65$ and $\mu=10^3$). The distribution of
the energies at temperature $T$ in the canonical ensemble is
$P(E)=(1/Z(\beta))e^{-\beta F(E)}$ \cite{ijmpb}. For $T_c<T<T_*$, the free
energy presents a local minimum (metastable state), a local maximum (unstable),
and a global minimum (fully stable). Since $T<T_t$ in the example of the figure, the
global minimum corresponds to the condensed phase and the local minimum to the
gaseous phase. The saddle point creates a barrier of free energy that hampers
the transition from the metastable state to the stable state. Since the barrier
of free energy is proportional to $N$, the  metastable states have a very long
lifetime. In the classical limit $\mu\rightarrow +\infty$, the free energy of
the global minimum tends to $-\infty$ (Dirac peak) while the local minimum and
the saddle point do not sensibly change. }
\label{metastable}
\end{center}
\end{figure}

The lifetime of a metastable state can be estimated by using an adaptation of the Kramers formula \cite{risken}. Let us first consider a system of self-gravitating Brownian particles in the canonical ensemble (fixed temperature $T$). For $\mu=10^3$, the series of equilibria is represented in Fig. \ref{ETcano3}.  The distribution of the smooth density $\rho({\bf r})$ at a fixed temperature $T$ is given by \cite{ijmpb,emergence}:
\begin{eqnarray}
\label{lifetime1}
P[\rho]=\frac{1}{Z(\beta)}e^{-\beta F[\rho]}\delta(M[\rho]-M).
\end{eqnarray}
For $T_c<T<T_*$, the free energy has a local minimum $F_{meta}$ (metastable state), a saddle point $F_{saddle}$ (unstable), and a global minimum $F_{stable}$ (fully stable). The saddle point creates a barrier of free energy that hampers the transition from the metastable state to the stable state (see Fig. \ref{metastable}). For a system initially prepared in the metastable state, the probability for a fluctuation to drive it to a state with density $\rho({\bf r})$ is $P[\rho]\sim e^{-\beta (F[\rho]-F_{meta})}$. If the fluctuations bring the system in a configuration $\rho_{saddle}({\bf r})$ corresponding to the saddle point of free energy, it will collapse to the stable state. Therefore, the lifetime of the metastable state may be estimated by $t_{life}\sim 1/P[\rho_{saddle}]$, i.e. $t_{life}\sim e^{\beta\Delta F}$ where $\Delta F=F_{saddle}-F_{meta}$ is the barrier of free energy between the metastable state and the saddle point. In the thermodynamic limit of Sec. \ref{sec_thermo} the free energy is proportional to $N$ so we can write $F[\rho]=N f[\rho]$ where $f[\rho]\sim 1$. Therefore, we obtain the estimate
\begin{eqnarray}
\label{lifetime2}
t_{life}\sim e^{N\beta\Delta f}.
\end{eqnarray}
Except in the vicinity of the critical point $T_c$ where $\Delta f\rightarrow 0$, the lifetime of a metastable state increases exponentially rapidly with the number of particles, as $e^N$,  and becomes infinite in the thermodynamic limit $N\rightarrow +\infty$. Therefore, for systems with long-range interactions, metastable states have considerable lifetimes and they can be regarded as stable states in practice \cite{metastable}. Close to the critical point the fluctuations are important and the collapse can take place slightly above the critical temperature $T_c$ (finite $N$ corrections to the onset of the gravitational collapse have been calculated in \cite{katzokamoto,metastable}). Similar results are obtained in the microcanonical ensemble except that the barrier of free energy $\Delta F/k_B T$ is replaced by a barrier of entropy $\Delta S$. In the microcanonical situation the transition is only induced by finite $N$ fluctuations since there is no coupling with a thermal bath in that case \cite{rieutord,art,metastable}.

It is interesting to consider the classical limit $\mu\rightarrow +\infty$ in order to make the link with the discussion given in the Introduction. For $\mu\rightarrow +\infty$, the transition temperature $T_t\rightarrow +\infty$ so the metastable states (LFEM) correspond to the gaseous states and the fully stable states (GFEM) correspond to the condensed states. Furthermore, in the $\mu\rightarrow +\infty$ limit,  the condensed states become equivalent to ``Dirac peaks'' and the global minimum of free energy tends to $-\infty$ (see the thick arrow in Fig. \ref{metastable}). By contrast, the gaseous metastable states and the saddle points of free energy do not change sensibly with $\mu$ when $\mu\gg 1$ since they are not affected by the exclusion constraint. Therefore, for classical point masses above $T_c$, the free energy has no global minimum (the free energy can be made arbitrarily small by creating a Dirac peak) but it has a local minimum corresponding to a metastable gaseous state, and saddle points. In a sense, the Dirac peak is the most probable state in the canonical ensemble since it has infinite free energy. However, since the metastable gaseous states have a very long lifetime they are also very relevant. Actually, they are more relevant than the Dirac peak because their basin of attraction is wider. Starting from a generic initial condition, the system will relax towards a metastable gaseous state, not towards a Dirac peak. Once in the metastable state, the system will stay there for a very long time because the formation of a Dirac peak requires very particular correlations that take a very long time to develop. For $N\gg 1$, the probability to form a Dirac peak is exponentially small, so this will not occur in practice. Therefore, for $T>T_c$, the system will be found in a gaseous metastable state even if there exist configurations (Dirac peaks) with lower free energy.\footnote{We note that an ordinary gas in a box which is fully stable in the absence of interaction is only metastable (strictly speaking) if we take gravity into account, even if its effect is completely negligible. However, both $N$ and $\Delta s$ are huge resulting in considerable lifetimes. It is normal that these ``metastable'' states behave as stable states in order to avoid a paradox.} By contrast, for $T<T_c$, there is no metastable state anymore and, in that case, the system will collapse and form a Dirac peak. The discussion is similar in the microcanonical ensemble, except that the ``Dirac peak'' is replaced by a ``binary star $+$ hot halo'', as discussed in the Introduction.

In conclusion, at high temperatures, the gaseous metastable states (local minima of free energy)
 are more physically relevant than the condensed states (global minima of free energy) or than the ``Dirac peaks''  for classical systems ($F\rightarrow -\infty$). Indeed, the system can remain frozen in a metastable gaseous phase for a very long time.  The time required for a
metastable gaseous system to collapse is in general tremendously long
and increases exponentially with the number $N$ of particles (thus,
$t_{life}\rightarrow +\infty$ in the thermodynamic limit $N\rightarrow
+\infty$). This is due to the long-range nature of the
gravitational potential. Therefore, metastable states are in reality
stable states. Condensed objects (e.g., planets, stars, white dwarfs, dark matter halos,...)  or singularities (Dirac peaks, black holes) only form
below a critical  temperature $T_{c}$
(Emden temperature), when the gaseous metastable phase
ceases to exist (spinodal point) \cite{post,crrs}. Similarly, at high energies,  globular clusters are found in long-lived gaseous metastable states while below a critical energy $E_c$ (Antonov energy) they undergo core collapse and ultimately form  a condensed object or a singularity (binary star $+$ a hot halo) \cite{cohn}.

\section{Caloric curves in other dimensions of space}
\label{sec_ccother}

In this section, we briefly discuss the caloric curves in other dimensions of
space and point out particular dimensions.

\subsection{The dimension $d=2$}
\label{sec_d2}

\begin{figure}
\begin{center}
\includegraphics[clip,scale=0.3]{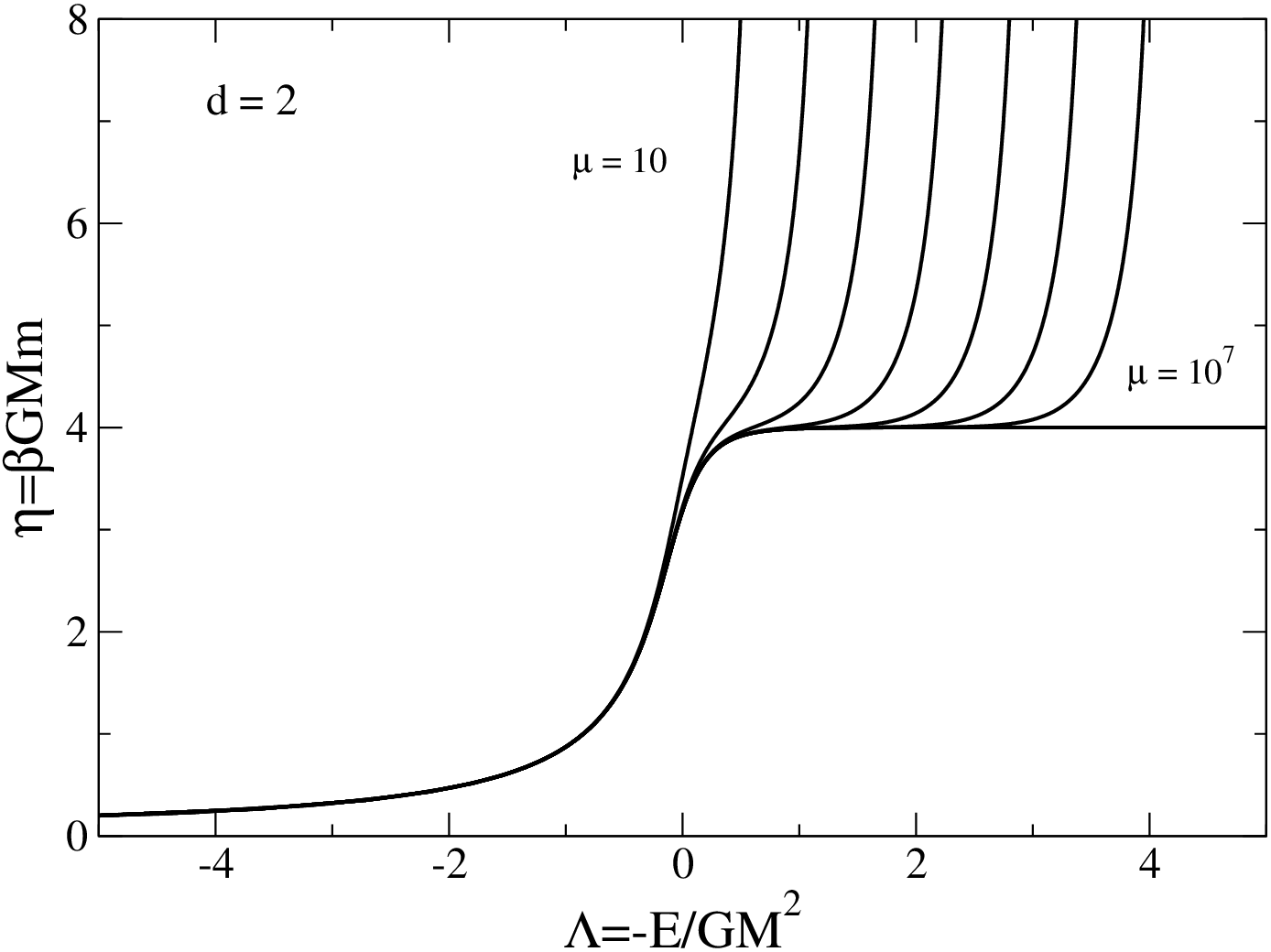}
\caption{Caloric curve in $d=2$ for
different values of the degeneracy parameter:
$\mu=10,100,10^3,10^4,10^5,10^6,10^7$. For $\mu\rightarrow +\infty$, we recover the
classical caloric curve displaying a critical temperature $T_{c}$. Below $T_{c}$
the system is expected to collapse and create a Dirac peak. When the particles experience
an exclusion constraint in position
space, the Dirac peak is replaced by a
``rocky core'' (homogeneous sphere).}
\label{calomultiD2}
\end{center}
\end{figure}

The caloric curve in $d=2$ is plotted in Fig. \ref{calomultiD2}. For
$\mu\rightarrow +\infty$, we recover the classical caloric curve of Fig. \ref{dim2}
displaying a critical temperature $k_{B}T_{c}=GMm/4$ \cite{sc}. Below
$T_{c}$ in the canonical ensemble, a classical gas experiences an isothermal collapse leading to
a Dirac peak \cite{sc}. If there is an exclusion constraint in position space,
the collapse stops when the system becomes degenerate. In that case, the
Dirac peak is replaced by a ``rocky core'' (homogeneous sphere) surrounded by a dilute
halo.\footnote{We note that the fluctuations of energy close to
the critical temperature $T_c$ are huge since the specific heat tends to zero
and $C=dE/dT=k_B\beta^2(\langle E^2\rangle-\langle E\rangle^2)$. This may lead
to interesting phenomena.} At $T=0$, we have a pure homogeneous sphere without
halo. This is
the ground state of the system  corresponding to
the vertical asymptotes in Fig.  \ref{calomultiD2}. The minimum energy $E_{min}$ is given by
Eq. (\ref{deg6}).  As we have seen in the Introduction, there is no collapse (gravothermal
catastrophe) in the microcanonical ensemble for classical isothermal spheres in $d=2$ \cite{sc}. There exist a stable equilibrium state at all energies. At low energies, classical isothermal spheres all have the same temperature $T_c$ (see Fig. \ref{dim2}). When the particles experience an exclusion constraint in position space, this is no more true. The temperature decreases to zero as we approach the minimum energy $E_{min}$ (see Fig. \ref{calomultiD2}).

\subsection{The dimension $d=1$}
\label{sec_d1}

\begin{figure}
\begin{center}
\includegraphics[clip,scale=0.3]{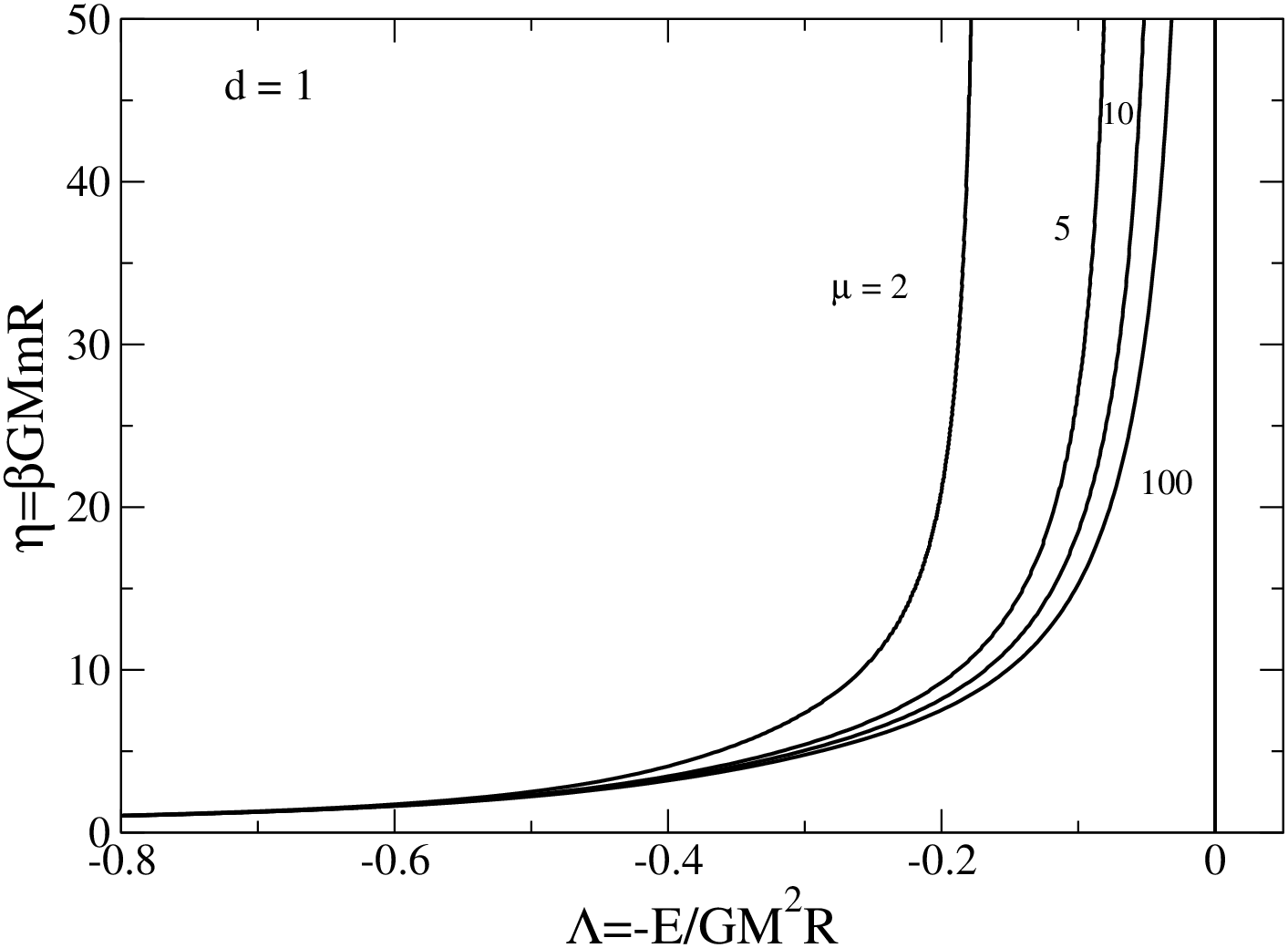}
\caption{Caloric curve in $d=1$ for different values of the degeneracy parameter (various system sizes).}
\label{calomultiD1}
\end{center}
\end{figure}

The caloric curve in $d=1$ is plotted in Fig. \ref{calomultiD1}. For $\mu\rightarrow +\infty$, we recover the classical caloric curve of Fig. \ref{dim1}.  The caloric curve $\beta(E)$ is monotonic so there is no phase transition. Therefore,
the change in the caloric curve due to the exclusion constraint in position space
is not very significant since an equilibrium state (global maximum of entropy or global minimum of
free energy) already exists for any accessible energy $E$ and any
temperature $T$ in the absence of small-scale regularization. The exclusion constraint, however,
changes the ground state of the system ($T=0$). The ground state  of the gas with
an exclusion constraint in position space is a homogeneous  sphere with energy $E_{min}$ given by Eq. (\ref{deg3}) instead of a Dirac peak with energy $E_{min}=0$ \cite{sc}. This corresponds to the asymptotes in
Fig. \ref{calomultiD1}.

\subsection{The dimension $d=10$}
\label{sec_d48}

\begin{figure}
\begin{center}
\includegraphics[clip,scale=0.3]{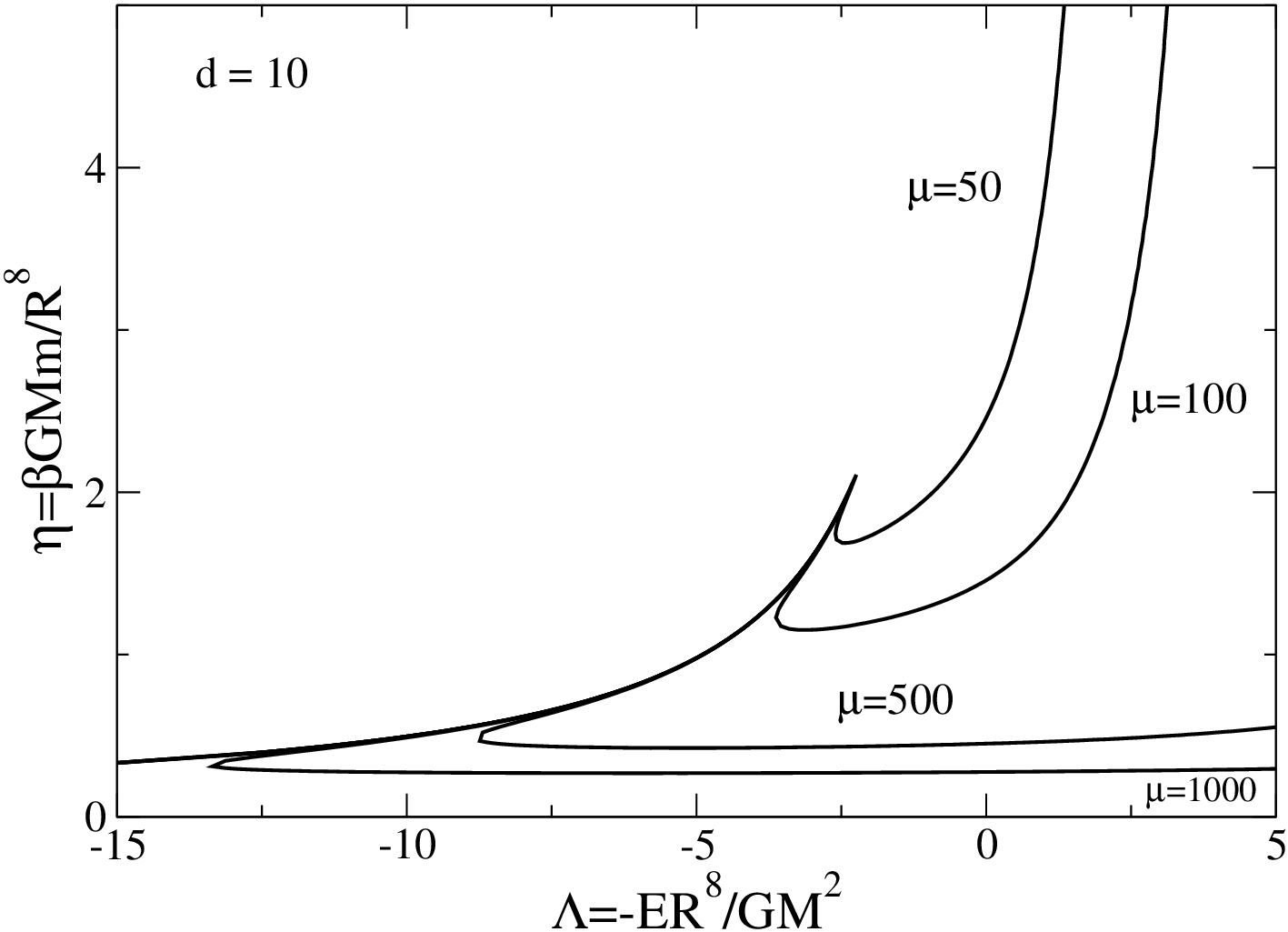}
\caption{Caloric curve in $d=10$ for different values of the degeneracy
parameter (various system sizes).}
\label{calod10}
\end{center}
\end{figure}

For $d\ge 10$, the classical spiral disappears as shown in Fig. \ref{calo2a10}. Therefore, the points of minimum energy and minimum temperature coincide. The series of equilibria in the presence of an exclusion constraint in position space are shown in Fig. \ref{calod10} for different values of $\mu$.

\section{Kinetic equations and theory of fluctuations}
\label{sec_kinetic}

Kinetic equations describing the evolution of self-gravitating systems in the
presence of microscopic  constraints changing the form of the entropy are given
in \cite{nfp,emergence} in both microcanonical and canonical ensembles. Here, 
we restrict ourselves to overdamped self-gravitating Brownian particles in the
canonical ensemble (fixed $T$) experiencing an exclusion constraint in position
space. The generalized Smoluchowski equation associated with the Fermi-Dirac
free energy
\begin{eqnarray}
&&F[\rho]=\frac{1}{2}\int\rho\Phi \, d{\bf r}-\frac{d}{2}Nk_BT\ln \left (\frac{2\pi k_B T}{m}\right )\nonumber\\
&+&k_BT\frac{\sigma_0}{m}\int \biggl\lbrace \frac{\rho}{\sigma_0}\ln\left (\frac{\rho}{\sigma_0}\right )
+\left (1-\frac{\rho}{\sigma_0}\right )\ln \left (1-\frac{\rho}{\sigma_0}\right )\biggr\rbrace\, d{\bf r}\nonumber\\
\label{kin1}
\end{eqnarray}
can be written as \cite{nfp,emergence}:
\begin{eqnarray}
\label{kin2}
\frac{\partial\rho}{\partial t}=\nabla\cdot \left\lbrack \frac{1}{\xi}(\nabla p+\rho\nabla\Phi)\right\rbrack,
\end{eqnarray}
where $p(\rho)$ is the barotropic pressure given by Eq. (\ref{eos2}). Using
\begin{eqnarray}
\label{kin3}
\nabla\frac{\delta F}{\delta\rho}=\frac{k_B T}{m}\frac{1}{\rho(1-\rho/\sigma_0)}\nabla\rho+\nabla\Phi,
\end{eqnarray}
we can rewrite Eq. (\ref{kin2}) as
\begin{eqnarray}
\label{kin4}
\frac{\partial\rho}{\partial t}=\nabla\cdot \left (\frac{1}{\xi}\rho\nabla\frac{\delta F}{\delta\rho}\right ).
\end{eqnarray}
Another possible form of the generalized Smoluchowski equation
is \cite{nfp,emergence}:
\begin{eqnarray}
\label{kin5}
\frac{\partial\rho}{\partial t}=\nabla\cdot \left\lbrack D\left (\nabla \rho+\beta\rho m (1-\rho/\sigma_0)\nabla\Phi\right )\right\rbrack,
\end{eqnarray}
with $D=1/(\xi\beta m)$. In terms of the free energy (\ref{kin1}), we get
\begin{eqnarray}
\label{kin6}
\frac{\partial\rho}{\partial t}=\nabla\cdot \left\lbrack D\beta\rho m (1-\rho/\sigma_0)\nabla\frac{\delta F}{\delta\rho}\right\rbrack.
\end{eqnarray}
These equations can be viewed as nonlinear mean field Fokker-Planck equations.\footnote{Generalized Fokker-Planck equations are associated with the notion of generalized thermodynamics \cite{nfp}. They arise when the transition probability from one state to another depends on the density of the arrival state due to microscopic constraints. This leads to generalized forms of entropy such as the Fermi-Dirac entropy in position space.} They respect the exclusion constraint $\rho({\bf r},t)\le\sigma_0$ at any time. In Eq. (\ref{kin2}), the pressure becomes infinite when $\rho\rightarrow \sigma_0$. In Eq. (\ref{kin5}), the mobility vanishes when $\rho\rightarrow \sigma_0$. These equations satisfy an $H$-theorem for the Fermi-Dirac free energy (\ref{kin1}) \cite{nfp,emergence}. As a result, they converge, for $t\rightarrow +\infty$, towards a (local) minimum of free energy at fixed mass. If several minima exist for the same values of the constraints, the selection depends on the initial condition and on a notion of basin of attraction. This is the case for our model of self-gravitating systems with an exclusion constraint in position space since it leads to canonical first order phase transitions in $d=3$ when the system is sufficiently large ($\mu\ge \mu_{CCP}=32.4$). The relaxation equations (\ref{kin2}) and (\ref{kin5}) could be used to describe an hysteresis cycle in the canonical ensemble by varying the temperature between $T_c$ and $T_*$ (see Fig. \ref{ETcano3summary}). This has been shown explicitly in \cite{crrs} for a gas of self-gravitating Brownian fermions.

Interestingly, the kinetic equations (\ref{kin2}) and (\ref{kin5}) are formally
similar to those describing the chemotaxis of bacterial populations. Indeed,
they can be viewed as generalized Keller-Segel models including an exclusion
constraint in position space (see Sec. 3.5 of \cite{nfp} and references
therein). These equations have been introduced phenomenologically in order to
prevent blow-up occurring in the ordinary Keller-Segel model. The kinetic 
equations (\ref{kin2}) and (\ref{kin5}) are also
similar to the relaxation equations proposed by Robert and Sommeria
\cite{rsmepp} (see also \cite{csr,stabvortex}) in their statistical mechanics of
2D turbulence where the coarse-grained vorticity must respect the constraint
$\overline{\omega}({\bf r},t)\le \sigma_0$ ($\sigma_0$ is the initial value
of the vorticity in the two levels case). Finally, the kinetic equations
(\ref{kin2}) and (\ref{kin5}) could  describe the dynamics of colloids at a
fluid interface driven by attractive capillary interactions (with possible
experimental applications) \cite{dominguez} when there is an excluded volume
around the particles.

The mean field Smoluchowski equation (\ref{kin2}) or (\ref{kin5}) is a deterministic equation valid in the $N\rightarrow +\infty$ limit. When the free energy $F[\rho]$ has several minima, and when $N$ is small, the system undergoes random transitions between these minima due to fluctuations. Such transitions can be described by the stochastic Smoluchowski equation \cite{hb5,emergence}. The stochastic
Smoluchowski equation corresponding to the form (\ref{kin2})
is \cite{emergence}:
\begin{eqnarray}
\label{kin7}
\frac{\partial\rho}{\partial t}=\nabla\cdot \left\lbrack \frac{1}{\xi}(\nabla p+\rho\nabla\Phi)\right\rbrack+\nabla\cdot \left (\sqrt{\frac{2k_B T\rho}{\xi}}{\bf R}\right ).
\end{eqnarray}
In terms of the free energy (\ref{kin1}), we get
\begin{eqnarray}
\label{kin8}
\frac{\partial\rho}{\partial t}=\nabla\cdot \left (\frac{1}{\xi}\rho\nabla\frac{\delta F}{\delta\rho}\right )+\nabla\cdot \left (\sqrt{\frac{2k_B T\rho}{\xi}}{\bf R}\right ).
\end{eqnarray}
The stochastic
Smoluchowski equation corresponding to the form (\ref{kin5})
is \cite{emergence}:
\begin{eqnarray}
\label{kin9}
\frac{\partial\rho}{\partial t}=\nabla\cdot \left\lbrack D\left (\nabla \rho+\beta\rho m (1-\rho/\sigma_0)\nabla\Phi\right )\right\rbrack\nonumber\\
+\nabla\cdot \left (\sqrt{2Dm\rho(1-\rho/\sigma_0)}{\bf R}\right ).
\end{eqnarray}
In terms of the free energy (\ref{kin1}), we get
\begin{eqnarray}
\label{kin10}
\frac{\partial\rho}{\partial t}=\nabla\cdot \left\lbrack D\beta\rho m (1-\rho/\sigma_0)\nabla\frac{\delta F}{\delta\rho}\right\rbrack\nonumber\\
+\nabla\cdot \left (\sqrt{2Dm\rho(1-\rho/\sigma_0)}{\bf R}\right ).
\end{eqnarray}

Eqs. (\ref{kin8}) and (\ref{kin10}) may be interpreted as stochastic Langevin equations for the smooth density field $\rho({\bf r},t)$. The corresponding Fokker-Planck equation for the probability density $P[\rho,t]$ of the density profile $\rho({\bf r},t)$ at time $t$ is
\begin{eqnarray}
\label{kin11}
&&\xi\frac{\partial P}{\partial t}[\rho,t]\nonumber\\
&=&-\int\frac{\delta}{\delta\rho({\bf r},t)}\left\lbrace \nabla\cdot \rho\nabla\left\lbrack k_B T\frac{\delta}{\delta\rho}+\frac{\delta F}{\delta\rho}\right\rbrack P[\rho,t]\right\rbrace\, d{\bf r}\nonumber\\
\end{eqnarray}
for Eq. (\ref{kin8}) and
\begin{eqnarray}
\label{kin12}
\frac{\partial P}{\partial t}[\rho,t]=-\int\frac{\delta}{\delta\rho({\bf r},t)}\biggl\lbrace \nabla\cdot D\beta m \rho(1-\rho/\sigma_0)\nonumber\\
\times\nabla\left\lbrack k_B T\frac{\delta}{\delta\rho}+\frac{\delta
F}{\delta\rho}\right\rbrack P[\rho,t]\biggr\rbrace\, d{\bf r}
\end{eqnarray}
for Eq. (\ref{kin10}). Their stationary solution is the canonical distribution (\ref{lifetime1}).
The form of the noise in Eqs. (\ref{kin8}) and (\ref{kin10}) can be obtained
from the theory of fluctuating hydrodynamics \cite{hb5,emergence}. It can also be
determined in order to recover the canonical distribution (\ref{lifetime1}) at
equilibrium. We note that the noise is multiplicative since it depends on
$\rho({\bf r},t)$. By a suitable rescaling (see Sec. \ref{sec_thermo}), one can
show that the noise term is of order $1/\sqrt{N}$. Taking $N\rightarrow +\infty$
amounts to neglecting the noise in Eqs. (\ref{kin8}) and (\ref{kin10}). In that
case, we recover the mean field Smoluchowski equation (\ref{kin4}) or
(\ref{kin6}) that relaxes towards a (local) minimum of free energy and remains
in that state for ever. When finite $N$ effects are taken into account, the
stochastic Smoluchowski equations (\ref{kin8}) and (\ref{kin10}) can describe
random transitions from a minimum of free energy to another. A detailed study of
these random transitions has been performed in \cite{ssp} for a 1D
self-gravitating system with a modified Poisson equation. In that case, the
phase transition is second order. The stochastic equations (\ref{kin8}) and
(\ref{kin10}) could be used to study random transitions in a self-gravitating
system with an exclusion constraint in position space presenting first order
phase transitions in $d=3$. In particular, in the region of phase transitions
$T_c<T<T_*$, we should be able to see a cycle of random events corresponding to
a succession of ``explosions'' and ``collapses''. These
stochastic equations could also be used to investigate the effect of
fluctuations close to $T_c$ in $d=2$ that are expected to be very important.
This will be considered in future works.

\section{Conclusion} \label{sec_conclusion}

We have considered a system of self-gravitating particles with an exclusion constraint in position space and we have studied the nature of phase transitions as a function of the size of the system and the dimension of space.  For $d\le 2$, there is no phase transition. For $d>2$, phase transitions can take place between a ``gaseous'' phase  unaffected by the exclusion constraint and  a ``condensed'' phase dominated by this constraint. For large systems there exist microcanonical and canonical first order phase transitions. For intermediate systems, only canonical first order phase transitions are present. For small systems there is no phase transition at all.

Clearly, the most interesting situation is the case of ``large'' systems in $d=3$ dimensions.
When an exclusion constraint is properly accounted for, there exist an equilibrium state 
(global maximum of entropy or global minimum of free energy) for each value of accessible energy and temperature. At high temperatures and high energies, the system is in a gaseous phase and the exclusion constraint is completely negligible. At some transition
temperature $T_{t}$ in the canonical ensemble, or at some transition energy $E_{t}$ in the microcanonical ensemble, a first order phase transition is expected to
occur and drive the system in a condensed phase.  However,
gaseous states are still metastable, and long-lived,  below this point so the first order phase transition does  not take place in practice when $N\gg 1$.
 Gravitational collapse rather occurs at a smaller critical
temperature $T_{c}$ (Emden temperature)  or at a smaller critical energy
$E_{c}$ (Antonov energy) at which the metastable branch disappears (spinodal point). This corresponds to a zeroth order phase transition. The end-state of the collapse is a compact object with a ``core-halo'' structure. Typically, it is made of a ``rocky core'' (homogeneous sphere) surrounded
by an  ``atmosphere''. This configuration is similar to the structure of a giant gaseous planet.
   The condensate results from the balance between
the gravitational attraction  and the pressure due to the close packing of the particles.

These results are similar to those obtained for the self-gravitating Fermi gas \cite{pt,ispolatov,rieutord,fermionsd,ijmpb} with, however, some differences. First of all, the form of the caloric curves and the values of the critical parameters differ quantitatively since the equilibrium states are different. In particular, in the self-gravitating Fermi gas, the completely degenerate configuration is a ``white dwarf'' equivalent to a polytrope $n=d/2$ while, in the present model, the completely degenerate configuration is a ``rocky core'' (homogeneous sphere) equivalent to a polytrope $n=0$. The  later is stable in any dimension of space while the former is stable only in dimensions $d<4$. Quantum mechanics (Pauli exclusion principle) is not able to stabilize matter against gravitational collapse in a space of dimension $d\ge 4$ \cite{wdd,fermionsd}. This implies that the dimension $d=3$ of our universe is very particular in this respect. For the
model studied in this paper, the small-scale constraint  stabilizes the system in any dimension of space. This is because the exclusion constraint acts directly in physical space, instead of phase space.

Additional differences between the present model and the self-gravitating Fermi gas model appear when the system is rotating. The rotating self-gravitating Fermi gas has been studied by Chavanis and Rieutord \cite{rieutord} and the rotating self-gravitating gas with an exclusion constraint in position space has been studied by Votyakov {\it et al.} \cite{votyakov1,votyakov2}\footnote{In their study, they considered only large excluded volumes corresponding to small values of $\mu$. As a result they did not ``see'' the dinosaur's necks and the corresponding microcanonical first order  phase transitions that appear at large $\mu$ (see \cite{grossdino} and Sec. 5.13 of \cite{ijmpb}).} who obtained different results. These differences can be understood as follows.  It is known that polytropic stars with an index $n<0.808$  bifurcate to non-axisymmetric configurations at high rotations (fission) while  polytropic stars with an index $n>0.808$ remain axisymmetric \cite{james}. As a result,  rapidly rotating self-gravitating systems with an exclusion constraint in position space ($n=0$) become non-axisymmetric and break into a {\it double cluster}  \cite{votyakov1,votyakov2} while rapidly rotating self-gravitating fermions ($n=3/2$) remain axisymmetric and develop a {\it cusp} at the equator when the Keplerian limit is reached \cite{rieutord}.

The model of self-gravitating fermions has application for white dwarfs, neutron stars, and dark matter halos made of massive neutrinos. The present model could describe (in an elementary way) the formation of planetesimals and gaseous planets in the solar nebula by gravitational collapse. On the other hand, the kinetic equations of Sec. \ref{sec_kinetic} can provide generalized models of chemotaxis in biology. They could also be used to describe the dynamics of colloids at a fluid interface driven by attractive capillary interactions when there is an excluded volume around the particles. We have also mentioned some analogies with the MRS theory of 2D turbulence. Therefore, the model studied in this paper presents a lot of applications in astrophysics, fluid mechanics, biology, and colloid science.

\appendix

\section{Computation of the entropy}
\label{sec_entropy}

In this Appendix, we establish the expressions (\ref{p17}) and (\ref{p17b}) of the entropy. The Fermi-Dirac entropy in position space is given by
\begin{equation}
S=-k_B\int C(\rho)\, d{\bf r}+\frac{d}{2}Nk_B\ln\left (\frac{2\pi k_B T}{m}\right )+\frac{d}{2}Nk_B
\label{entropy1}
\end{equation}
with
\begin{equation}
\label{entropy2}
C(\rho)=\frac{\sigma_0}{m}\left\lbrack\frac{\rho}{\sigma_0}\ln\left (\frac{\rho}{\sigma_0}\right )+\left (1-\frac{\rho}{\sigma_0}\right )\ln\left (1-\frac{\rho}{\sigma_0}\right )\right\rbrack\, d{\bf r}.
\end{equation}
We have to compute
\begin{equation}
\label{entropy3}
S_1=-k_B\int C(\rho)\, d{\bf r}.
\end{equation}
Using Eqs. (\ref{p3}) and (\ref{entropy2})  we find that
\begin{equation}
\label{entropy4}
S_1/k_B=N\ln k+\int\frac{\rho}{m}\psi\, d{\bf r}+\frac{\sigma_0}{m}\int \ln\left (1+\frac{1}{k}e^{-\psi}\right )\, d{\bf r}.
\end{equation}
Since $\psi=\beta m(\Phi-\Phi_0)$, we get
\begin{equation}
\label{entropy5}
S_1/k_B=N\ln k+2\beta W-\beta M \Phi_0+\frac{\sigma_0}{m}\int \ln\left (1+\frac{1}{k}e^{-\psi}\right )\, d{\bf r},
\end{equation}
where $W$ is the potential energy. The central potential may be obtained from the relation
\begin{equation}
\label{entropy6}
\Phi_0=\Phi(R)-\frac{\psi(\alpha)}{\beta m},
\end{equation}
where $\Phi(R)$ is given by Eqs. (\ref{p12})-(\ref{p13}). Therefore
\begin{eqnarray}
\label{entropy7}
S_1/k_B=N\ln k+2\beta W+\epsilon\frac{1}{d-2} \frac{\beta GM^2}{R^{d-2}}+N\psi(\alpha)\nonumber\\
+\frac{\sigma_0}{m}\int \ln\left (1+\frac{1}{k}e^{-\psi}\right )\, d{\bf r}
\end{eqnarray}
with $\epsilon=1$ if $d\neq 2$ and  $\epsilon=0$ if $d=2$.

According to the virial theorem (see Appendix \ref{sec_virial}), we have
\begin{equation}
\label{entropy8}
d\frac{k_B T}{m}\sigma_0 \int \ln\left (1+\frac{1}{k}e^{-\psi}\right )\,
d{\bf r}-{\cal V}_d=dp(R)V_d R^d.
\end{equation}
As a result
\begin{equation}
\label{entropy9}
\frac{S_1}{Nk_B}=\ln k+\frac{2\beta W}{N}+\frac{\epsilon}{d-2}\eta+\psi(\alpha)+\frac{\beta}{dN}{\cal V}_d+\frac{\beta}{N}p(R)V_dR^d.
\end{equation}
According to Eqs. (\ref{eos2}) and (\ref{p3}), the pressure on the box can be written
as
\begin{equation}
\label{entropy10}
p(R)=\frac{\sigma_0}{\beta m} \ln\left \lbrack 1+\frac{1}{k}e^{-\psi(\alpha)}\right \rbrack.
\end{equation}
Substituting Eq. (\ref{entropy10}) in Eq. (\ref{entropy9}) and recalling that $V_d=S_d/d$, we get
\begin{eqnarray}
\label{entropy11}
\frac{S_1}{Nk_B}=\ln k+\frac{2\beta W}{N}+\frac{\epsilon}{d-2}\eta+\psi(\alpha)+\frac{\beta}{dN}{\cal V}_d\nonumber\\
+\mu \ln\left \lbrack 1+\frac{1}{k}e^{-\psi(\alpha)}\right \rbrack.
\end{eqnarray}

For $d\neq 2$, using Eq. (\ref{virial2}), we obtain
\begin{eqnarray}
\label{entropy12}
\frac{S_1}{Nk_B}=\ln k+\frac{d+2}{d}\frac{\beta W}{N}+\frac{1}{d-2}\eta+\psi(\alpha)\nonumber\\
+\mu \ln\left \lbrack 1
+\frac{1}{k}e^{-\psi(\alpha)}\right \rbrack.
\end{eqnarray}
Using Eq. (\ref{phen9}), we find that
\begin{eqnarray}
\label{entropy13}
\frac{S_1}{Nk_B}=\ln k-\frac{d+2}{d}\eta\Lambda+\frac{1}{d-2}\eta+\psi(\alpha)\nonumber\\
+\mu \ln\left \lbrack 1+\frac{1}{k}e^{-\psi(\alpha)}\right \rbrack-\frac{d+2}{2}.
\end{eqnarray}
Finally, the total entropy can be written as Eq. (\ref{p17}).

For $d=2$, using Eq. (\ref{virial3}), we obtain
\begin{equation}
\label{entropy14}
\frac{S_1}{Nk_B}=\ln k+\frac{2\beta W}{N}+\frac{\eta}{4}+\psi(\alpha)+\mu \ln\left \lbrack 1+\frac{1}{k}e^{-\psi(\alpha)}\right \rbrack.
\end{equation}
Using Eq. (\ref{phen9}), we find that
\begin{equation}
\label{entropy15}
\frac{S_1}{Nk_B}=\ln k-2\eta\Lambda+\frac{\eta}{4}+\psi(\alpha)+\mu \ln\left \lbrack 1+\frac{1}{k}e^{-\psi(\alpha)}\right \rbrack-2.
\end{equation}
Finally, the total entropy can be written as Eq. (\ref{p17b}).

\section{Virial theorem}
\label{sec_virial}

The virial of the gravitational force is defined by
\begin{equation}
\label{virial1}
{\cal V}_d=\int\rho {\bf r}\cdot \nabla\Phi\, d{\bf r}.
\end{equation}
We can show that \cite{virial1}:
\begin{equation}
\label{virial2}
{\cal V}_d=-(d-2)W,\qquad (d\neq 2),
\end{equation}
\begin{equation}
\label{virial3}
{\cal V}_2=\frac{GM^2}{2},\qquad (d=2),
\end{equation}
where $W$ is the potential energy. Substituting the condition of hydrostatic equilibrium (\ref{eos1})
in Eq. (\ref{virial1}), and integrating by parts, we get
\begin{equation}
\label{virial5}
d\int p\, d{\bf r}-{\cal V}_d=\oint p{\bf r}\cdot d{\bf S}
\end{equation}
which is the general expression of the virial theorem in a box. If $p$ is uniform on the box
with the value $p_b$, this equation can be rewritten as
\begin{equation}
\label{virial6}
d\int p\, d{\bf r}-{\cal V}_d=d p_b V.
\end{equation}
For the equation of state (\ref{eos2}), using Eq. (\ref{p3}), we find that
\begin{equation}
\label{virial8}
\int p\, d{\bf r}=\frac{\sigma_0}{\beta m}\int_0^{\alpha}\ln\left (1+\frac{1}{k}e^{-\psi}\right )S_d \left (\frac{R}{\alpha}\right )^d \xi^{d-1}\, d\xi.
\end{equation}
Therefore, the virial theorem leads to the identity
\begin{eqnarray}
\label{virial9}
d\frac{\sigma_0}{\beta m}S_d \left (\frac{R}{\alpha}\right )^d\int_0^{\alpha}\ln\left (1+\frac{1}{k}e^{-\psi}\right ) \xi^{d-1}\, d\xi-{\cal V}_d\nonumber\\
=d p(R) V_d R^d,
\end{eqnarray}
where $V_d=S_d/d$ is the volume of a unit sphere in $d$ dimensions.

\section{The equation for the mass profile}
\label{sec_massprofile}

For spherically symmetric distributions, using the Gauss theorem, the (generalized) Smoluchowski-Poisson system can be reduced to a single equation for the mass profile $M(r,t)=\int_0^r\rho(r',t)S_d {r'}^{d-1}\, dr'$ \cite{sc}. For Eq. (\ref{kin5}) corresponding to an exclusion constraint in position space, we get
\begin{eqnarray}
\label{kin6b}
\xi\frac{\partial M}{\partial t}=\frac{k_B T}{m}\left (\frac{\partial^2M}{\partial r^2}-\frac{d-1}{r}\frac{\partial M}{\partial r}\right )\nonumber\\
+\frac{GM}{r^{d-1}}\frac{\partial M}{\partial r}\left (1-\frac{1}{S_d\sigma_0 r^{d-1}}\frac{\partial M}{\partial r}\right ).
\end{eqnarray}
In the classical limit $\sigma_0\rightarrow +\infty$, we recover the results of \cite{sc}.

\end{document}